\documentclass[%
 reprint,
 amsmath,amssymb,
 aps,
]{revtex4-2}

\usepackage{graphicx}
\usepackage{dcolumn}
\usepackage{bm}
\usepackage{comment}
\graphicspath{{figs/}}
\usepackage{tikz}
\usetikzlibrary{shapes}
\usetikzlibrary{positioning}

\begin{document}

\preprint{APS/123-QED}

\title{Investigating the origins of fluctuation forces on plates immersed in turbulent flows}

\author{Daniel Putt}
\affiliation{Department of Mechanical Engineering, University of Houston, Houston 77004, USA}

\author{Rodolfo Ostilla Mónico}
\affiliation{Department of Mechanical Engineering, University of Houston, Houston 77004, USA}
\affiliation{Escuela de Ingenier\'ia, Universidad de C\'adiz, Spain}%


\date{\today}

\begin{abstract}
A net force can arise on objects which lie in systems with complex energy partitions, even if the system is on average stationary. These forces are usually called fluctuation forces, as they arise due to the objects modifying the character of the fluctuations within the system. We continue the investigation of Spandan \emph{et al.}, \textit{Sci. Adv., 6(14), eaba0461} (2020), who found an attractive fluctuation force between two parallel square plates in homogeneous isotropic turbulence (HIT). We conduct simulations which systematically vary the plate size and Reynolds number.  At $Re_\lambda=100$ small plates show a monotonic force dependence, with a maximum force for the smallest plate separations, while medium and large plates show a non-monotonic behaviour of the force with maximum attractive force at intermediate separations. We find that energy-related statistics cannot explain the dependence on plate separation of the force, but that statistics related to vorticity do show qualitative variations around the plate separation corresponding to the maximum force. This suggests that the role of plates in affecting intense vorticity structures is critical to the behaviour of the force. By decreasing $Re_\lambda$, we show that removing vortex stretching decreases the attractive force, but does not completely eliminate it, and find that the local maximum at intermediate distances becomes a local minimum. This confirms that the attractive force is related to vorticity, while suggesting that a second mechanism is present- supporting the proposal for a two-fold origin from earlier work: the plates both restrict the presence of energy structures in the slit and pack intense vortical structures which stretch each other causing the pressure to drop. 
\end{abstract}

\maketitle

\onecolumngrid

\section{Introduction}
\label{sec:intro}

Long-range forces can arise when fluctuating fields are contained between surfaces. The classical Casimir force is the paradigmatic example of such forces \cite{casimir1948}. This force arises when two metal plates are placed in a vacuum a few micrometers apart. The two plates are found to attract each other, even in the absence of electrical charges on the plates \cite{lamoreaux1997}. This is due to the quantum fluctuations of the electromagnetic field present in a vacuum. A single plate does not affect these fluctuations, but when two plates are placed nearby to each other, many modes of fluctuation of the field are restricted. The energy of the system then becomes a function of plate separation- and because of this there must a force force arises that propels the plates to minimize the energy \cite{casimir1948}.

The classical Casimir force arises in a vacuum, but Casimir-like forces can arise in other media \cite{fisher1978}. For this to happen, it is essential that the underlying fluctuations have certain properties, as otherwise no force would arise. The basic ingredient is that the fluctuating field results in a non-trivial spatial dependence of energy \cite{kardar99,vella2017}. It is often a challenge to find these forces in Nature as there are not many media where they appear, and when they do, they are often not easy to measure. 
The first medium where they were postulated to exist was in binary mixtures close to their critical point \cite{fisher1978}. The long-range order which results from closeness to a critical point makes the otherwise random thermal fluctuations into a fluctuating field which satisfies the required properties. Experimental observation of the force had to wait several decades, but the hypothesis was eventually confirmed \cite{hertlein2008}. 
More recent experiments have uncovered Casimir-like forces in a wide variety of systems, such as run-and-tumble particles \cite{ray2014,ni2015} and  colloidal spheres in mixtures \cite{hanke1998,hertlein2008,gambassi2009}. In particular, using numerical simulations we showed in Ref.~\cite{spandan2020fluctuation} the existence of an attractive turbulent fluctuation force in homogeneous isotropic turbulence (HIT), an idealized turbulent state which is numerically simulated using a triply-periodic computational domain \cite{ishihara2009}. The presence of this force in two-dimensional HIT was later confirmed experimentally \cite{davoodianidalik2022fluctuation}.

This finding becomes more relevant when one takes into account recent work that shows that a large class of biological fluids comprising microbial suspensions exhibit striking analogies with turbulent flows \cite{wensink_meso-scale_2012,bratanov2015new,kokot2017active,mickelin2018anomalous,alert2022active}. 
Other examples of active flows include artificial self-propelled particles \cite{nishiguchi2015mesoscopic,karani2019tuning}, which also show fluctuation-force types of interaction \cite{ni2015tunable}. However, coarse-grained models of these types of flows contain a number of parameters and unknown quantities \cite{marchetti2013,wensink_meso-scale_2012}. As a result, when fluctuation forces are found, they show very complicated behaviour as a function of plate distance \cite{ni2015}. In contrast, the fluctuation forces found in HIT have a simpler dependence on plate distance \cite{spandan2020fluctuation}, and in place of the complex and not fully understood energy transfer mechanisms present in active flows or colloidal spheres, the energy transfer mechanisms in hydrodynamic turbulence are well studied \cite{urielfrisch, jim12, marusic2019}. 

But beyond the fact that Casimir-like forces exist in HIT, not much more is known. In Ref.~\cite{spandan2020fluctuation}, we proposed that the force arises due to a complex interaction between the energy-containing scales and the dissipative scales which resulted on a force that behaves non-monotonically depending on the plate distance. 
When the distance between the plates was changed, the energy-containing structures in HIT were modified affecting the overall pressure on the plates. In addition, at intermediate plate distances, the intense vorticity structures (worms) are forced to interact in close vicinity between the plates, affecting the pressure distributions in the slit and increasing the attractive force between the plates. We proposed that the combination of these two effects caused a non-monotonic attractive force with a complex Reynolds number dependence. A somewhat similar mechanism was experimentally found for two-dimensional turbulence, despite the fact that two-dimensional HIT has a reverse energy cascade from small to large scales, and no vortex stretching -the effect hypothesized to cause the drop in pressure- can be found. Ref.~\cite{davoodianidalik2022fluctuation} showed that the force was generated by a restriction of the length-scales coherent structures could take. This resulted in a resonance phenomenon at the flow forcing scale which lead to complex short-range interactions, an energy partition, and the generation of the fluctuation force.

The fact that the force was found when vortices were not stretching each other at close distances leaves open several questions which we intend to explore in this manuscript. The non-monotonicity of the force hits at the fact that two competing effects happen in the flow, and by exploring a larger parameter space than in Ref.~\cite{spandan2020fluctuation} we hope to fully reveal them. In Ref.~\cite{spandan2020fluctuation}, the plate size was fixed, and the Reynolds number was varied. Due to the forcing method used, this in effect meant that the ratio between the sizes of the energy containing eddies and the plates were kept constant. Furthermore, all the cases simulated contained fully developed turbulence, where a length-scale separation between the pressure sources and the energy containing scales was present. The question of the minimum scale separation required to produce a non-monotonic force was also left unanswered.

To answer these questions, we conduct two new simulation campaigns. First, we vary the plate sizes relative to the energy-containing structures, while keeping the Reynolds number (and hence the smallest dissipative structures fixed). By doing this, we show that a minimum plate size is needed for the force to show non-monotonicity. By analyzing in detail the flow statistics related to energy and vorticity, we show the link between non-monotonicity and vortex stretching. Second, we progressively reduce the Reynolds number down to a point where there is no length-scale separation between the energy containing structures and the dissipative structures, to analyze what is the minimum scale separation for the non-monotonicity of the fluctuation force to arise. Through this, we shed further light on the general behavior of the fluctuation force: we find that the increased attraction at intermediate plate distances is due to the vortex stretching mechanism, which appears once the Reynolds number is large enough. With this new simulations, we give further evidence to support many of the speculative statements in Ref.~\cite{spandan2020fluctuation}.

The manuscript is organized as follows: in Section \ref{sec:methods} we discuss the numerical methods used. We present detailed results and discussion for different plate sizes in Section \ref{sec:size}, and the results for varying Reynolds number in Section \ref{sec:rey}. We finish off by summarizing our findings and giving an outlook for further research in Section \ref{sec:conc}. 

\section{Methods}
\label{sec:methods}

For all simulations, we directly simulate homogeneous isotropic turbulence using the incompressible Navier-Stokes equations in a triply periodic cube with a periodic length $\mathcal{L}$:

\begin{equation}
     \displaystyle\frac{\partial\textbf{u}}{\partial t} + \textbf{u}\cdot\nabla\textbf{u} = -\rho^{-1}\nabla p +\nu\nabla^2 \textbf{u}+ \textbf{f} 
 \label{eq:ns}
\end{equation}

\begin{equation}
 \nabla\cdot \textbf{u}=0
 \label{eq:compress}
\end{equation}

\noindent where $\textbf{u}$ is the velocity, $t$ is time, $\rho$ is the fluid density, $\nu$ is the fluid kinematic viscosity and $\textbf{f}$ is a body force which is composed of the immersed boundary method (IBM) forcing used to simulate the plates ($\textbf{f}_s$) and a random forcing used to force the flow ($\textbf{f}_f$). Two rigid square plates of length $l_p$ and zero thickness are placed inside the computational domain parallel to each other at a distance of $d$. A schematic of the resulting system is shown in Figure \ref{fi:sch}. We have chosen to limit ourselves to the Navier-Stokes equations, and do not consider any extensions similar to those used to study active media \cite{wensink_meso-scale_2012}. The reasons for this choice are explained in Appendix \ref{sec:app}.

In the same manner as Ref.~\cite{spandan2020fluctuation}, the system of equations \ref{eq:ns}-\ref{eq:compress} is solved using an energy-conserving second-order centered finite difference scheme with fractional time stepping. An explicit low-storage third-order Runge-Kutta scheme is used to discretize the nonlinear terms, while an implicit Crank-Nicholson scheme is used for the viscous terms. As mentioned above, the flow is forced through a large-scale force vector $\textbf{f}_f$, which forces all modes whose wavenumber $\kappa$ is smaller than $\kappa_f$. In practice this is taken as $\kappa_f/\kappa_1=2.3$ where $\kappa_{1}=2\pi/\mathcal{L}$ is the base wavenumber in any direction. The instantaneous magnitude and direction of this force is calculated based on random processes which drive the time evolution of these selected modes based on a target energy flux $\epsilon^*$ and a force correlation time $T_L$. Additional details on the forcing scheme and its corresponding parameters can be found in the study by Eswaran and Pope \cite{eswaran1988examination} and the study by Chouippe and Uhlmann \cite{chouippe2015forcing}. We note that this method was chosen over other HIT forcing methods, such as a forcing proportional to the existing velocity field, because it has been shown to avoid artifacts when combined with the immersed boundary method \cite{chouippe2015forcing}. Further confidence in our choice of forcing was given by showing that the fluctuation forces persists over several eddy turnover times once the forcing is turned off, dropping off as the fluid becomes less energized and fluctuations decrease \cite{spandan2020fluctuation}, even if we note that the energy injection will still be anisotropic due to the characteristics of the flow between the plates and could still play a role.

Depending on the choice of $\epsilon^*$ and $T_L$, we obtain flow fields of varying turbulent intensities. To characterise the turbulence, we conduct a simulation with no plates present to obtain the equivalent Taylor-Reynolds number $Re_\lambda=u^\prime \lambda/\nu$, where $u^\prime$ is the root-mean-square velocity in one direction, $\lambda=u^\prime\sqrt{15\nu/\epsilon}$ the Taylor microscale and $\epsilon$ the actual time-averaged energy dissipation of the flow. Table \ref{tab:my_label} shows the simulated (non-dimensional) values of $\epsilon^*$, $\kappa_f$, and $T_L$, and the resulting $Re_\lambda$ for the simulations presented below. The table also includes the average Kolmogorov length-scale $\eta_K=(\nu^3/\epsilon)^{1/4}$ and the average integral, or decorrelation length-scale $L=k^{3/2}/\epsilon$ of the simulations, where $k$ is the flow kinetic energy $k=\frac{1}{2}\rho(u_x^2+u_y^2+u_z^2)$. 

The spatial discretization of the domain is performed using a cubic uniform grid with $360^3$ points for all $Re_\lambda$ considered in Table \ref{tab:my_label}. This resolution ensures that the flow is well resolved as $\kappa_{\max} \eta_K > 2$, where $\kappa_{\max}$ is the maximum wavenumber in the flow in a single direction. Earlier simulations  for $Re_\lambda=140$ taken from Ref.~\cite{spandan2020fluctuation} have a resolution of $480^3$. The time step is dynamically chosen so that the maximum Courant-Friedrich-Lewy condition number (CFL) is $1.2$. 

The influence of rigid plates on the surrounding fluid is simulated through the force $\textbf{f}_s$, calculated using an immersed boundary method (IBM) based on the moving least squares (MLS) approximation \cite{spandan2017parallel}. The IBM has the benefit of not needing to recreate or update the mesh, since there is a translation operation between the Eulerian mesh and the immersed body. This method is also useful due to the ease at which different objects can be placed in the domain. While we have kept our objects rigid and in place for this paper, the MLS formulation of IBM also allows for deformations in the object during simulations, and avoids mesh regeneration. Furthermore, it allows the simulation of objects of zero thickness as the ones used here. The immersed plates have been discretized using $\sim 10^4$ triangular computational elements with low skewness. The normal forces acting on the immersed rigid plates are computed from the pressure interpolated on the individual triangular elements on both sides.

\newcommand{\Depth}{4}
\newcommand{\Height}{4}
\newcommand{\Width}{4}
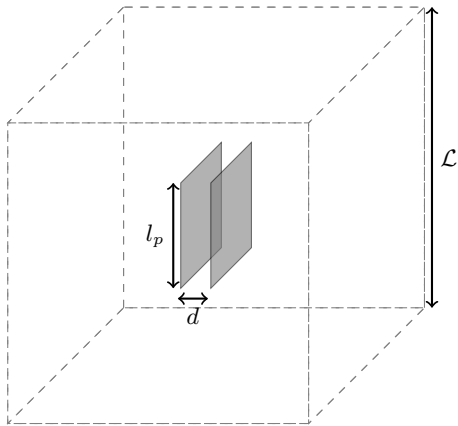
\begin{figure}
\begin{tikzpicture}
\coordinate (O) at (0,0,0);
\coordinate (A) at (0,\Width,0);
\coordinate (B) at (0,\Width,\Height);
\coordinate (C) at (0,0,\Height);
\coordinate (D) at (\Depth,0,0);
\coordinate (E) at (\Depth,\Width,0);
\coordinate (F) at (\Depth,\Width,\Height);
\coordinate (G) at (\Depth,0,\Height);
\coordinate (As) at (1.8,1.3,2.7);
\coordinate (Bs) at (1.8,2.7,2.7);
\coordinate (Cs) at (1.8,2.7,1.3);
\coordinate (Ds) at (1.8,1.3,1.3);
\coordinate (As2) at (2.2,1.3,2.7);
\coordinate (Bs2) at (2.2,2.7,2.7);
\coordinate (Cs2) at (2.2,2.7,1.3);
\coordinate (Ds2) at (2.2,1.3,1.3);

\coordinate (Ov) at (1,1,1);
\coordinate (Av) at (1,3,1);
\coordinate (Bv) at (1,3,3);
\coordinate (Cv) at (1,1,3);
\coordinate (Dv) at (3,1,1);
\coordinate (Ev) at (3,3,1);
\coordinate (Fv) at (3,3,3);
\coordinate (Gv) at (3,1,3);

\draw[gray,dashed,thin] (O) -- (C) -- (G) -- (D) -- cycle;
\draw[gray,dashed,thin] (O) -- (A) -- (E) -- (D) -- cycle;
\draw[gray,dashed,thin] (O) -- (A) -- (B) -- (C) -- cycle;
\draw[gray,dashed,thin] (D) -- (E) -- (F) -- (G) -- cycle;
\draw[gray,dashed,thin] (C) -- (B) -- (F) -- (G) -- cycle;
\draw[gray,dashed,thin] (A) -- (B) -- (F) -- (E) -- cycle;


\draw[black,fill=gray,opacity=0.6] (As) -- (Bs) -- (Cs) -- (Ds) -- cycle;
\draw[black,fill=gray,opacity=0.6] (As2) -- (Bs2) -- (Cs2) -- (Ds2) -- cycle;

\draw[black,thick,<->] (1.8,1.2,2.8) -- (2.2,1.2,2.8);
\fill[black](2,1.2,2.8) node [scale=1,anchor= north]{$d$};

\draw[black,thick,<->] (1.7,2.7,2.7) -- (1.7,1.3,2.7);
\fill[black](1.7,2,2.7) node [scale=1,anchor= east]{$l_p$};

\draw[black,thick,<->] (4.1,0,0) -- (4.1,4,0);
\fill[black](4.1,2,0) node [scale=1,anchor= west]{$\mathcal{L}$};

\end{tikzpicture}
\caption{Schematic showing the computational domain, and the three geometrical parameters. The dashed gray lines indicate the boundaries of the periodic cube. The two plates lie within the periodic domain.}
\label{fi:sch}
\end{figure}

After a start-up phase, consisting usually of two or three large-eddy turnover times (defined as $T_e = u^{\prime 2}/\epsilon$), the forces on the plates and other statistics are computed and averaged. Unless stated otherwise, temporal convergence of the forces and other statistics is assured by running the simulations until the forces originating from the hydrodynamic pressure on both plates are equal (but oppositely signed) to within 3\%. We also check for statistical convergence by dividing the force time series in two, and ensuring that the average values in both halves match within $5\%$. This defines the magnitude of the error bars. In practice, this means a run time of $T_e\approx300$.

\begin{table}[]
    \centering
    \begin{tabular}{|c|c|c|c|c|c|}
    \hline
    $\kappa_f/\kappa_{1}$ & $\epsilon^*\mathcal{L}^4/\nu^3$ & $T_L\nu/\mathcal{L}^2$ & $Re_{\lambda}$ & $\eta_K/\mathcal{L}$ & $L/\mathcal{L}$ \\ \hline
    2.3 & $5.64\cdot10^3$ & $4.49\cdot10^{-3}$ & 7.6 & $3.7\cdot 10^{-2}$ & $1.8\cdot 10^{-1}$  \\
    2.3 & $4.51\cdot10^4$ & $2.24\cdot10^{-3}$ & 15 & $2.2\cdot 10^{-2}$& $3.1\cdot 10^{-1}$  \\
    2.3 & $1.85\cdot10^5$ & $1.40\cdot10^{-3}$ & 22 & $1.6\cdot 10^{-2}$& $3.7\cdot 10^{-1}$  \\
    2.3 & $2.89\cdot10^6$ & $5.61\cdot10^{-4}$ & 42 & $7.9\cdot 10^{-3}$& $5.1\cdot 10^{-1}$  \\
    2.3 & $4.50\cdot10^7$ & $2.24\cdot10^{-4}$ & 73 & $3.9\cdot 10^{-3}$ & $6.0\cdot 10^{-1}$  \\
    2.3 & $3.61\cdot10^8$ & $1.12\cdot10^{-4}$ & 100 & $2.4\cdot 10^{-3}$& $6.2\cdot 10^{-1}$  \\
    \hline 
    \end{tabular}
    \caption{Forcing parameters and the resulting flow statistics for all cases simulated.}
    \label{tab:my_label}
\end{table}

\section{Results and Discussion}

\subsection{Mechanisms generating the attractive force}
\label{sec:size}

For the first set of simulations, we fix $Re_\lambda=100$, vary the plate size, and analyse how the fluctuation force and other flow statistics change as a function of plate separation. We simulate three plate sizes: $l_p/\mathcal{L}=0.1$, $l_p/\mathcal{L}=0.175$ and $l_p/\mathcal{L}=0.25$. We also simulated a few cases with $l_p/\mathcal{L}=0.4$, but preliminary flow visualizations showed interference effects appearing due to the periodic images. Therefore, we do not show these results here.

\begin{figure}
    \centering
    \includegraphics[width=.45\textwidth]{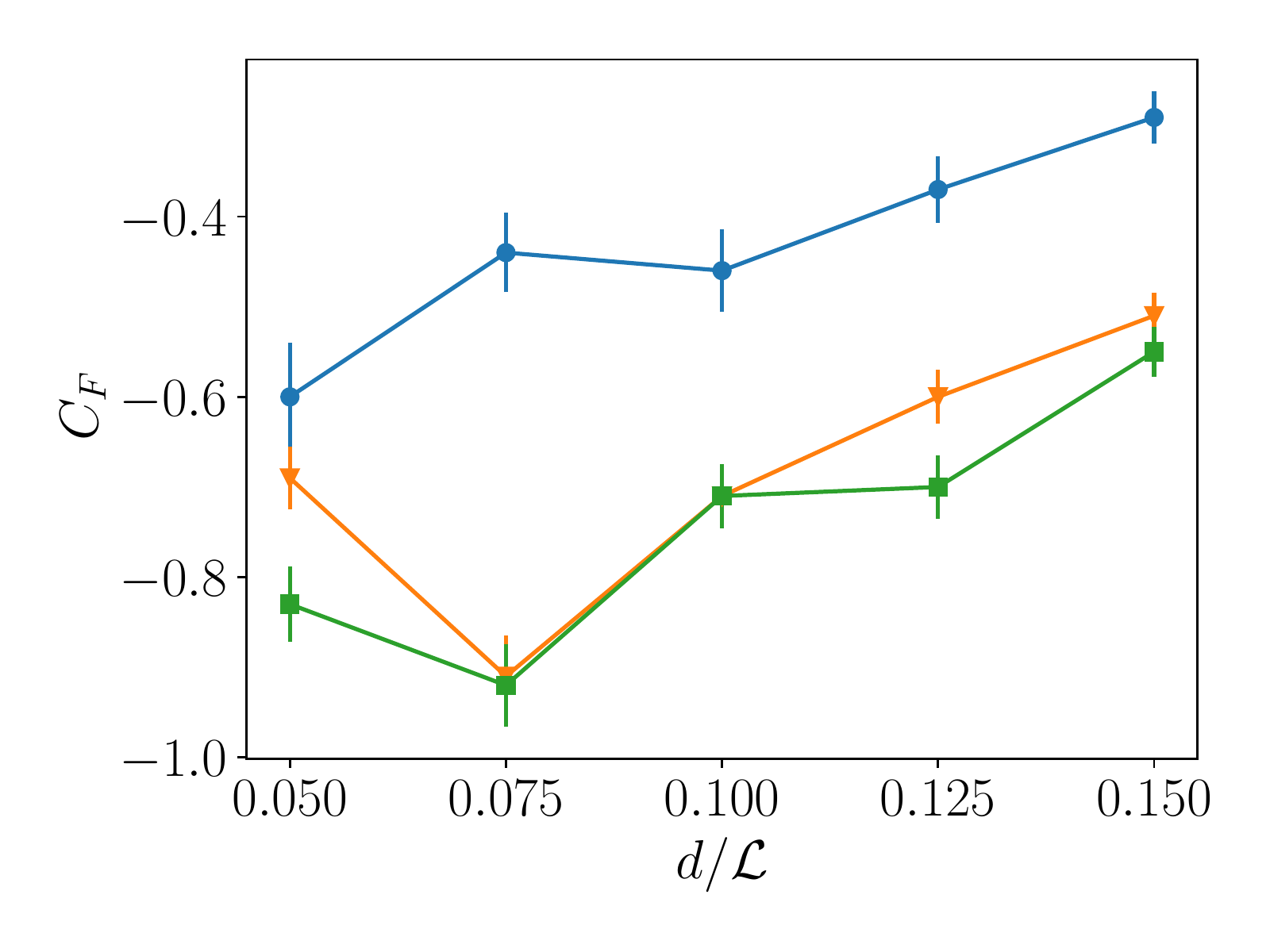}
    \includegraphics[width=.45\textwidth]{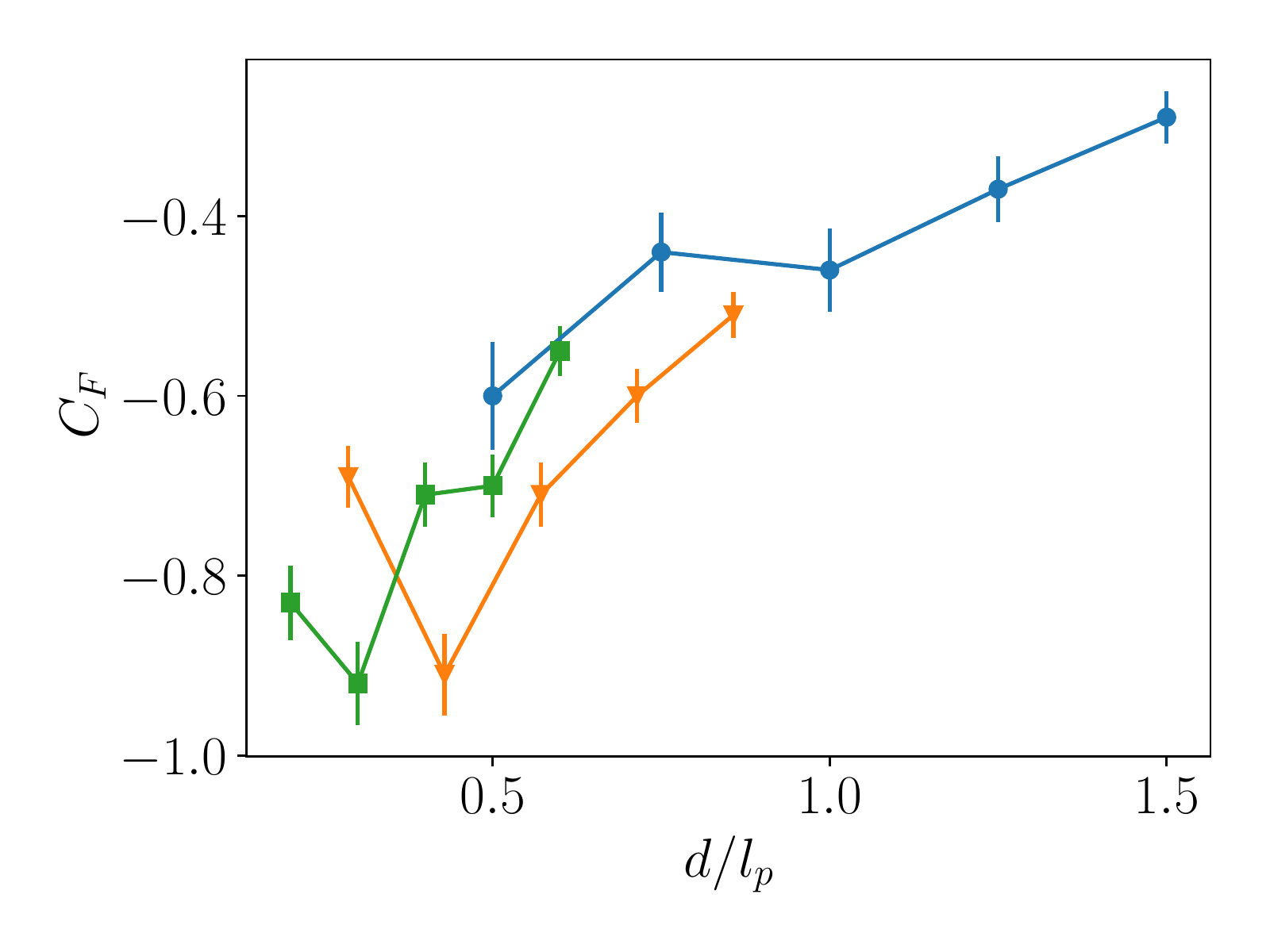}
    \caption{Left: Non-dimensionalized force coefficient as a function of non-dimensional plate distance $d/\mathcal{L}$. Right: As in left, with distance re-scaled using the plate length. Symbols: blue circles are $l_p/\mathcal{L}=0.1$, orange triangles are $l_p/\mathcal{L}=0.175$ and green squares are $l_p/\mathcal{L}=0.25$.}
    \label{fig:cf_platesize}
\end{figure}

We begin by analyzing the instantaneous non-dimensional force coefficient $C_F$, defined as $C_F=F/(\frac{1}{2}\rho u^{\prime2} l_p^2)$, as a function of the non-dimensional plate distance for the three plate sizes considered. As per convention, $F$ is the average normal force on both plates, and is negative when attractive, and positive when repulsive \cite{spandan2020fluctuation}. In Figure \ref{fig:cf_platesize} we analyze the temporal behaviour of $C_F$ for a sample case with $l_p/\mathcal{L}=0.175$ and $d/\mathcal{L}=0.1$. On the left panel, we show the time signal of $C_F(t)$, as well as the running average. The force can be seen to vary between negative and positive values up to more than ten times its mean value ($\sim 0.71$). We can also see that we require a long running time, of $t/T_e\approx 300$ to obtain an adequate value for the averaged $C_F$. In the right panel, we show a probability distribution function of the standardized $C_F(t)$ on a single plate, and of the sum of the force on the two plates. Both curves have a weak positive skew, with a skewness coefficient of $\sim 0.4$, and have long tails, with a Fisher's kurtosis of $\sim 1.1$. We also note that the cross-correlation coefficient between the force on the plates for this case is $0.32$, and is generally between $0.25$ and $0.35$, depending on the plate separation.

\begin{figure}
    \centering
    \includegraphics[width=.45\textwidth]{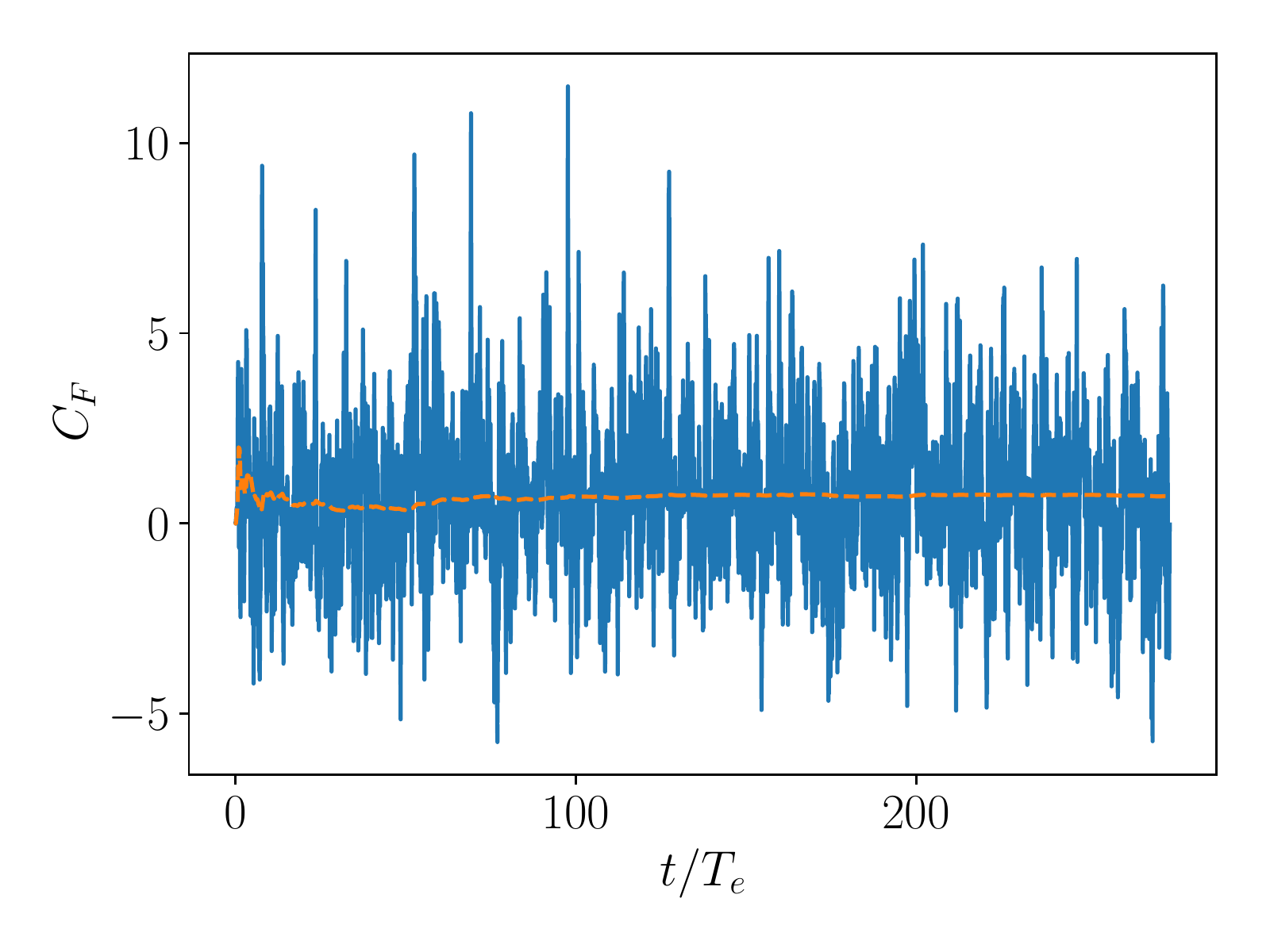}
    \includegraphics[width=.45\textwidth]{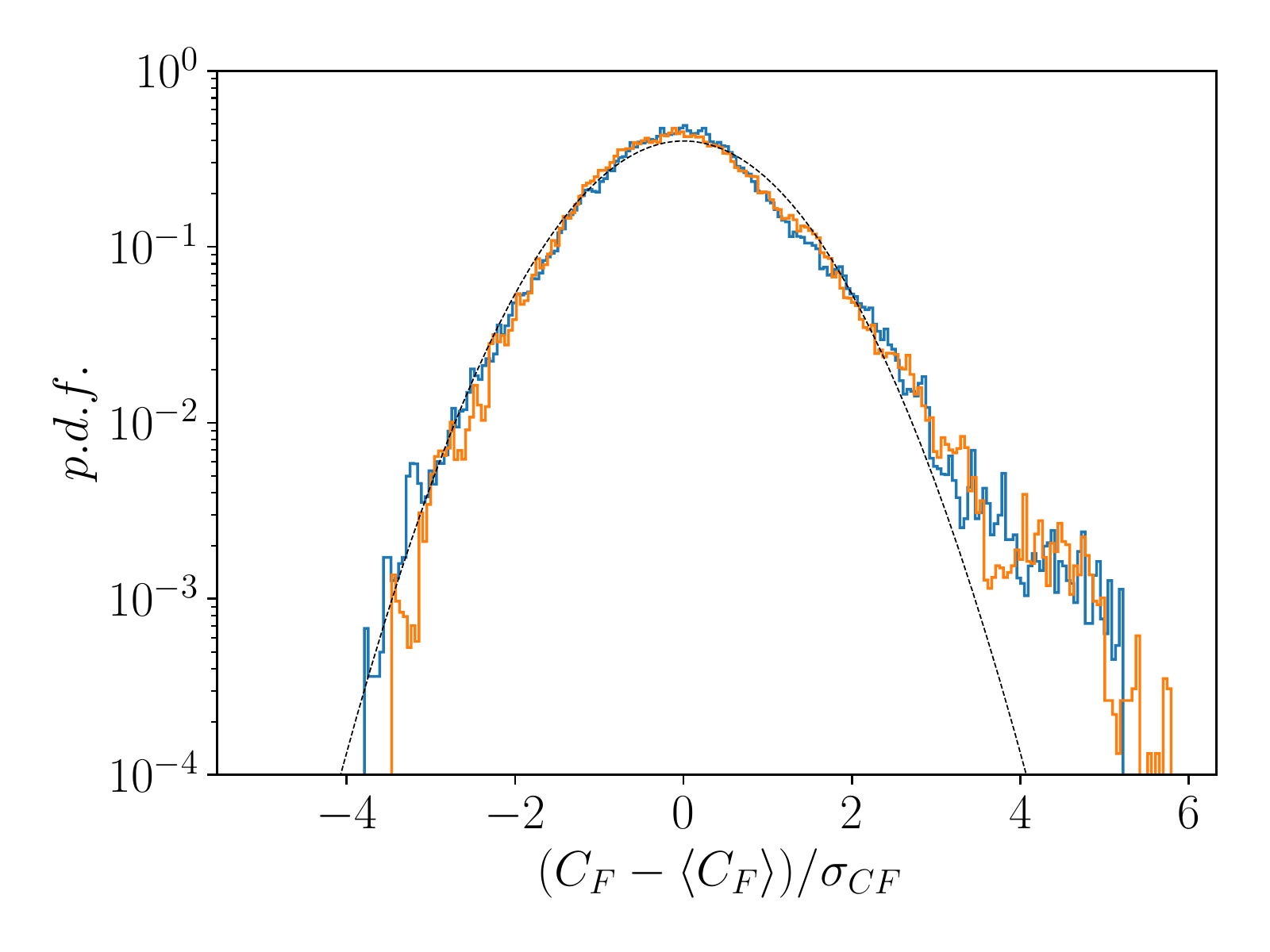}
    \caption{Left: Instantaneous value of the non-dimensional attractive force in the system (blue) as well as the running average (orange dashed) for $d/\mathcal{L}=0.1$. Right: Probability distribution function for the feature-scaled force coefficient on a single plate (blue) and on both plates (orange) for $d/\mathcal{L}=0.1$. The dashed black line denotes a normal distribution. }
    \label{fig:cf_statistics}
\end{figure}

\begin{figure}
    \centering
    \includegraphics[trim={5.5cm 0 5.5cm 0cm}, clip,width=.32\textwidth]{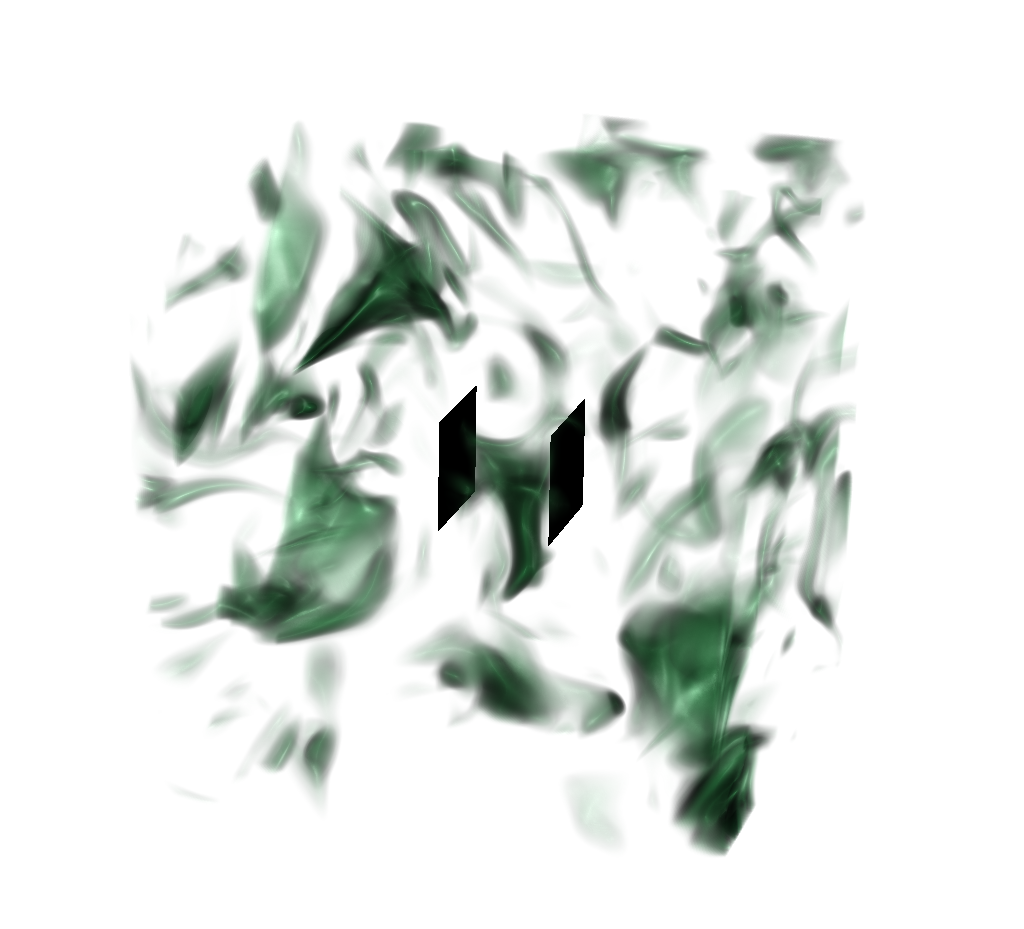}
    \includegraphics[trim={5.5cm 0 5.5cm 0cm}, clip,width=.32\textwidth]{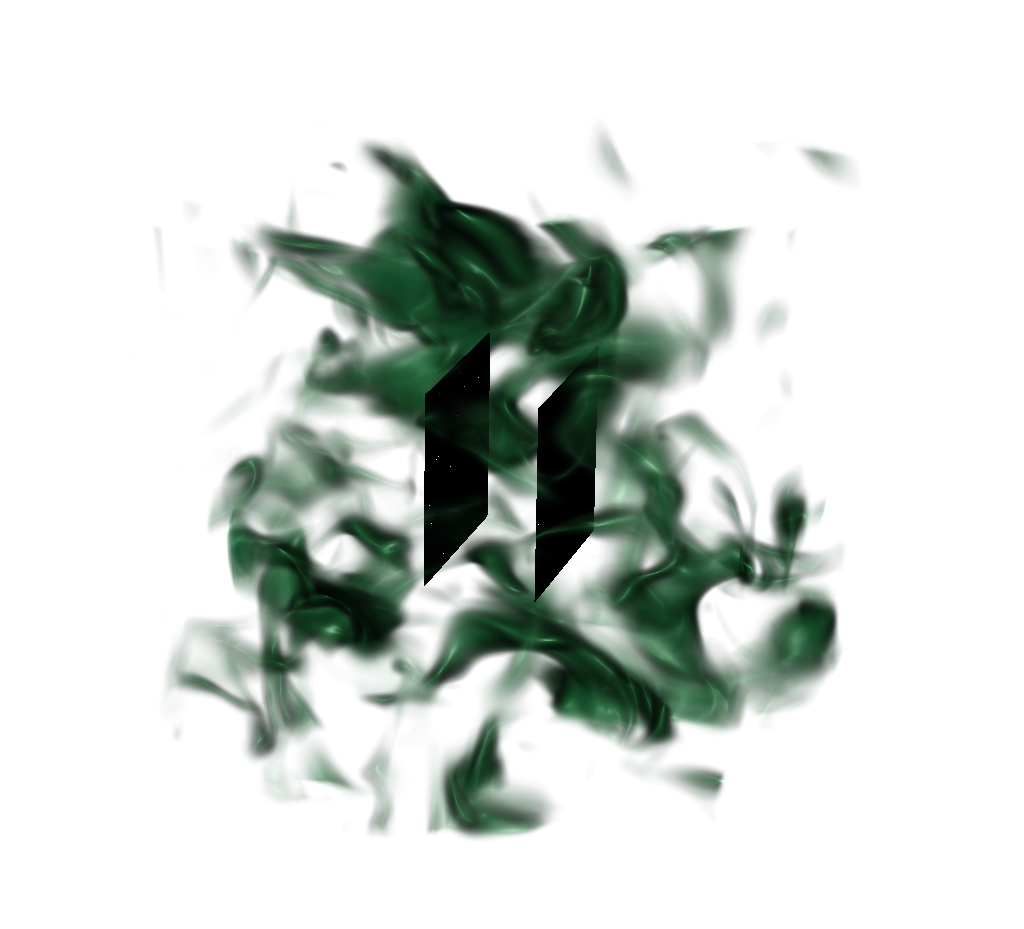}
    \includegraphics[trim={5.5cm 0 5.5cm 0cm}, clip,width=.32\textwidth]{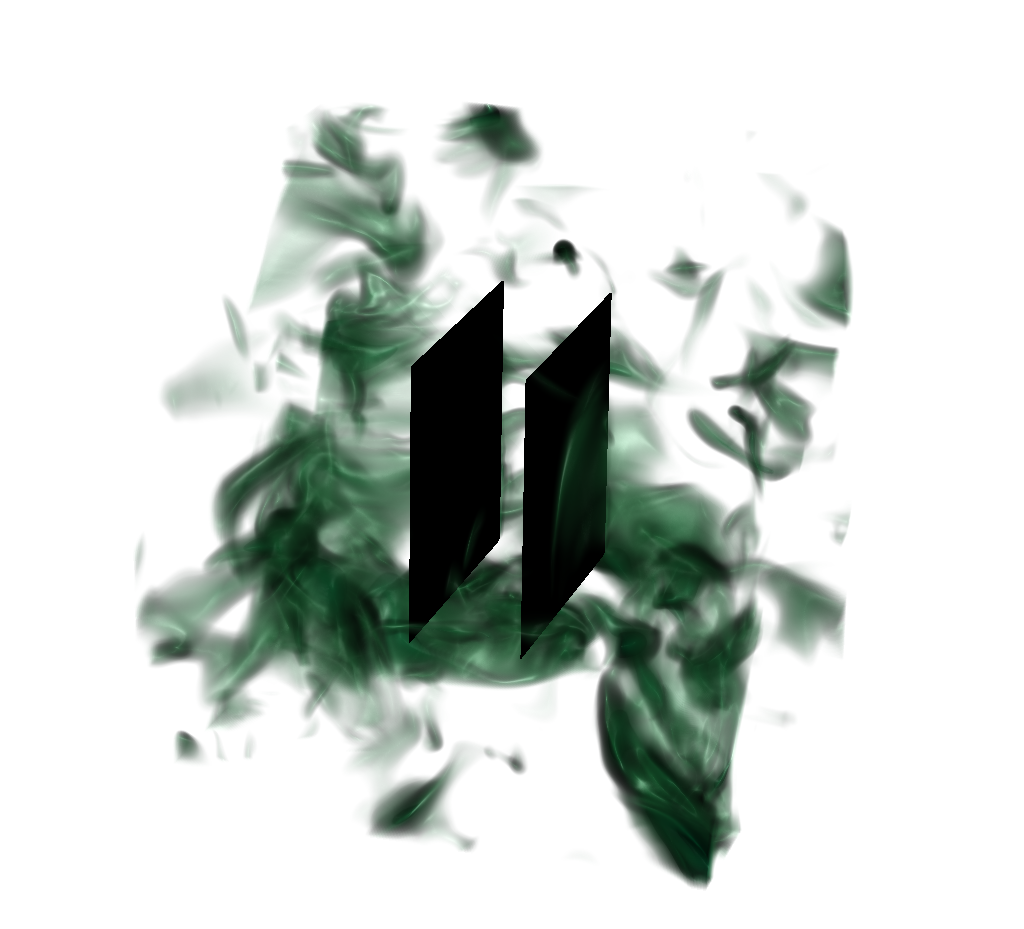}\\
    \includegraphics[trim={5.5cm 0 5.5cm 0cm}, clip,width=.32\textwidth]{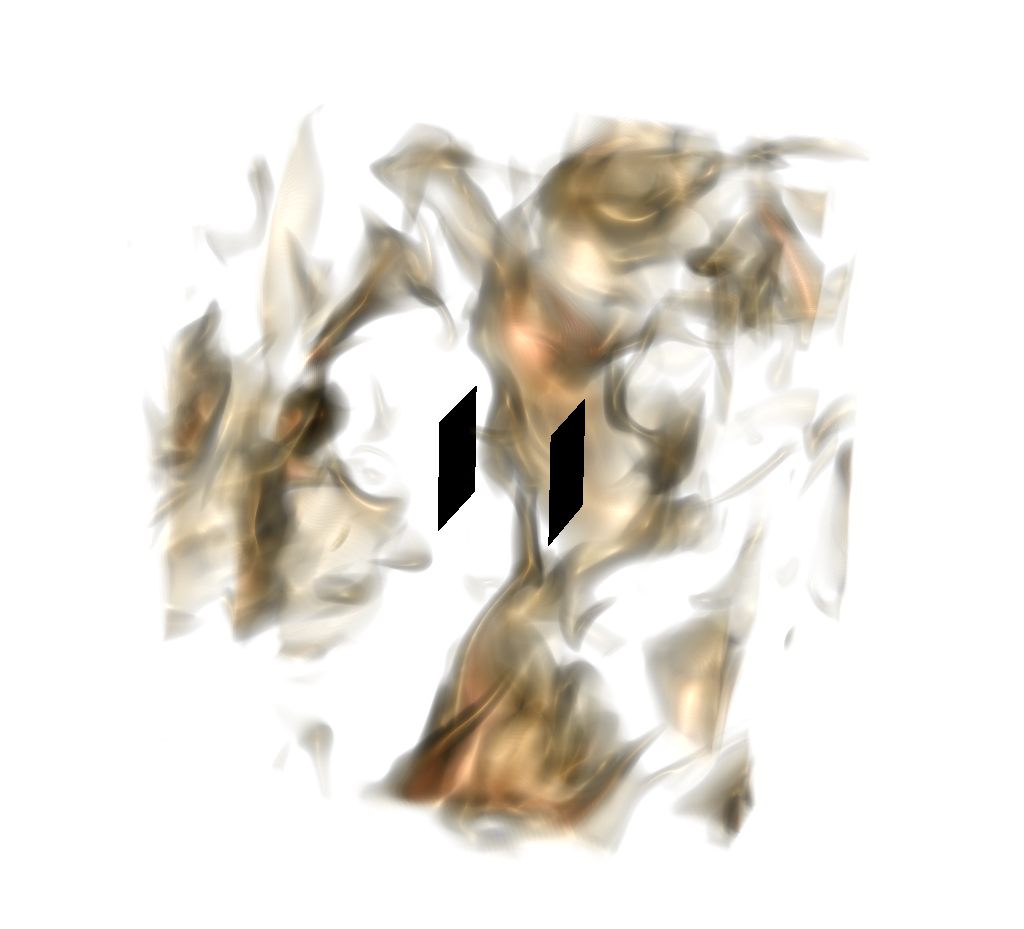}
    \includegraphics[trim={5.5cm 0 5.5cm 0cm}, clip,width=.32\textwidth]{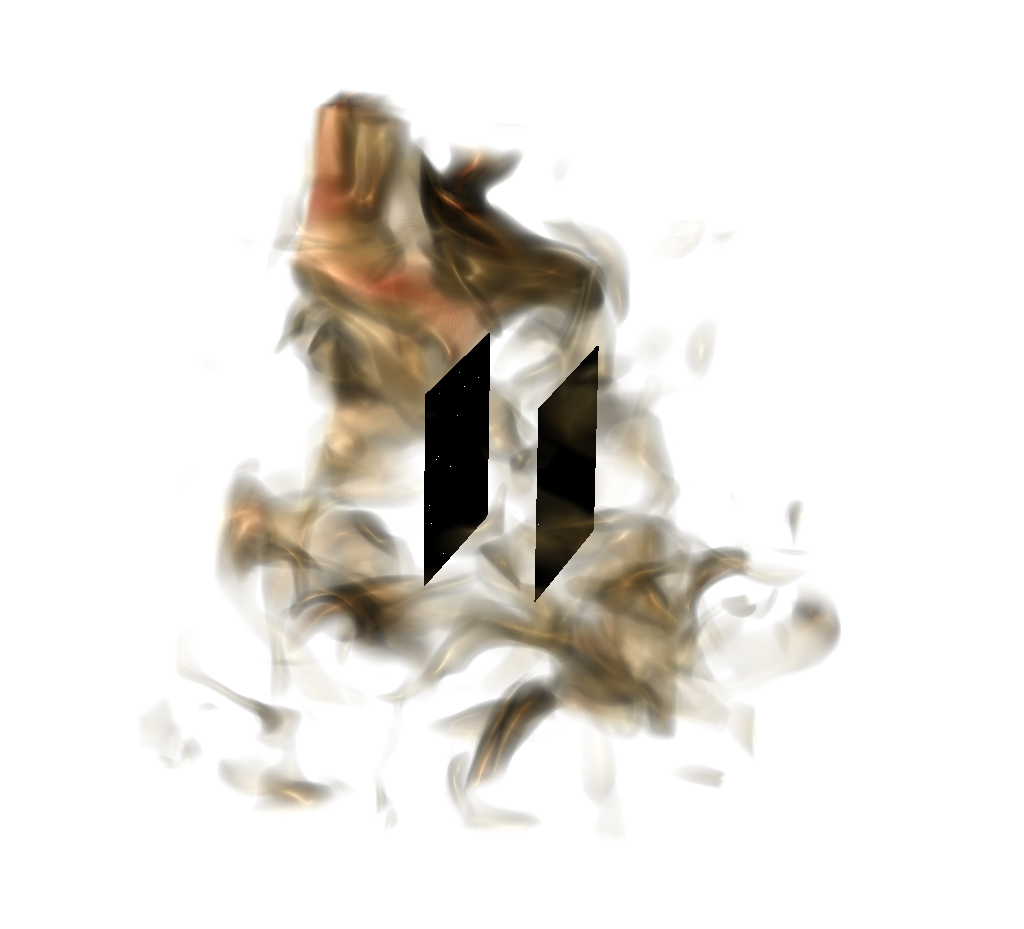}
    \includegraphics[trim={5.5cm 0 5.5cm 0cm}, clip,width=.32\textwidth]{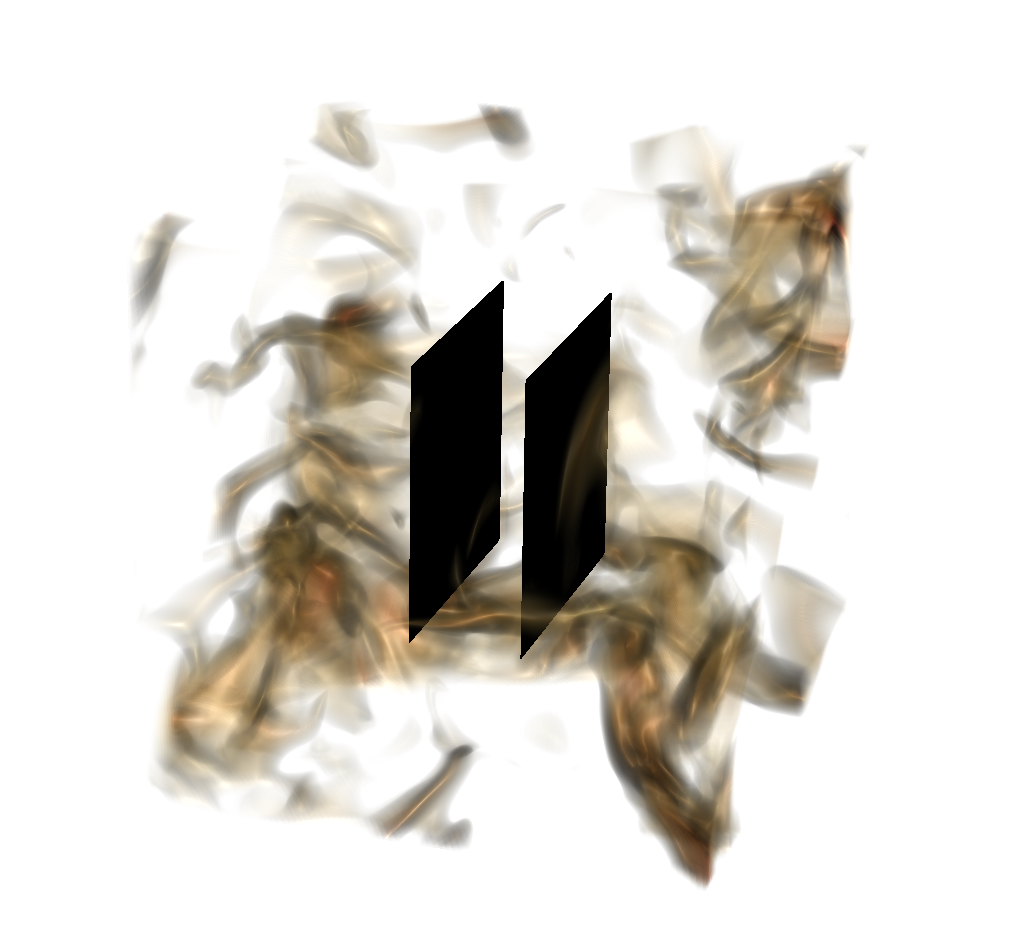}
    \caption{Top row: Volume visualization of the pressure source term $Q$ at $Re_\lambda=100$, $d/\mathcal{L}=0.1$ and three plate sizes: $l_p/\mathcal{L}=0.1$ (left), $l_p/\mathcal{L}=0.175$ (middle) and $l_p/\mathcal{L}=0.25$ (right). Regions of  positive $Q$ are shown in green, while regions of zero or negative $Q$ are left transparent. Bottom row: Volume visualization of the kinetic energy $k$ for the same cases. Regions of high $k$ are shown in orange/brown, while regions of low $k$ are left transparent. For clarity only a section of the computational domain is shown.}
    \label{fig:vis_platesizes}
\end{figure}

Having established the statistical properties of the temporal behaviour of the force, we now turn to an instantaneous visualization of the flow in Figure \ref{fig:vis_platesizes}. For clarity, we show only an eighth of the domain (a cube of side $\mathcal{L}/2$ centered on the plates). The top row panels show the pressure source term $Q$, defined as $\nabla^2(p/\rho) = (\frac{1}{2}\omega^2-\sigma^2)=2Q$, where $\omega$ is the fluid vorticity and $\sigma$ is the fluid strain \cite{pumirpressure}, for the three plate sizes simulated at a plate separation of $d/\mathcal{L}=0.1$. The pressure source term is commonly used to visualize vortices and highlights the presence of tubular high-vorticity structures \cite{hunt1988eddies}. These have a radius which is usually estimated through the Kolmogorov length scale $\eta_K$, and a length estimated as the integral length-scale $L$. For $Re_\lambda=100$, these take the values of $\eta_K/\mathcal{L}=2.4\cdot 10^{-3}$ and $L/\mathcal{L}=6.2\cdot10^{-1}$, which approximately matches what is seen in the visualizations. 

In Ref.~\cite{spandan2020fluctuation}, it was hypothesized that the plates (of size $l_p/\mathcal{L}=0.25$) packed the intense vortical structures in the slit, forcing them to interact at close quarters. This interaction would cause a pressure drop between the plates, and this was linked to an increase of the average attractive fluctuation force at medium plate separations and hence to the force non-monotonicity. In this manuscript, we intend to explore this link further. By visually comparing the proportions of vortical structures and plates of varying $l_p$, we can hypothesize that the small plates will have limited packing capacities, due to to their reduced size. In contrast, we can expect that the largest plates pack several structures in the slit, resulting in strong interactions between the vortices, as was observed in Ref.~\cite{spandan2020fluctuation}. The capacity of the medium sized plates to pack interacting vortical structures lies somewhere between the small and large plates.

The other cause for the force hypothesized in Ref.~\cite{spandan2020fluctuation} was the exclusion of energy structures. To better visualize the effect of plate size on this phenomena, the bottom row of Figure \ref{fig:vis_platesizes} shows the kinetic energy $k$ for the same simulation at the same time instant. The energy-containing structures in HIT are much larger than their vortical counterparts, as they approximately extend a decorrelation scale $L$ in all directions, even if they show features at multiple length-scales. Therefore, as hinted by the visualizations, even the smallest plates will make a difference on the energy distribution of the flow, as their presence would perturb the behaviour of the energy-containing scales in the slit due to the no-slip condition on their surfaces, which causes regions of low $k$ to appear close to the plates. These regions can be thought of as rough analogs to boundary layers, even if there is no mean flow. Their effect will be explored later, when $Re_\lambda$ is varied.

Because the two hypothesized causes of the attractive force will behave rather differently with plate size, by varying $l_p/\mathcal{L}$ we can disentangle them, and further support or falsify the explanations brought forward in Ref.~\cite{spandan2020fluctuation}. Moving to quantitative analysis, we plot the averaged non-dimensional force coefficient $C_F$, in the left panel of Figure \ref{fig:cf_platesize}. For the small plate ($l_p/\mathcal{L}=0.1$), the force loses its non-monotonic character and is maximum for $d/\mathcal{L}=0.05$. We note that the computed force appears to be slightly larger for $d/\mathcal{L}=0.1$ than for $0.075$, but the differences between both data points are well contained within error bars.  Meanwhile, the force is non-monotonic for both the medium ($l_p/\mathcal{L}=0.175$) and large ($l_p/\mathcal{L}=0.25$) plates, with a maximum attractive force at $d/\mathcal{L}=0.075$, consistent with the value obtained in Ref.~\cite{spandan2020fluctuation}. The $C_F(d)$ curves for the medium and large plates also appear remarkably close to each other in the region $d/\mathcal{L} \geq 0.075$.

These results hint that as long as the plates are large enough, changing the plate size does not alter the underlying physical processes of energy exclusion and vortex interaction. We confirm this by showing in the right panel the same $C_F$ data against the re-scaled plate separation in terms of the plate size $d/l_p$. We can observe that when comparing left and right panels, the $C_F$ curves for the medium and large plates show a much worse collapse. In the left panel, the force maximum in $d/\mathcal{L}$ units is found for $d/\mathcal{L}=0.075$, consistent with Ref.~\cite{spandan2020fluctuation}, but it is located at different values of $d/l_p$ in the right panel. This allows us to confirm that if the force is non-monotonic, the location of the force maximum does not dependent on $l_p$, i.e.~the plate size does not introduce a new length-scale into the problem.

To further examine the effect of plate size on the mechanisms at play, and to help us unravel the causes behind the generation of the force, we can analyze the behaviour of a series of flow quantities around the plates. These quantities will be spatially averaged on squares of side $0.8l_p$ co-centered with the plates to avoid edge effects, and later symmetrized around the plane of symmetry of the system. 

\begin{figure}
    \centering
    \includegraphics[width=.32\textwidth]{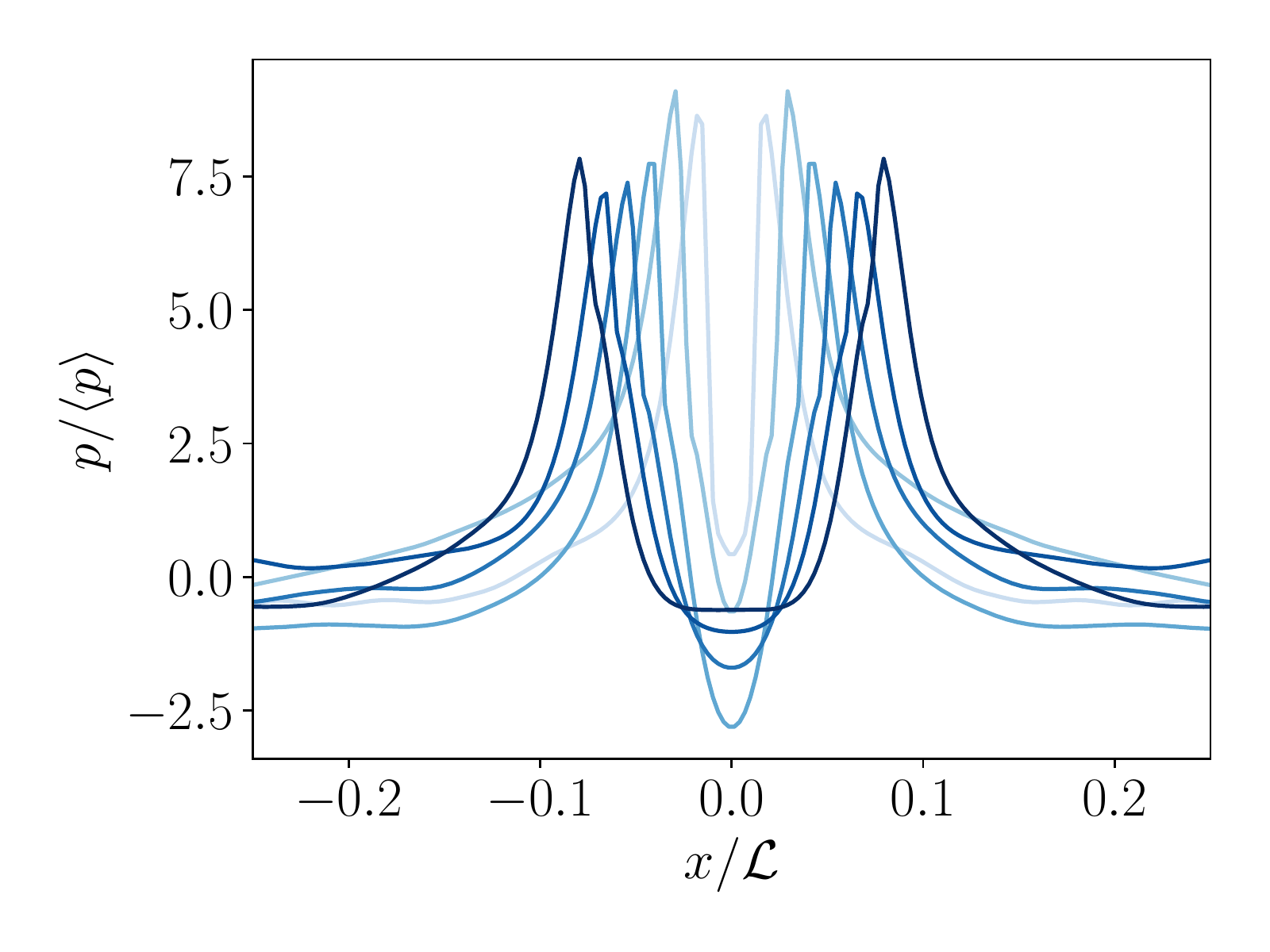}
    \includegraphics[width=.32\textwidth]{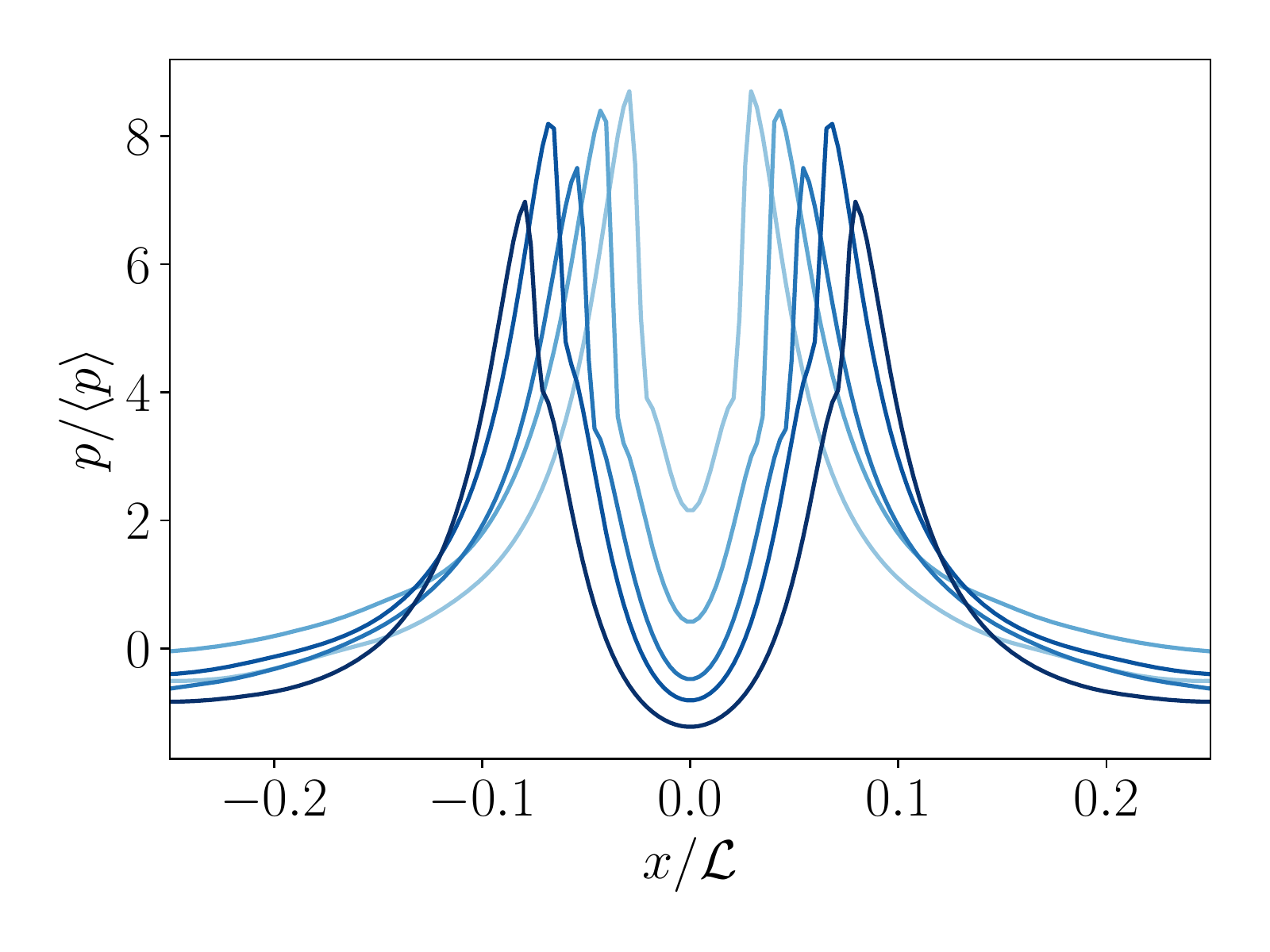}
    \includegraphics[width=.32\textwidth]{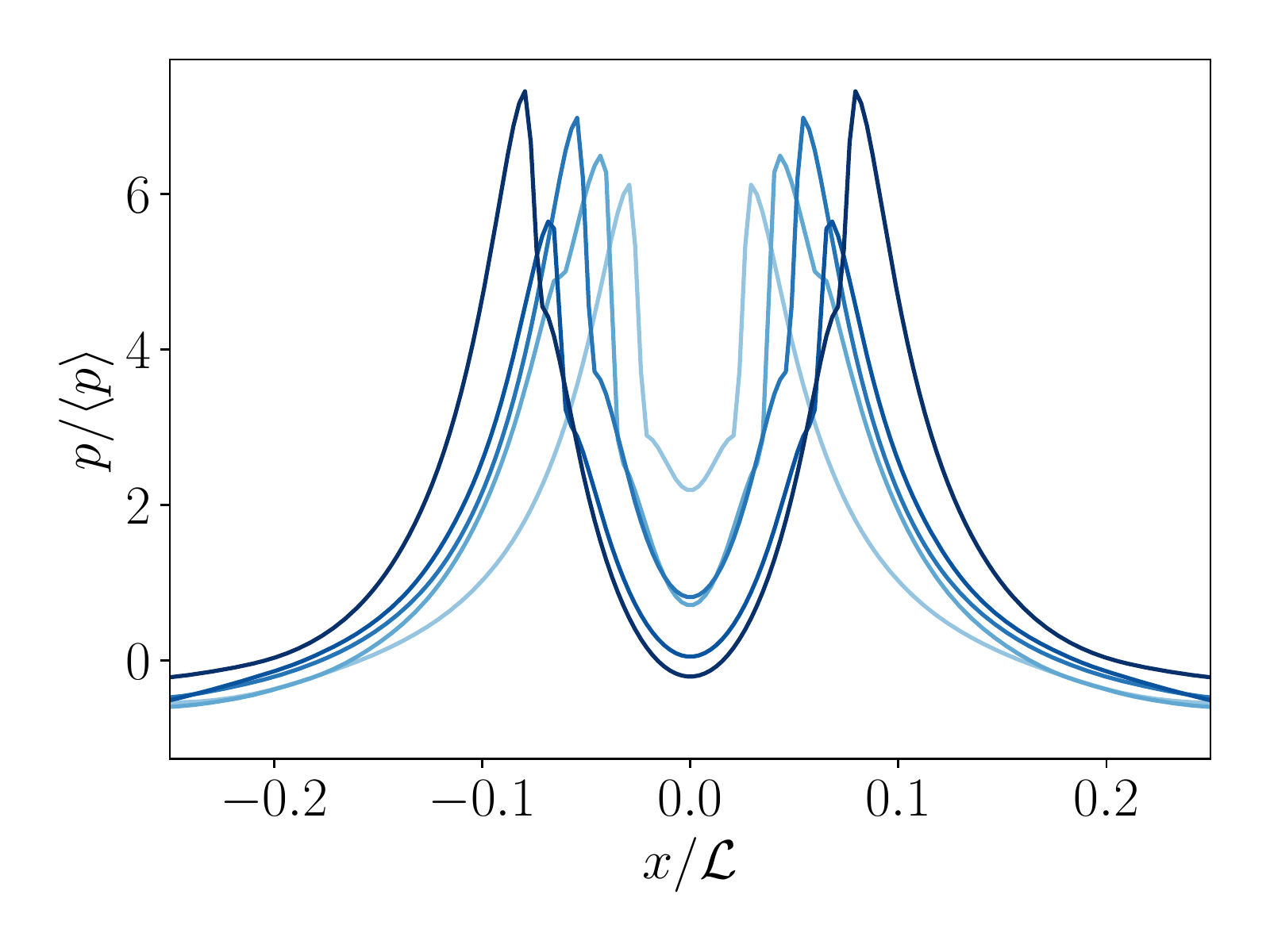}\\
    \includegraphics[width=.32\textwidth]{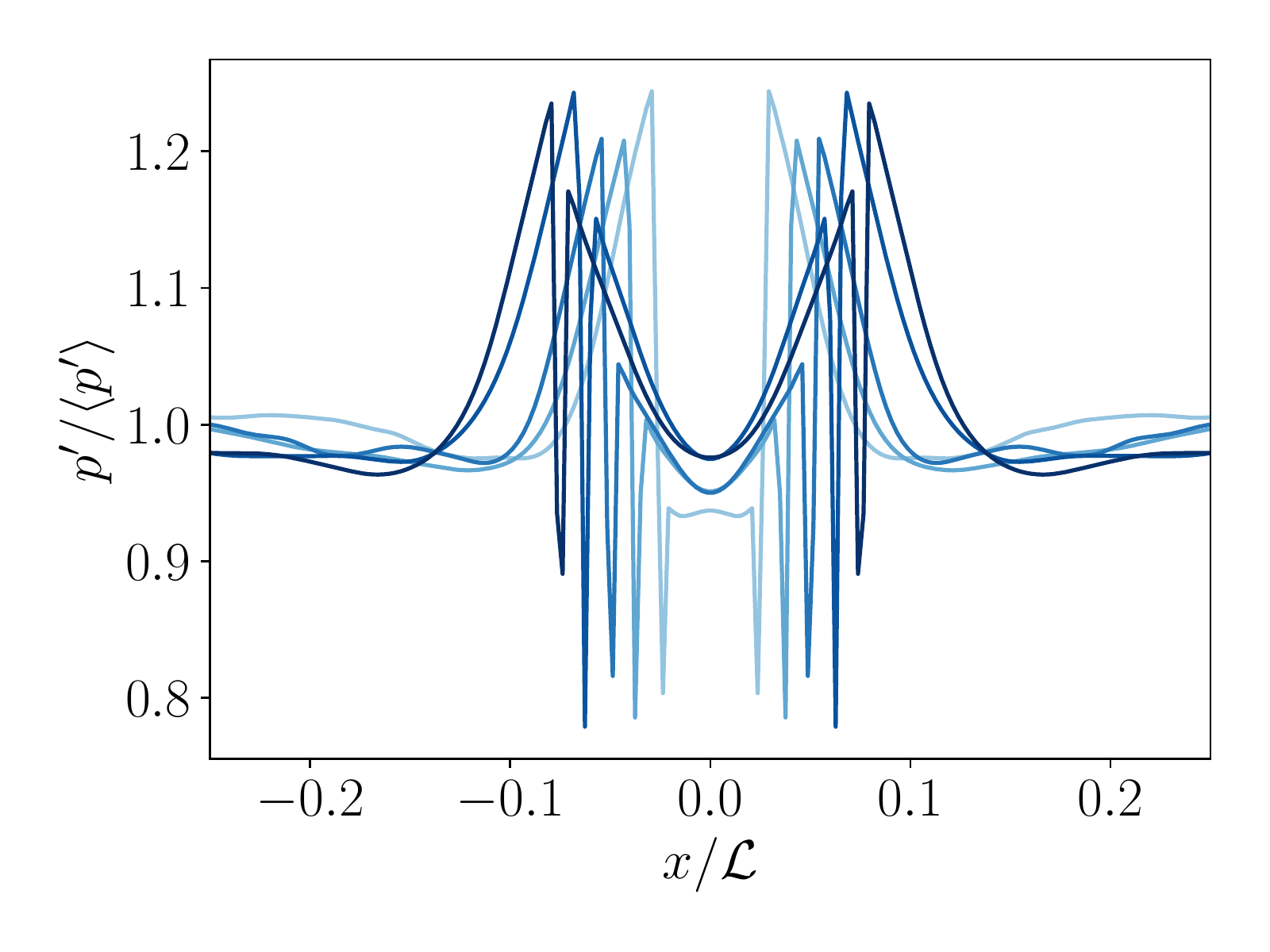}
    \includegraphics[width=.32\textwidth]{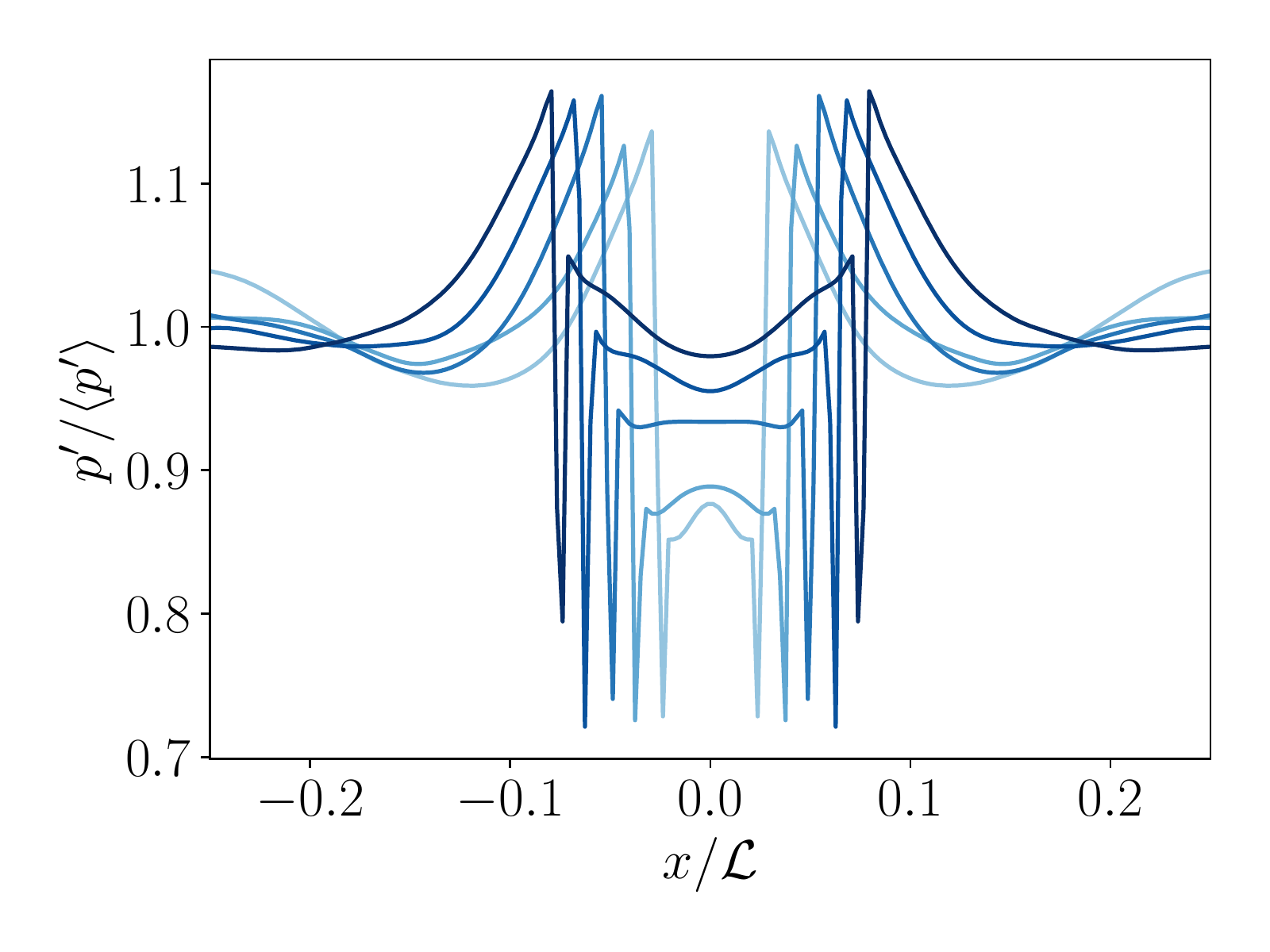}
    \includegraphics[width=.32\textwidth]{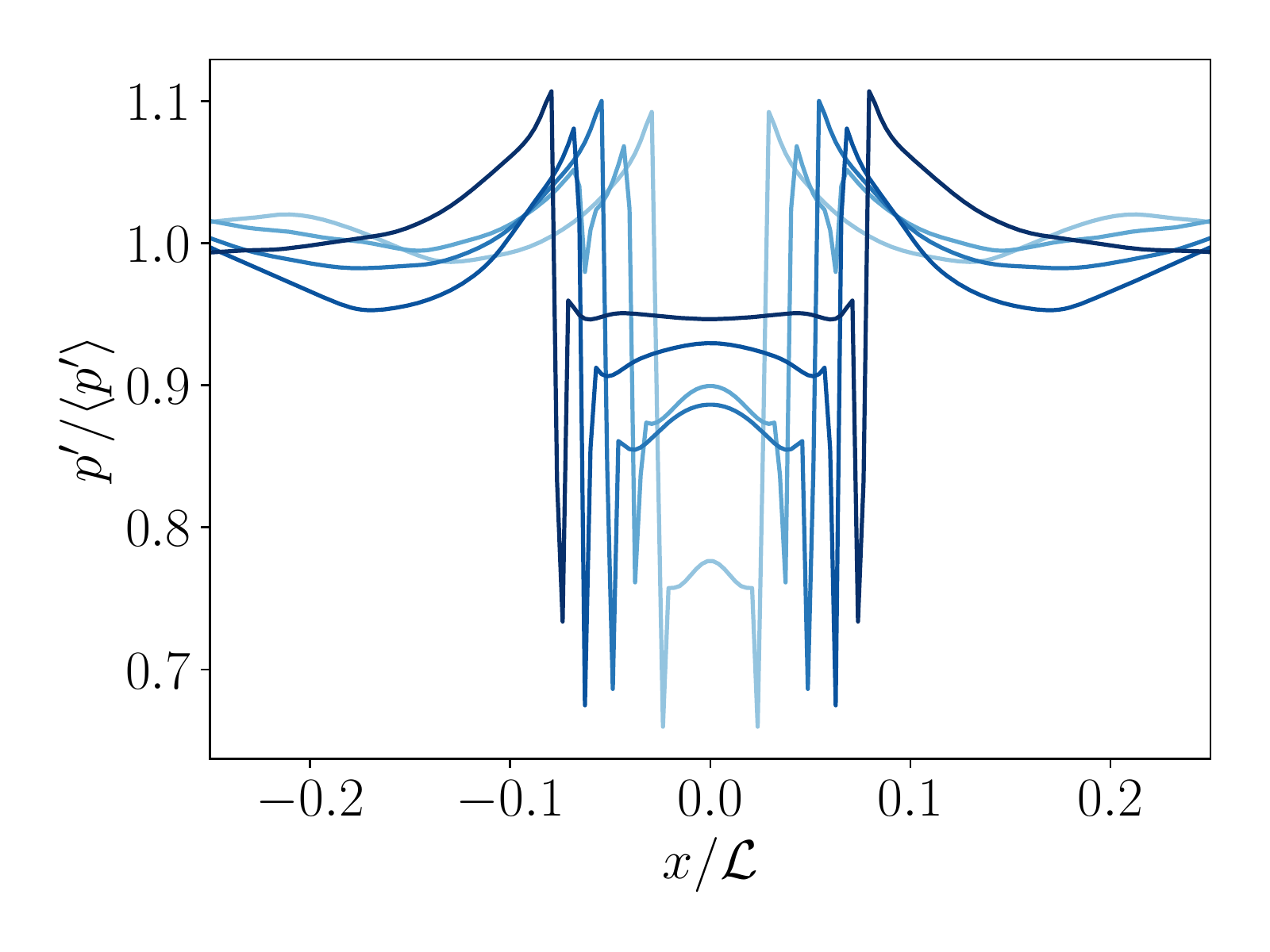}
    \caption{Flow statistics for all cases simulated in this section with $Re_\lambda=100$. The two rows represent mean pressure (top) and pressure fluctuations (bottom), while the three columns represent changing plate size: $l_p/\mathcal{L}=0.1$ (left) $0.175$ (middle) and $0.25$ (right). The lines are coded from light to dark blue depending on plate separation. }
    \label{fig:stats_pr_platesize}
\end{figure}

We start with the behaviour of both the mean and the fluctuations of pressure, shown in Figure \ref{fig:stats_pr_platesize}. We can observe a qualitative difference between the mean pressure curves for the small plate, and for the medium and large plates. For the small plates, the mean pressure has almost returned to the average value at large plate separations, and shows a flat region for the largest plate separations. Meanwhile, for the medium and especially for the large plates, the mean pressure does not recover its average value outside the plates even for the largest separations. The small plates behave as ``individuals'' for much smaller values of plate separation, even if there is a net attractive force. 

Turning to the pressure fluctuations, we observe that the level of fluctuations between the plates is on average lower than the level outside at small plate separations. On the other hand, a local maxima of fluctuations can be seen for some cases. For the smallest plate size, it is only really observed at the smallest plate distance $d/\mathcal{L}=0.05$, and is absent from the other graphs. On the other hand, a pressure fluctuation enhancement is present for the medium and for the large plate in the slit. The effect of plate size is very apparent: the larger the plate, the larger the pressure fluctuations. This increase in pressure fluctuations was already observed in Ref.~\cite{spandan2020fluctuation}, and was associated to increased vortex activity and used to help explain why the attractive force is greater at larger plate separations for the $l_p/\mathcal{L}=0.25$ plate. Here, we will attempt to find evidence to further support or falsify this statement.

\begin{figure}
    \centering
    \includegraphics[width=.32\textwidth]{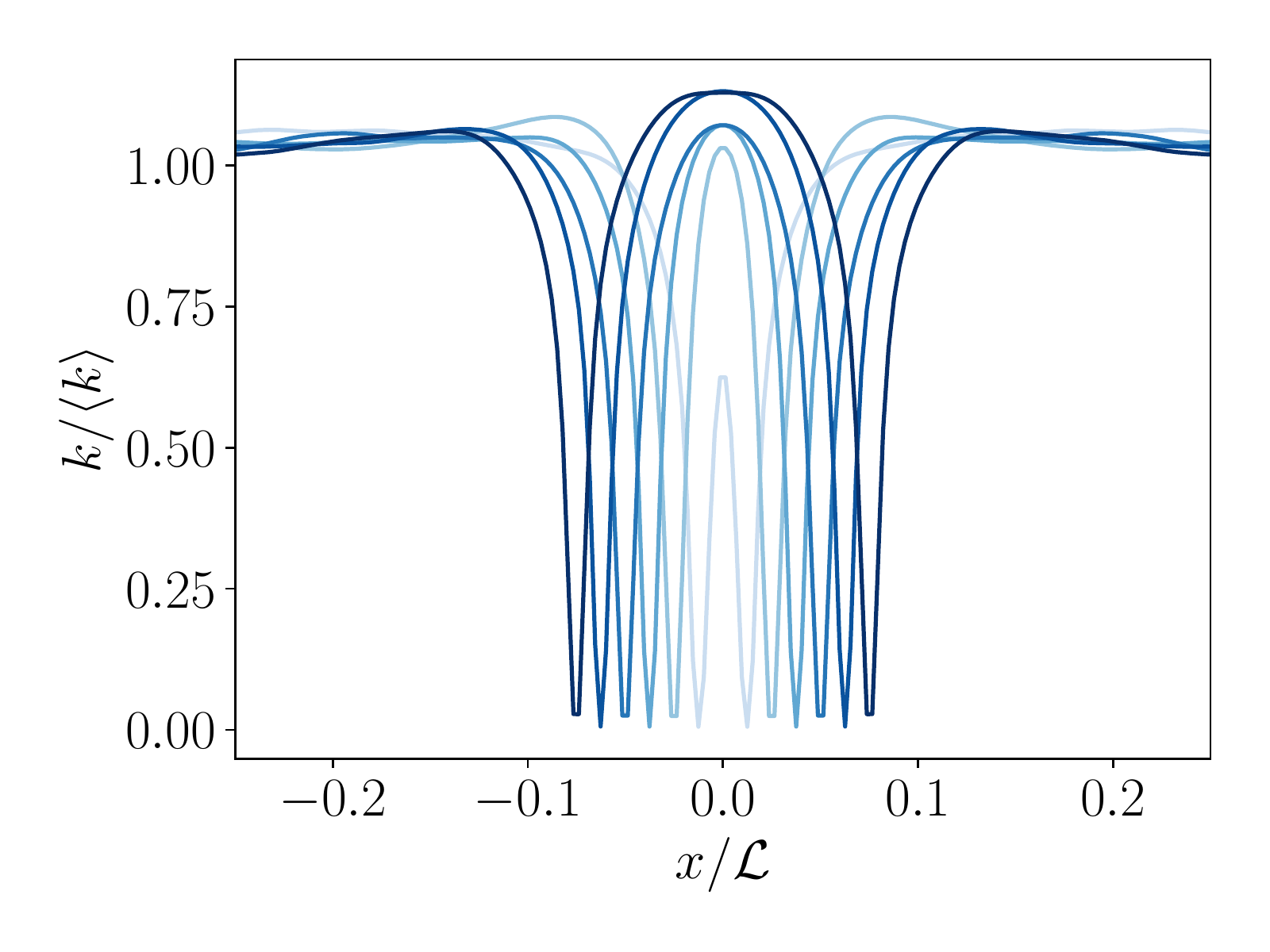}
    \includegraphics[width=.32\textwidth]{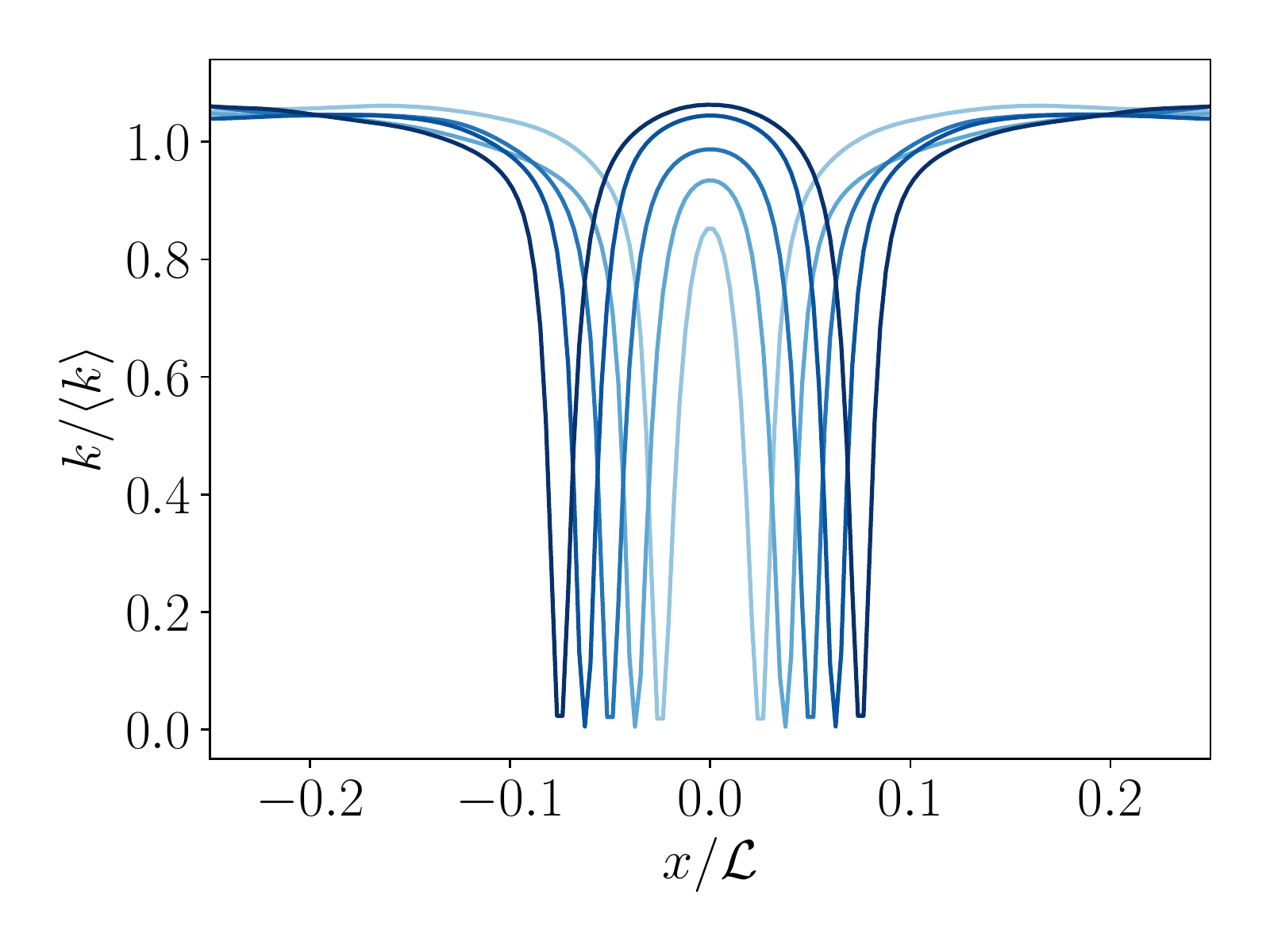}
    \includegraphics[width=.32\textwidth]{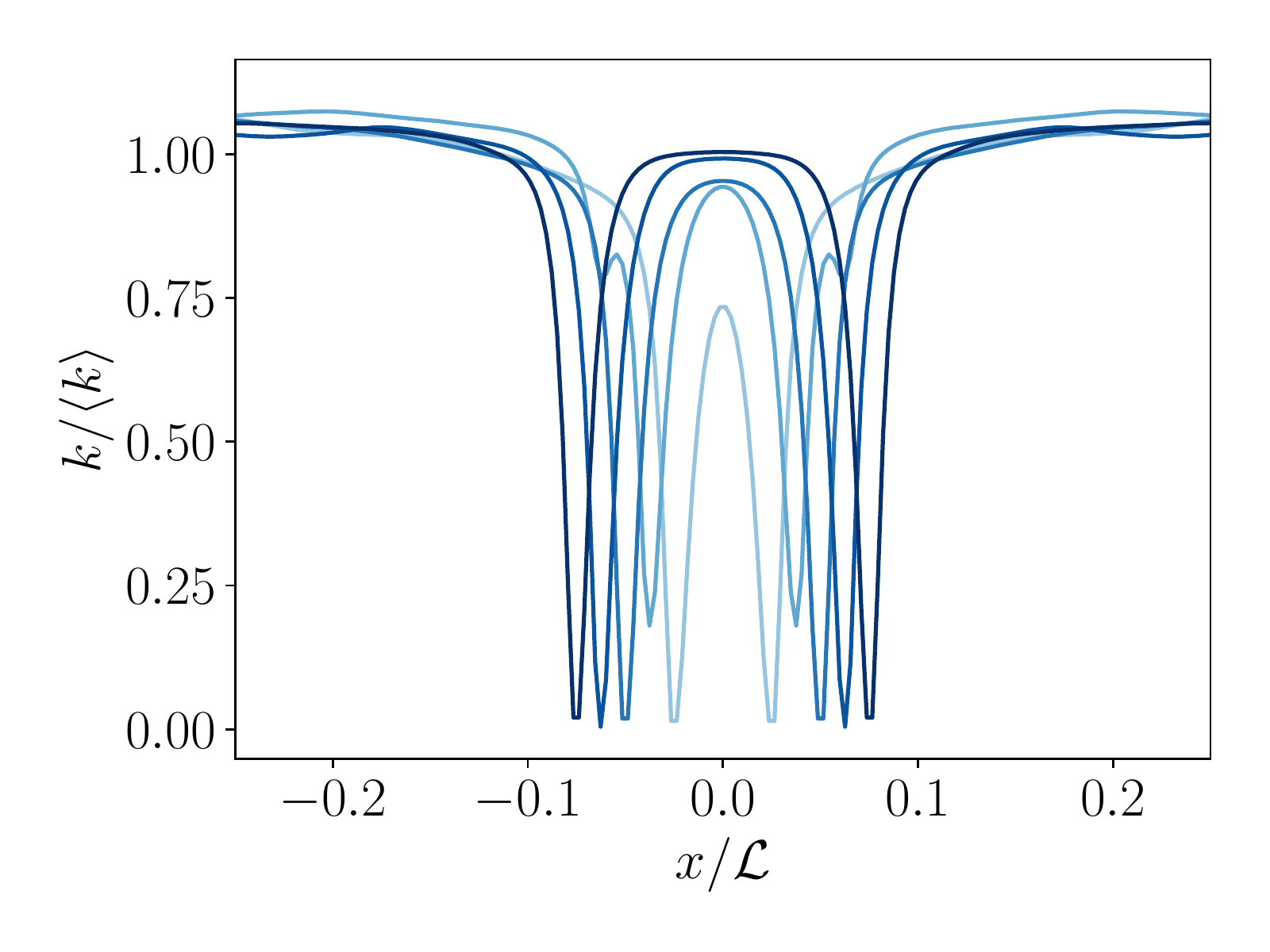}\\
    \includegraphics[width=.32\textwidth]{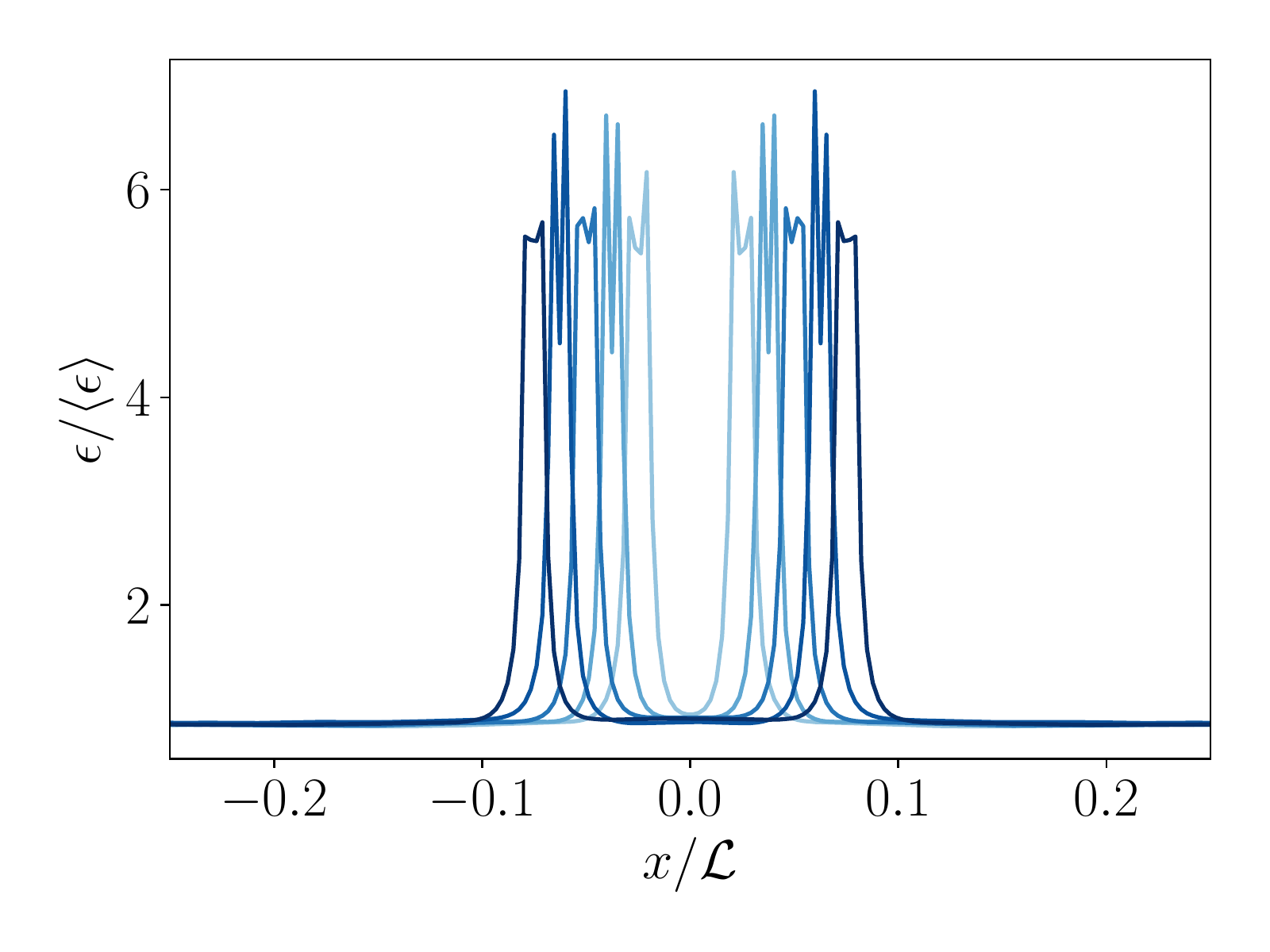}
    \includegraphics[width=.32\textwidth]{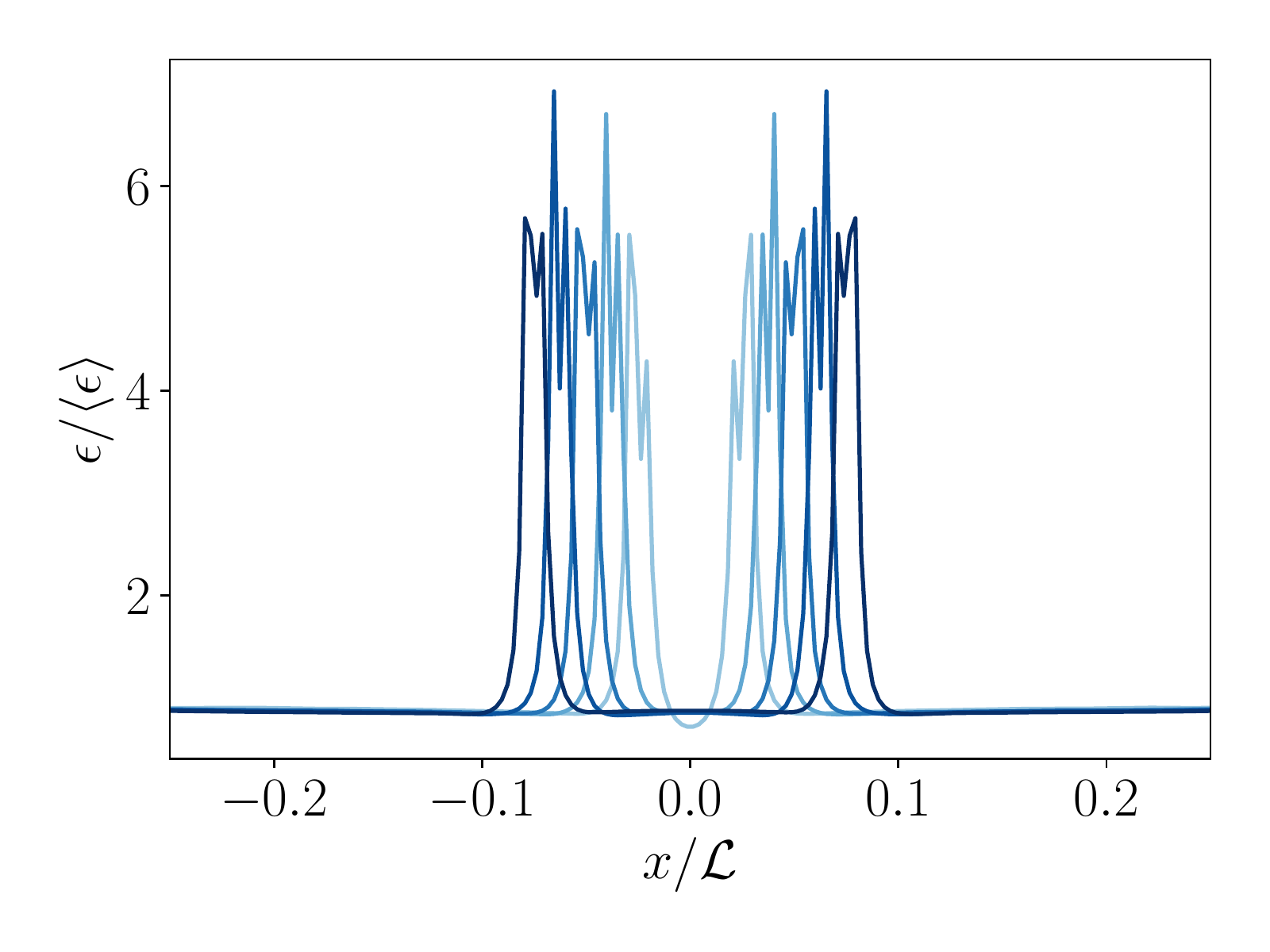}
    \includegraphics[width=.32\textwidth]{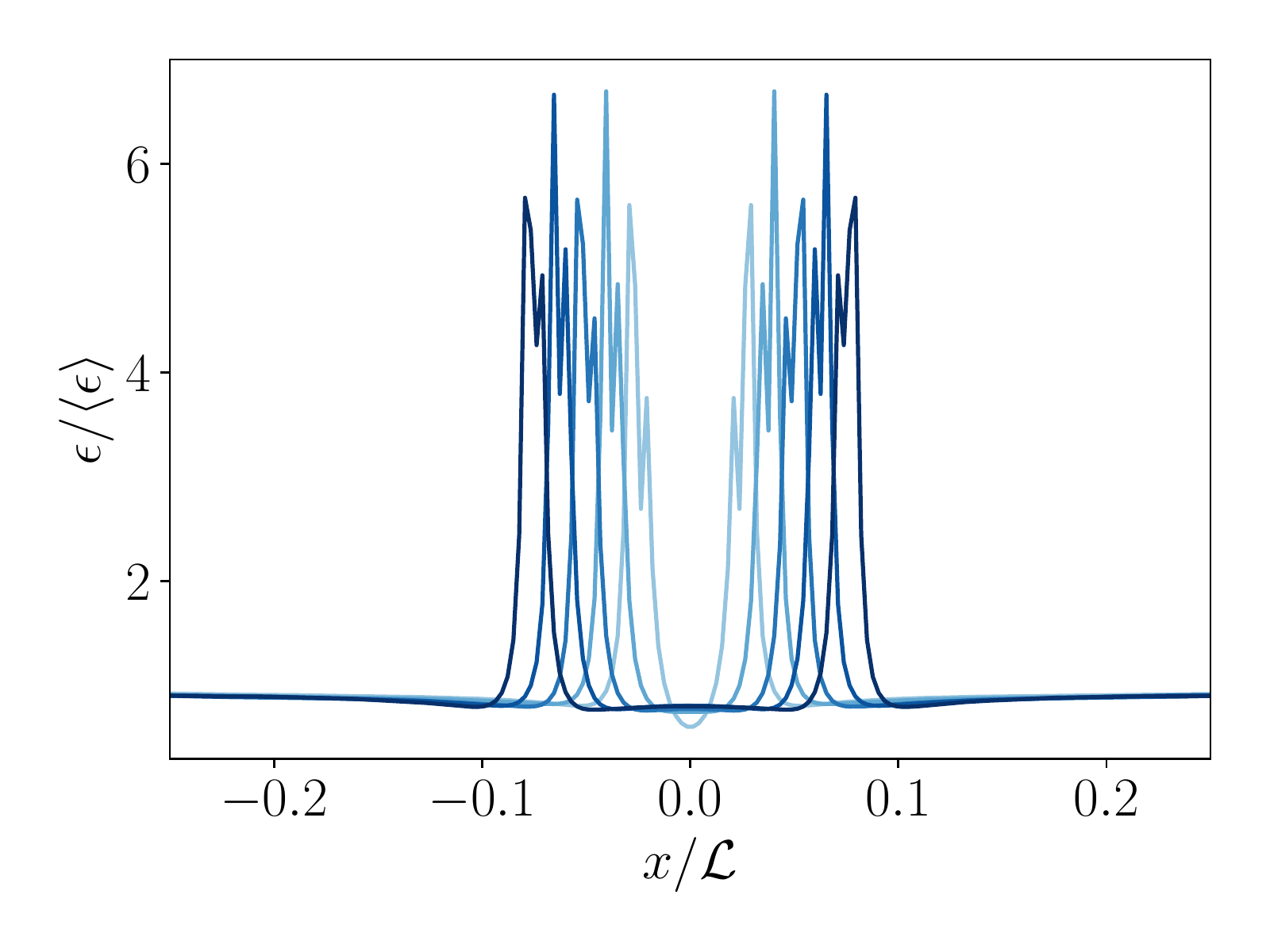}\\
    \includegraphics[width=.32\textwidth]{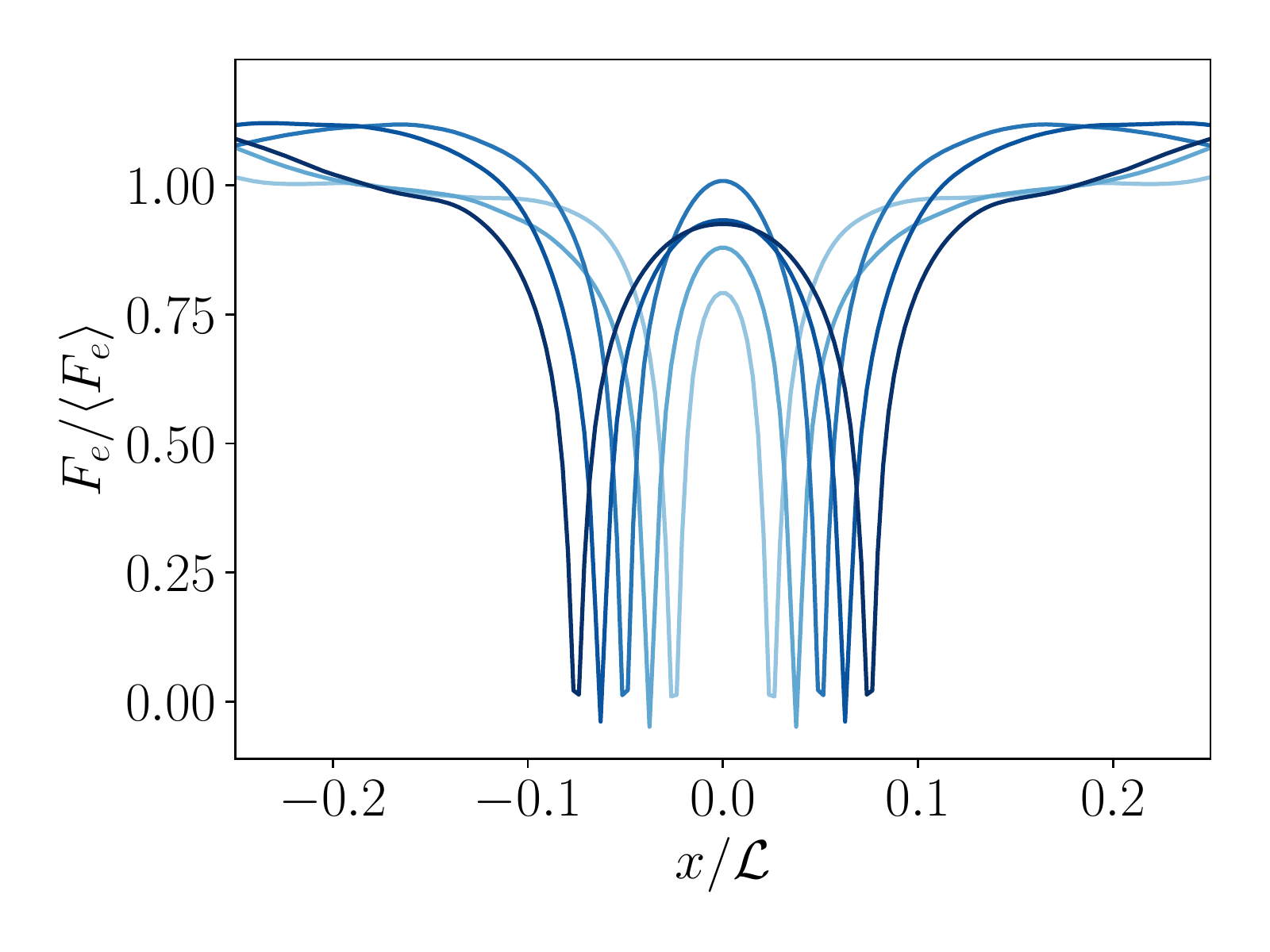}
    \includegraphics[width=.32\textwidth]{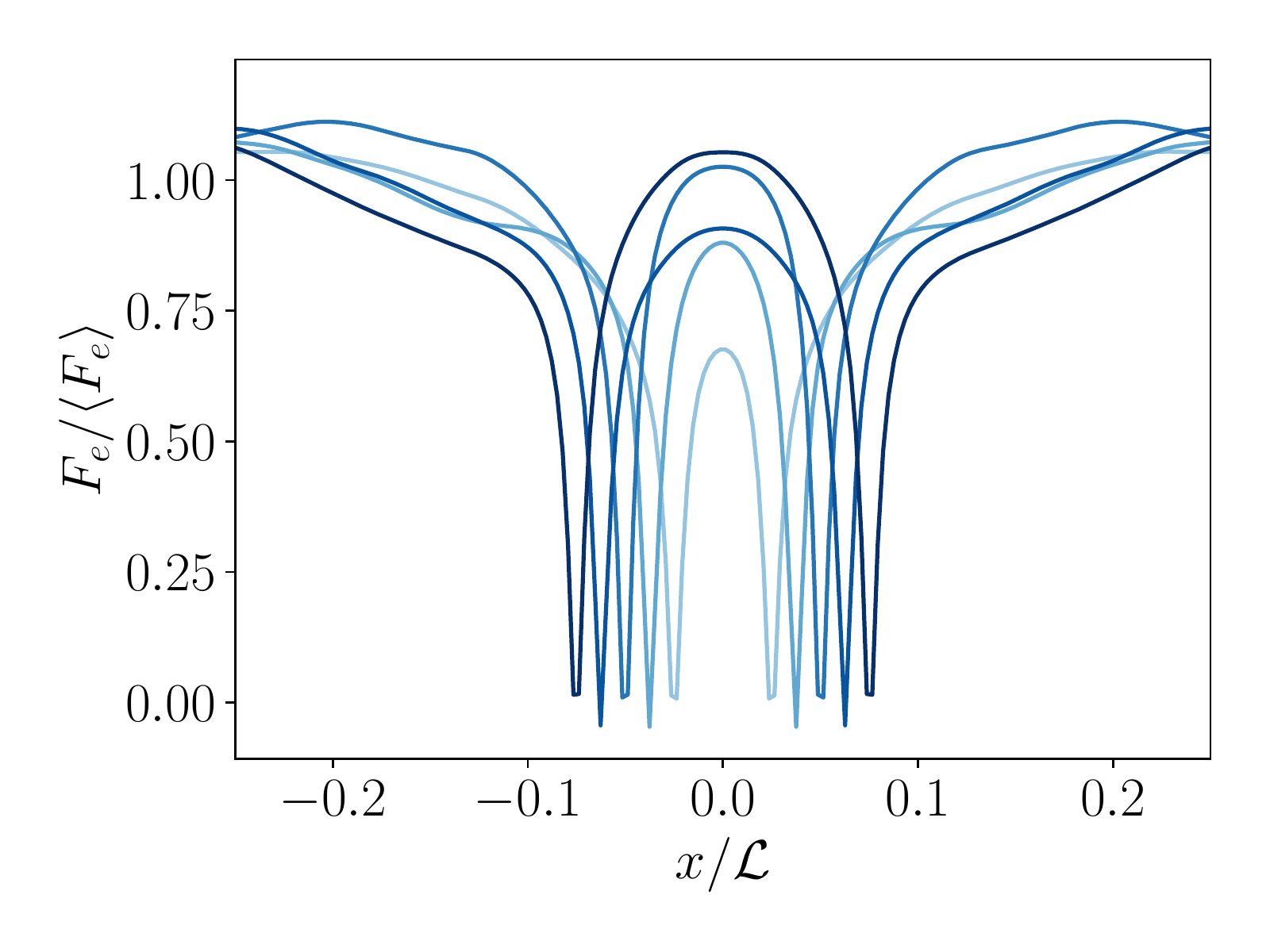}
    \includegraphics[width=.32\textwidth]{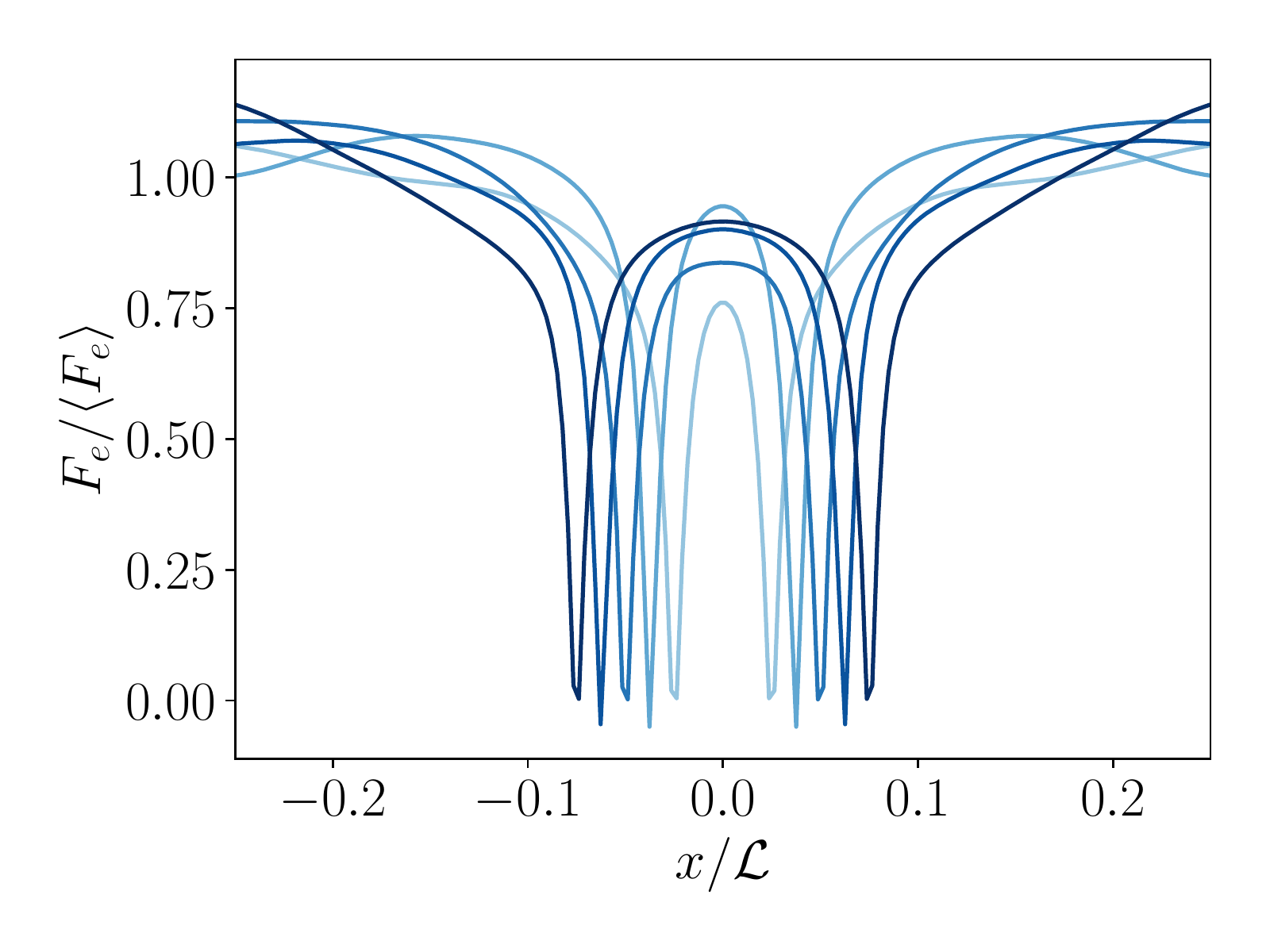}
    \caption{Flow statistics for all cases simulated in this section with $Re_\lambda=100$. The three rows represent kinetic energy (top), pressure source term $Q$ (middle) and pressure fluctuations (bottom), while the three columns represent changing plate size: $l_p/\mathcal{L}=0.1$ (left) $0.175$ (middle) and $0.25$ (right). The lines are coded from light to dark blue depending on plate separation. }
    \label{fig:stats_energ_platesize}
\end{figure}

We now turn to the energy-based quantities of the flow. We first focus on the averaged kinetic energy $k$, shown in the top row of Fig.~\ref{fig:stats_energ_platesize}. This quantity is of importance to the generation of a fluctuation force, because as mentioned in the introduction, the primary mechanism in generating fluctuation forces in non-equilibrium systems is the modification or exclusion of energy containing structures. In the top row of Figure \ref{fig:stats_energ_platesize}, we observe a similar behaviour for all three plate sizes. As could be expected, the presence of the plates causes the kinetic energy between the two plates to be reduced, and s the plates are moved further apart, the energy increases until the collective effect of the plates disappears, and they behave as `individuals'. In conventional fluid mechanics terms, the no-slip condition on the plates is also a $k=0$ boundary condition, and this causes a pseudo-boundary layer to be present in the flow where the value of $k$ slowly increases to the free-stream value.

When approaching this from the existing statistical mechanics literature on fluctuation forces, one could also say that the energy-containing structures do not ``fit'' between the plates, and this causes the energy in the slit to be lower than outside. And once the plates behave as individuals, the energy in the slit becomes approximately equal to the energy outside the slit, there is no significant exclusion, and this coincides with the force becoming increasingly small. With this explanation, i.e.~if the fluctuation force was due to the diminished energy alone, it could theoretically be derived by analyzing how the energy in the slit changes as a function of plate distance. However, in a manner consistent with Ref.~\cite{spandan2020fluctuation}, the $k$ profiles for all plate sizes remain qualitatively similar, and there is no change in the profiles of kinetic energy close to the maximum force for the medium and large plates. This means that is no reason to expect the force to be non-monotonic from analyzing the kinetic energy profiles alone.

We extend our analysis of energy-based statistics by looking at the average dissipation $\epsilon$, and the average energy injection from the random force $F_e=\textbf{u}\cdot\textbf{f}_f$, shown in the middle and bottom rows of Fig.~\ref{fig:stats_energ_platesize}. We can observe that even if there is a peak of dissipation close to the plates, the average energy dissipation in the slit increases quite rapidly to the value outside, varying only significantly for the medium and large plates at the shortest plate separation. On the other hand, the energy injected, shown in the bottom row of Fig.~\ref{fig:stats_energ_platesize}, shows no discernible patterns, even if it is clear that the injected energy approaches the value of $F_e$ outside the plates rather rapidly. 

\begin{figure}
    \centering
    \includegraphics[width=.32\textwidth]{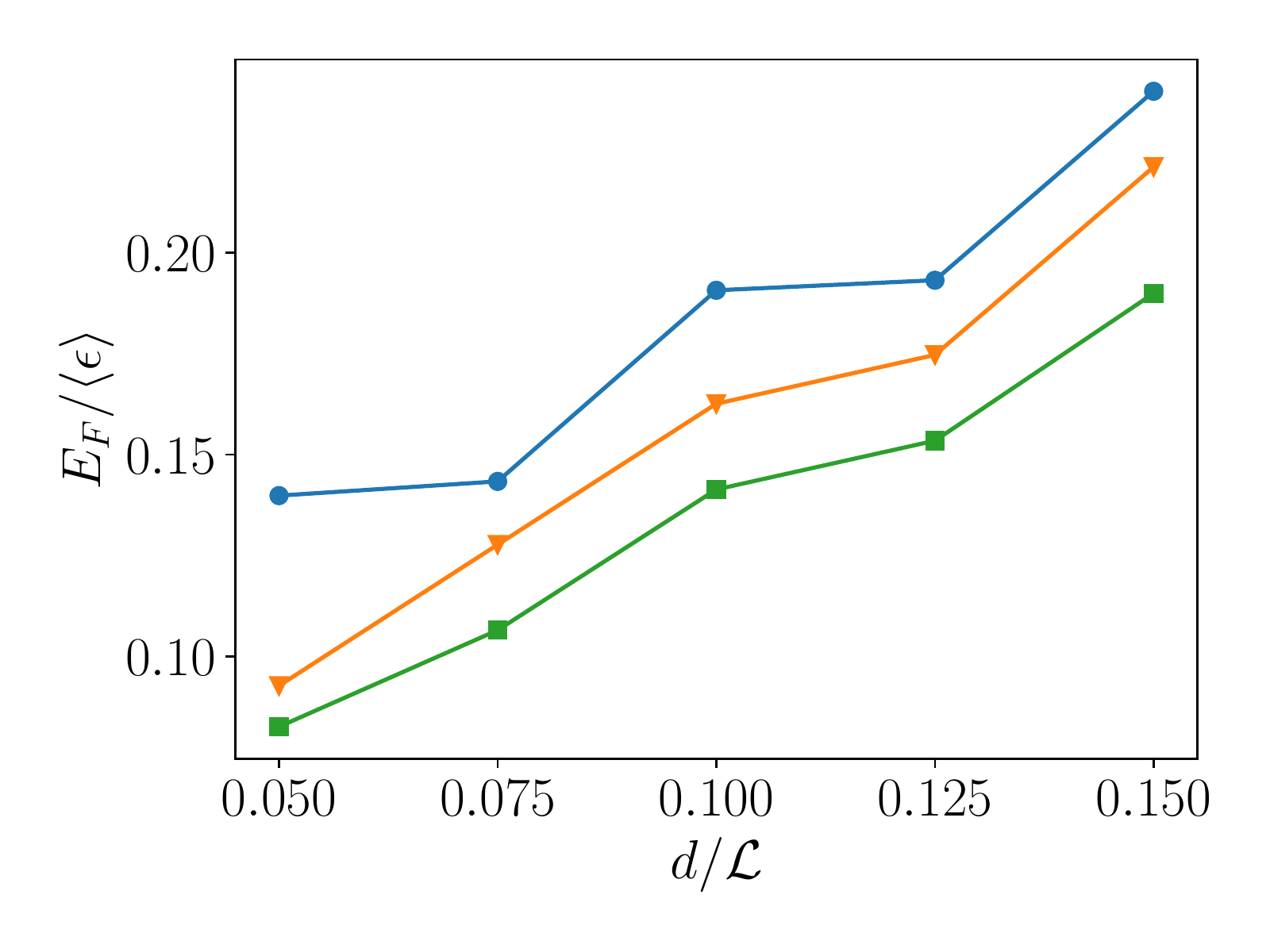}
    \includegraphics[width=.32\textwidth]{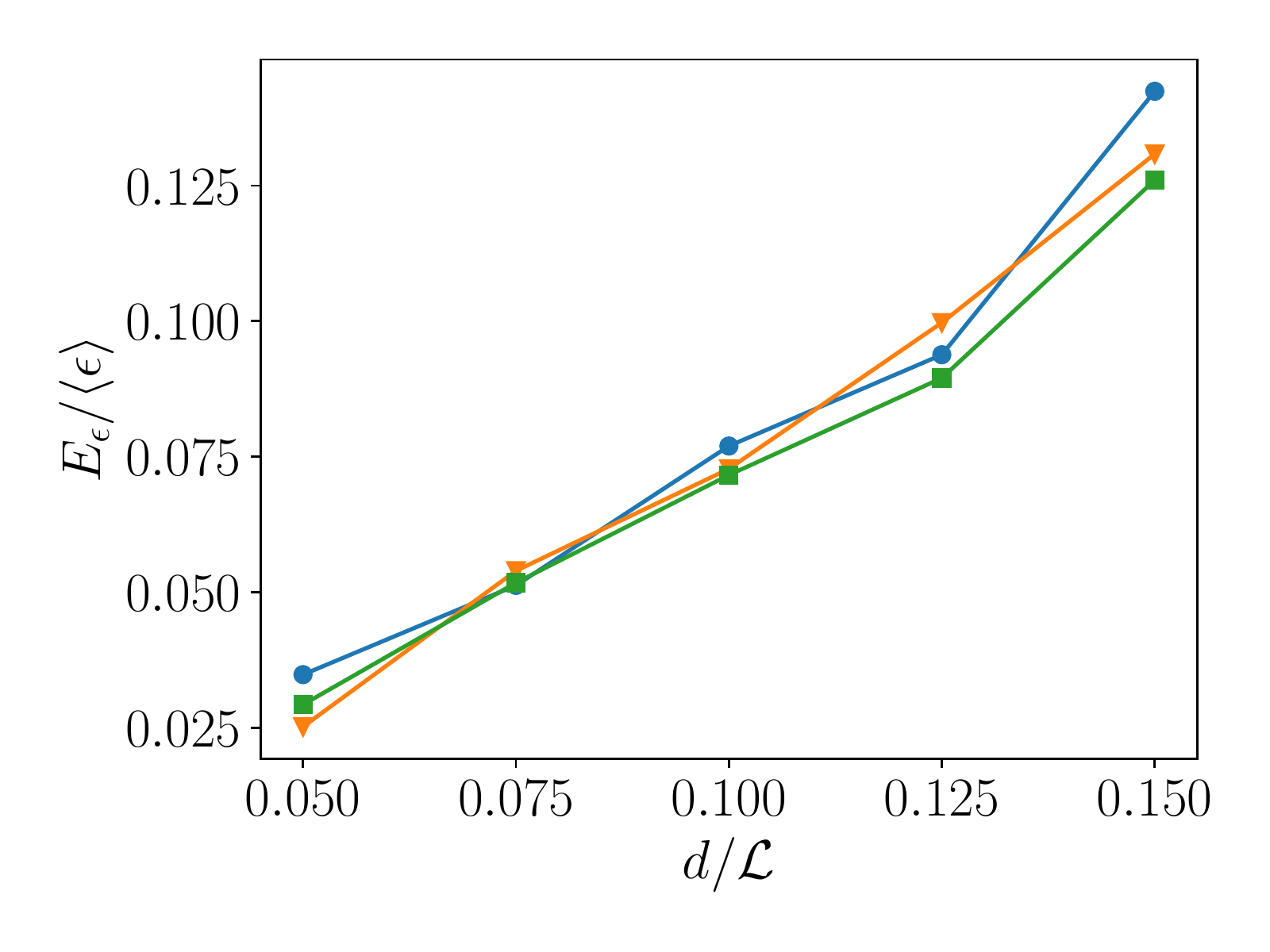}
    \includegraphics[width=.32\textwidth]{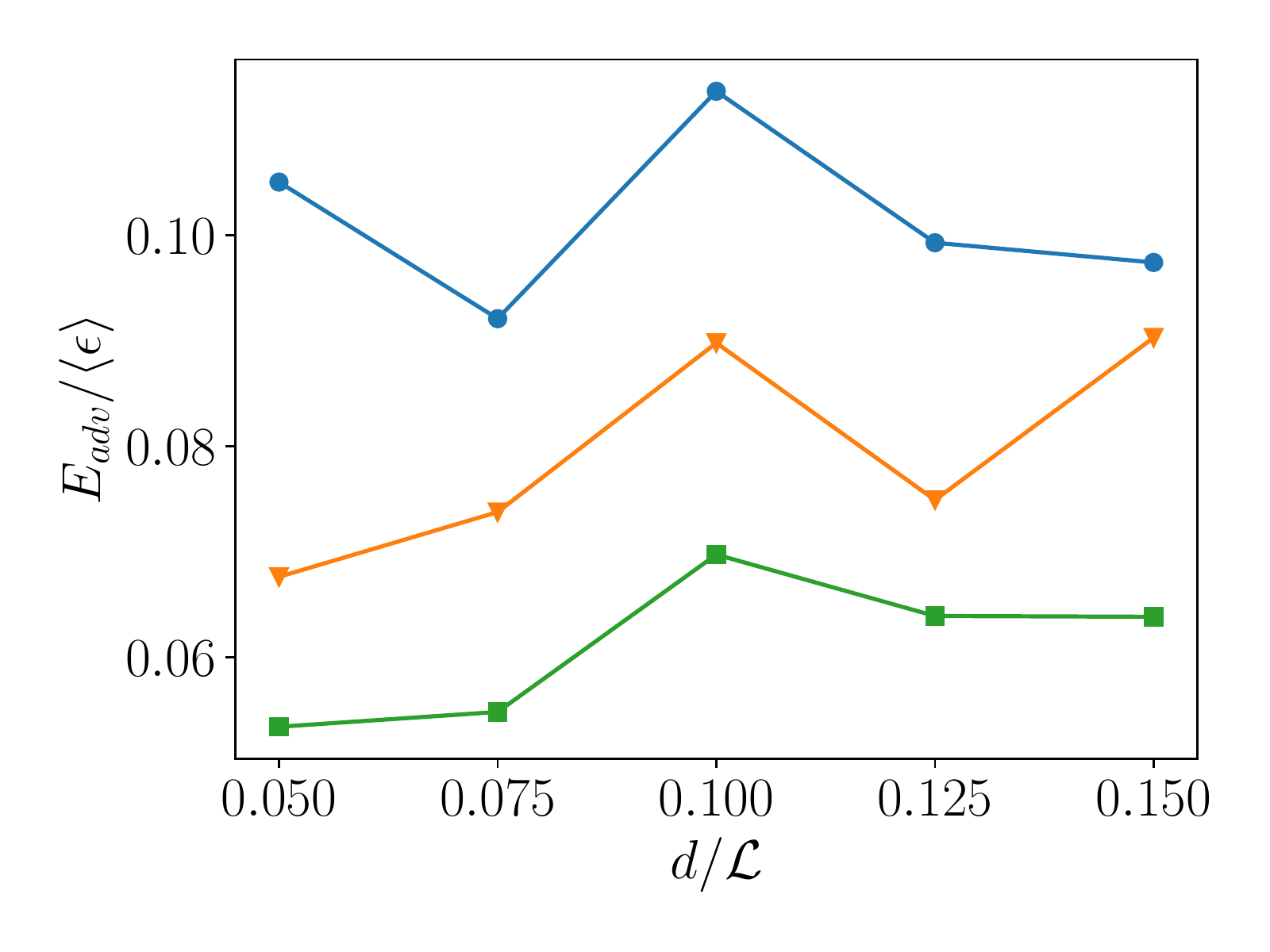}
    \caption{Energy balance in the slit for all cases simulated in this section with $Re_\lambda=100$. The left panel is the energy injected by the force in the slit, the middle panel is the energy dissipated in the slit, and the third panel is the difference between both which indicates the additional energy advected to compensate for dissipation. Symbols: blue circles are $l_p/\mathcal{L}=0.1$, orange triangles are $l_p/\mathcal{L}=0.175$ and green squares are $l_p/\mathcal{L}=0.25$. }
    \label{fig:stats_energ_balances}
\end{figure}

Taken together, these statistics imply that there is some sort of transport from the region outside the plates to the slit. We can quantify this transport by using a balance of energy. Using $E_{adv}$ for the advection of energy from the outside the slit to the slit, one can say that in the statistically steady state:

\begin{equation}
    E_{adv} = \int_{\Omega} \epsilon  dV - \int_\Omega F_e dV = E_\epsilon - E_F
\end{equation}

\noindent where $\Omega$ is the domain between the plates, i.e. the slit. In Figure \ref{fig:stats_energ_balances}, we show $E_{adv}$ for all cases considered, as well as $E_\epsilon$ and $E_F$, i.e.~the average energies injected by the force and dissipated by viscosity in the slit. We can see how there is a linear monotonic increase of the dissipation and of the energy injected with plate size, something one could expect as the integration volume becomes larger. We can see how on average these two tend to almost compensate, and the dependence of $E_{adv}$ on plate separation is not considerable. We can rationalize this by thinking that energy comes in from the outside to compensate for the increased dissipation at the plates, which is relatively independent of $d$, while the energy dissipation in the slit is relatively compensated by the force injection.

\begin{figure}
    \centering
    \includegraphics[width=.32\textwidth]{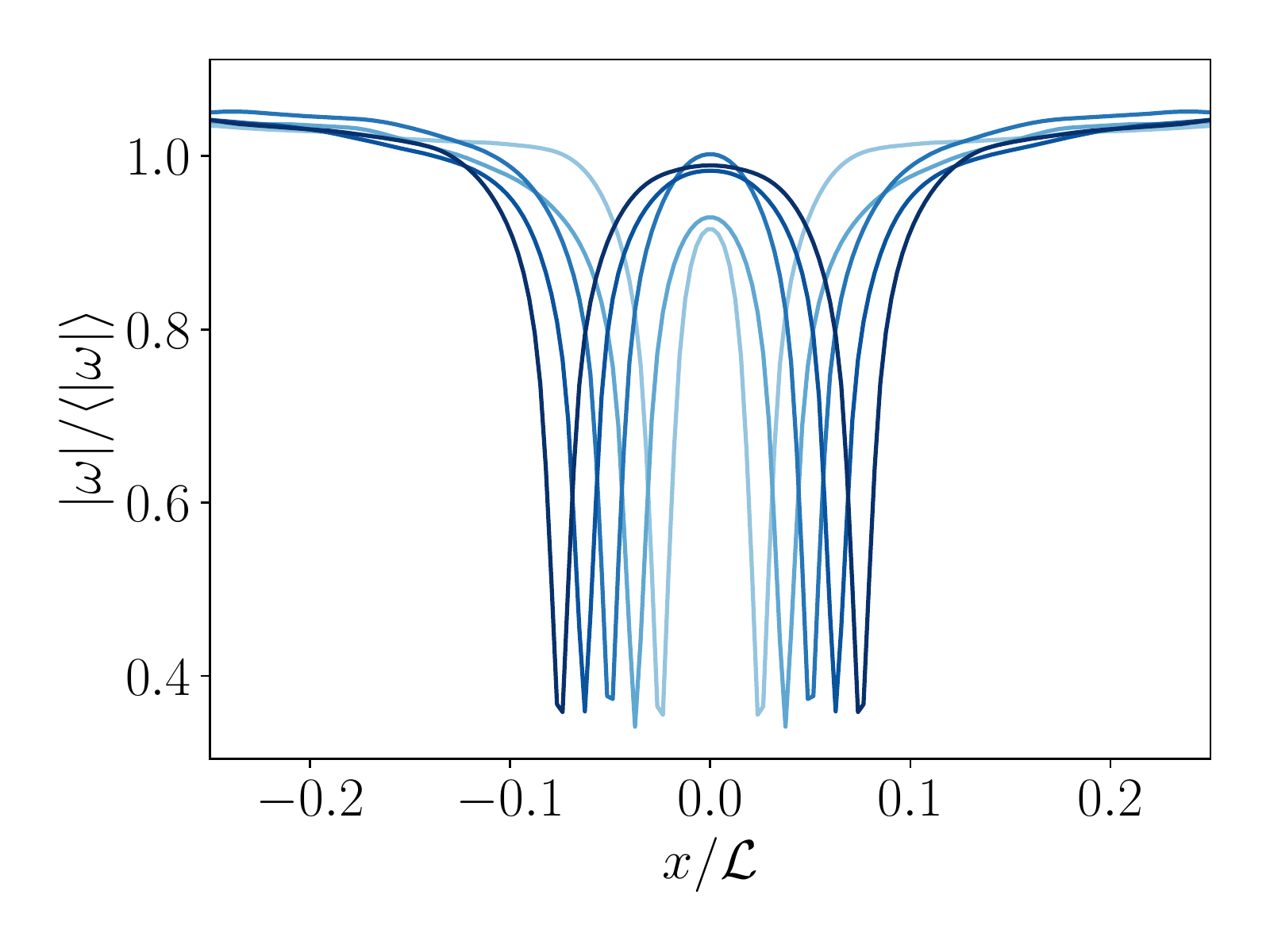}
    \includegraphics[width=.32\textwidth]{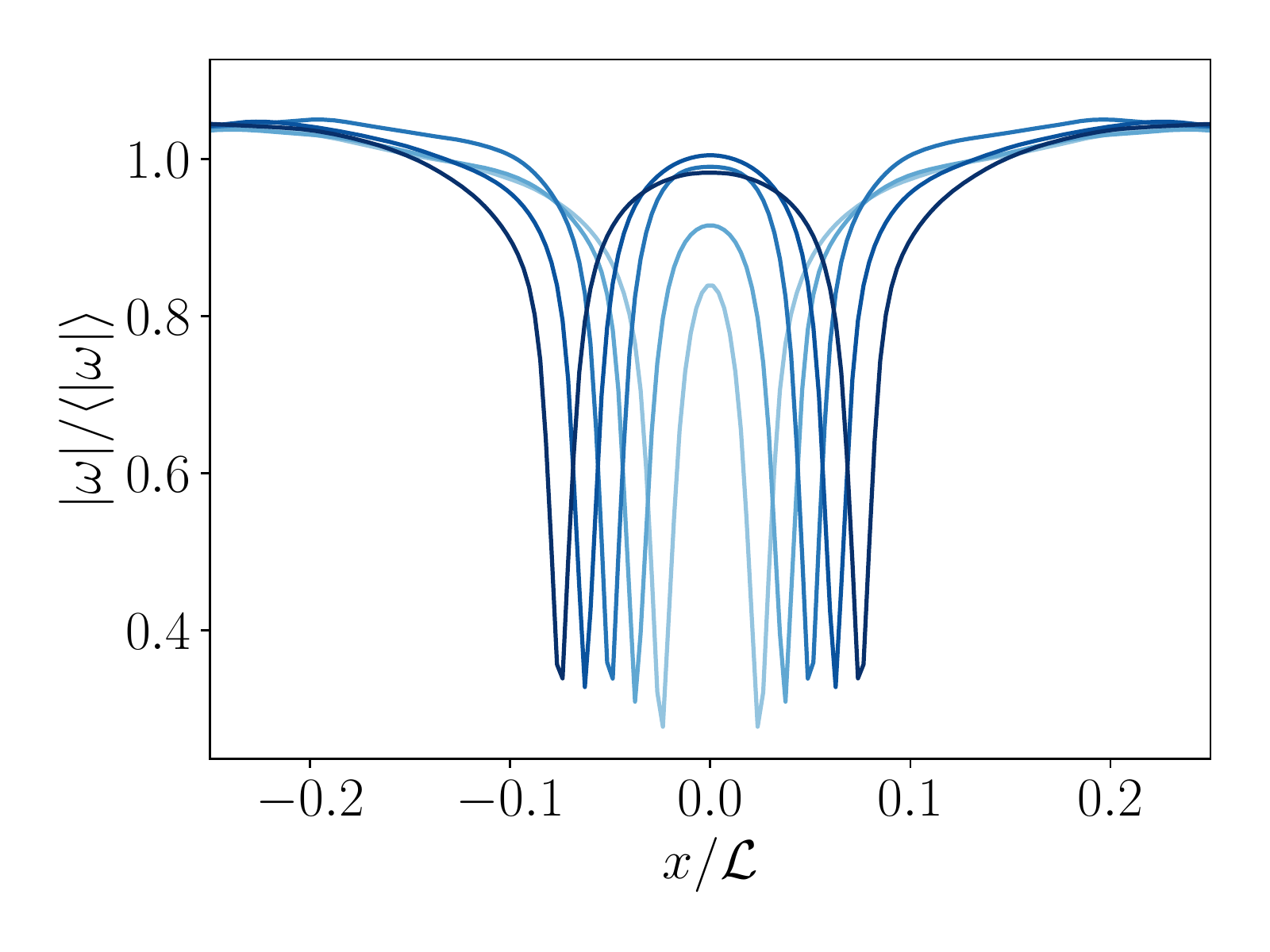}
    \includegraphics[width=.32\textwidth]{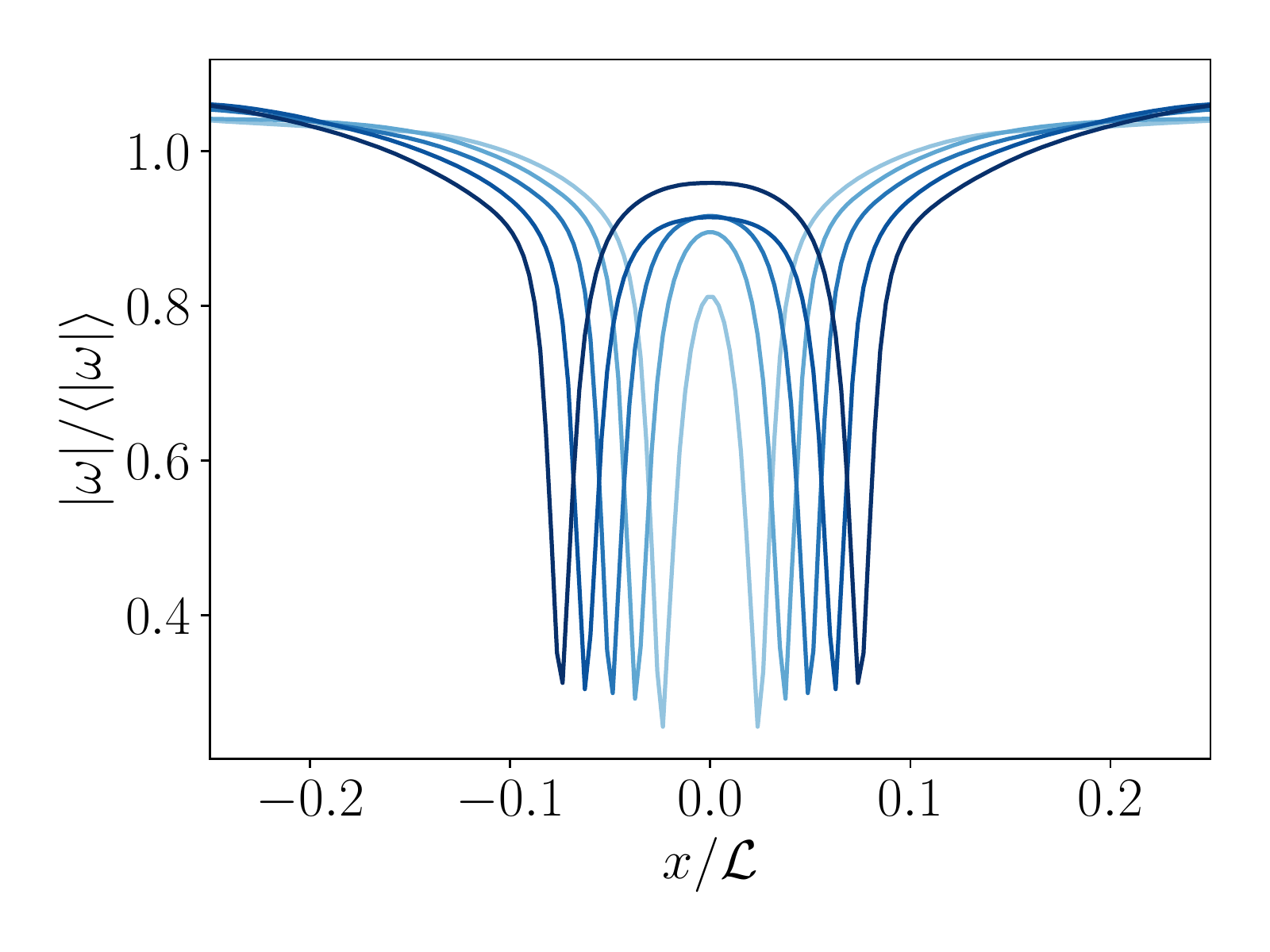}\\
    \includegraphics[width=.32\textwidth]{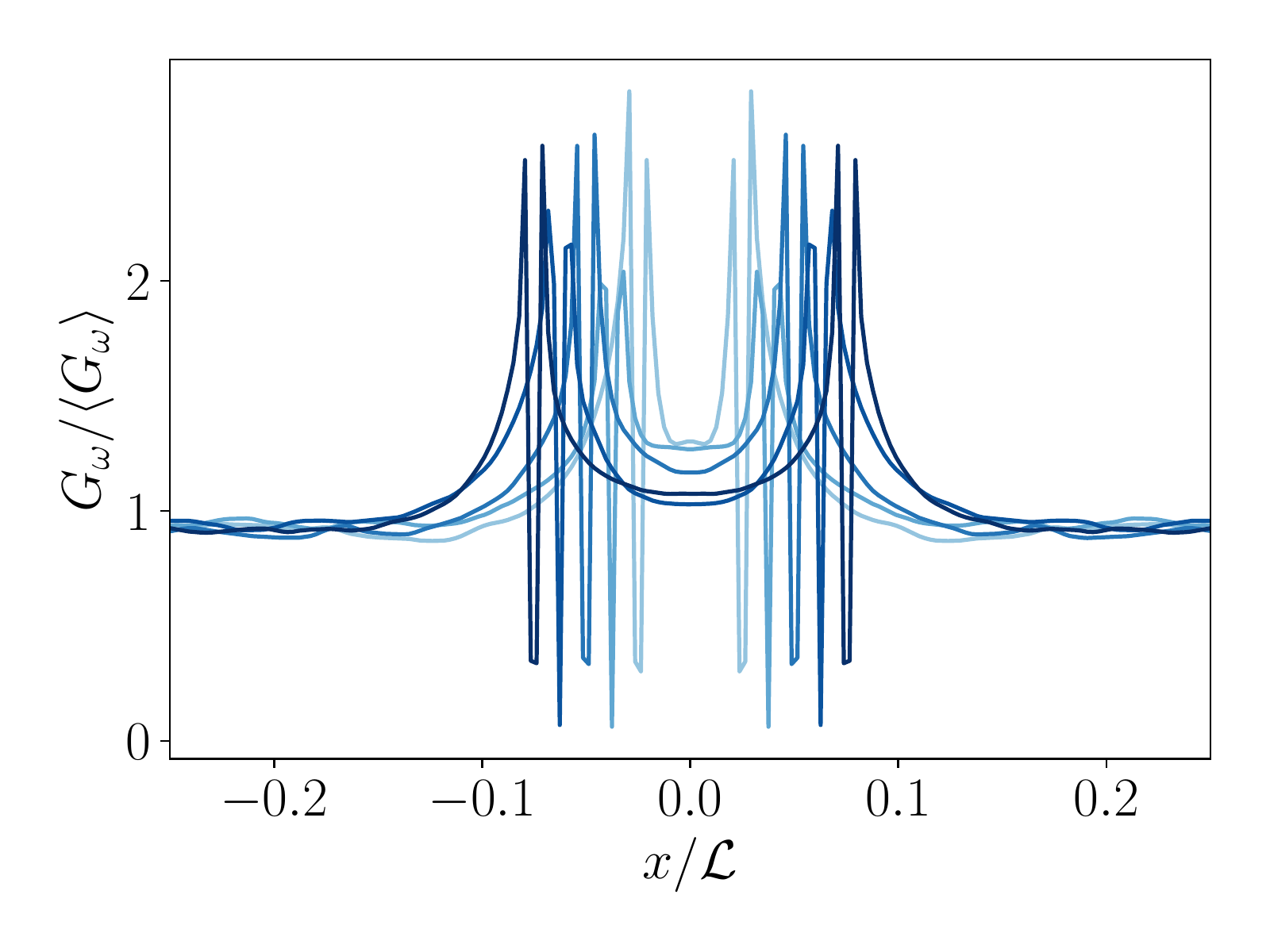}
    \includegraphics[width=.32\textwidth]{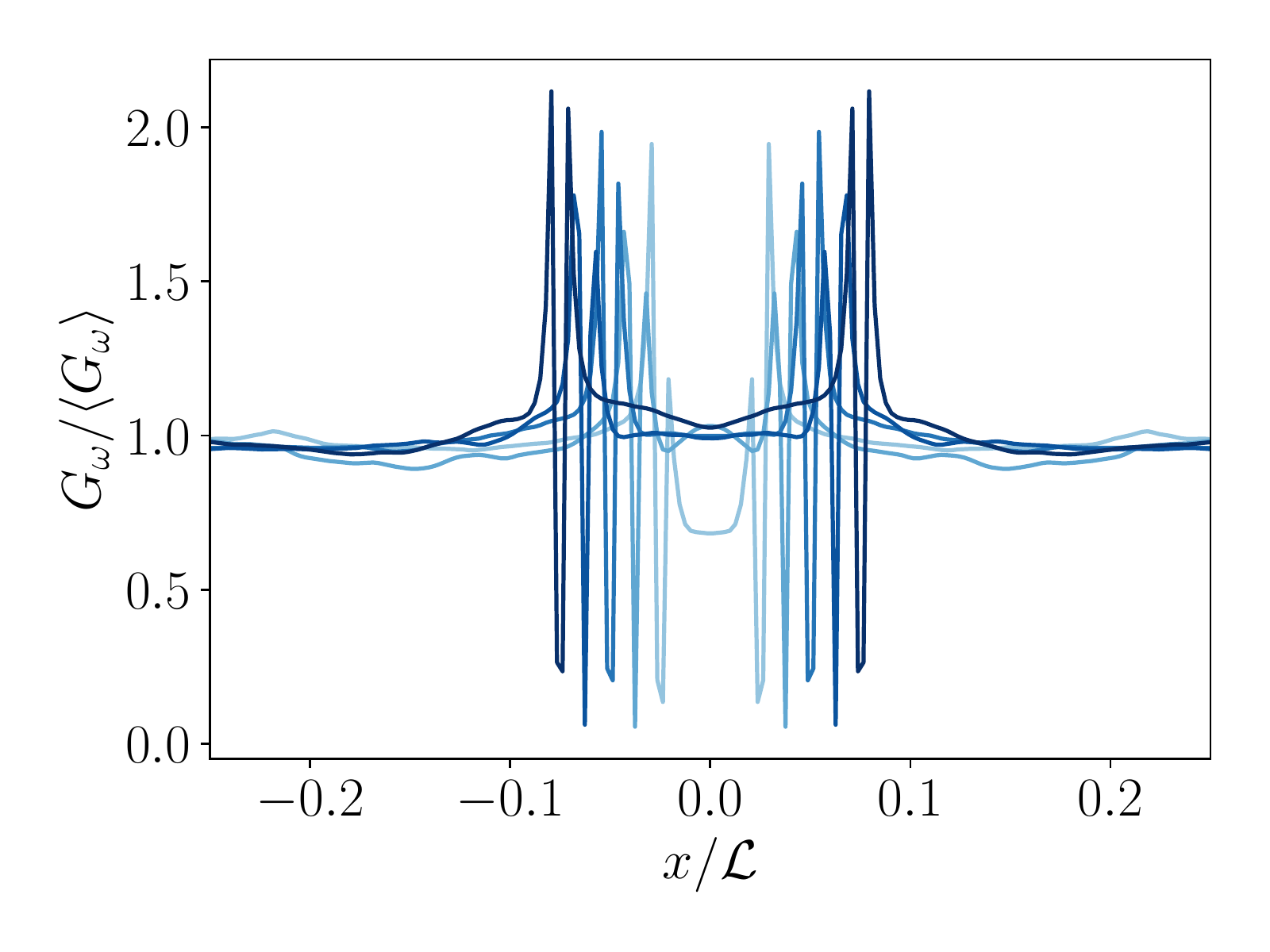}
    \includegraphics[width=.32\textwidth]{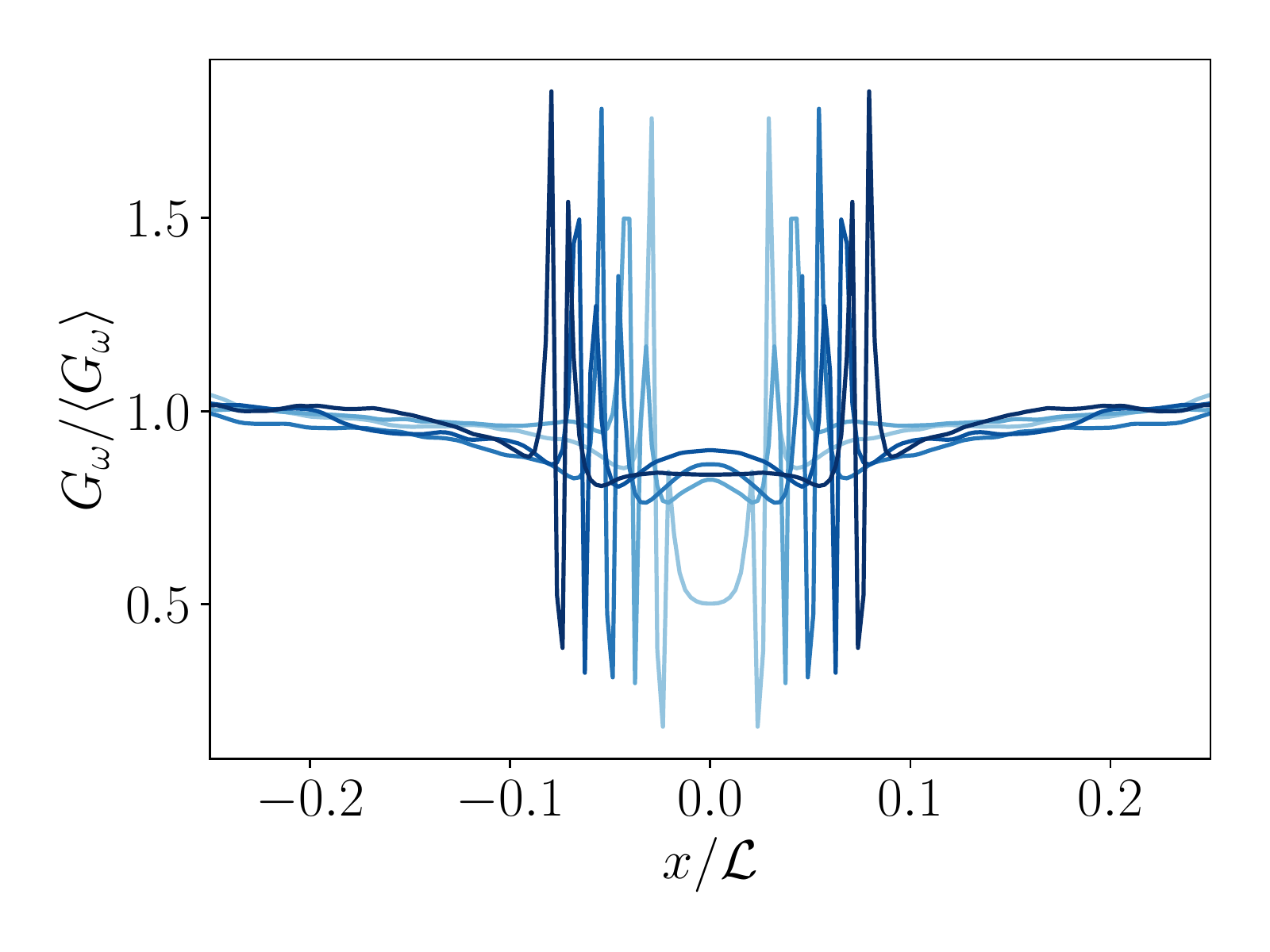}\\
    \includegraphics[width=.32\textwidth]{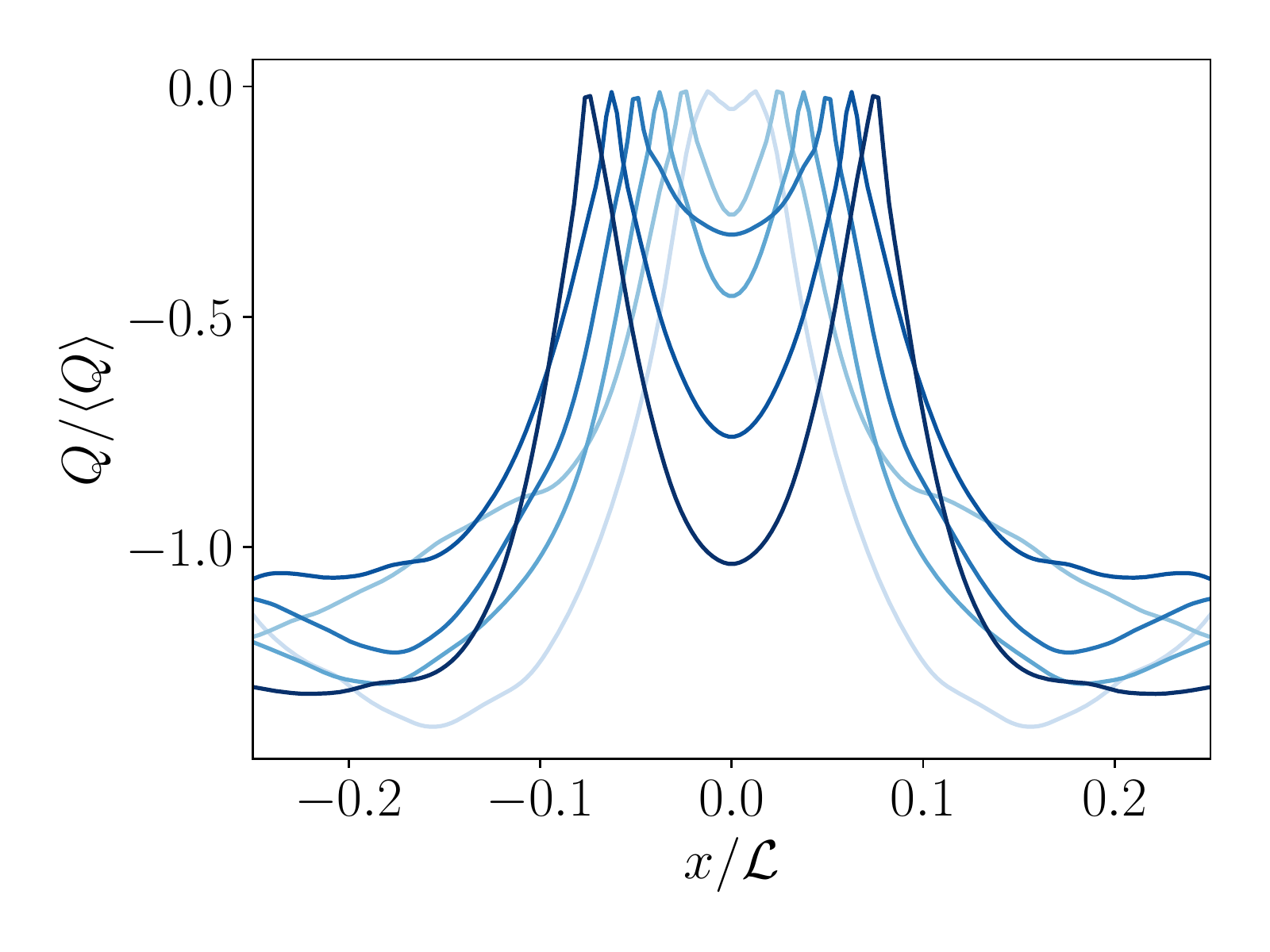}
    \includegraphics[width=.32\textwidth]{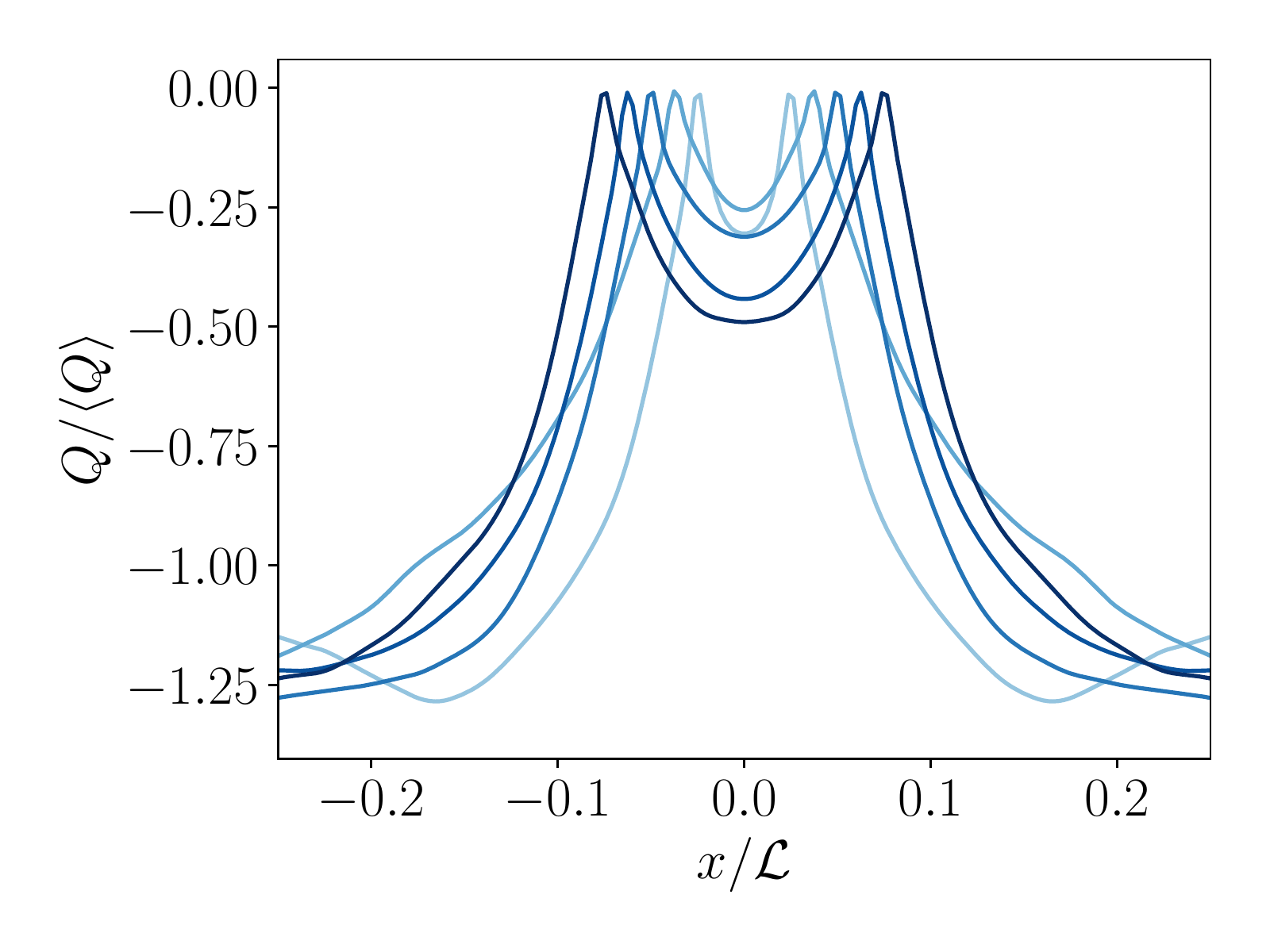}
    \includegraphics[width=.32\textwidth]{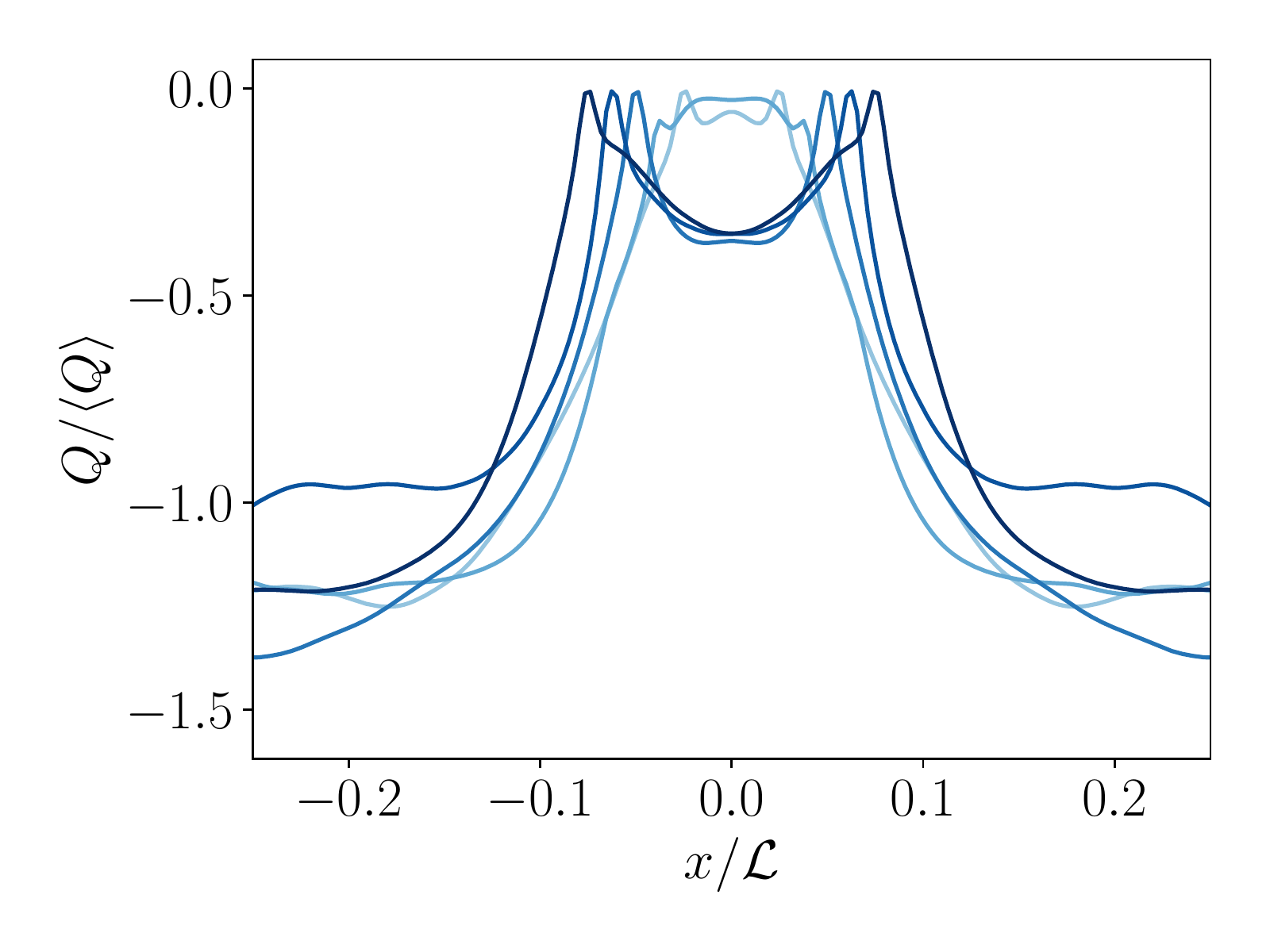}
    \caption{Flow statistics for all cases simulated in this section with $Re_\lambda=100$. The three rows represent vorticity modulus (top), vortex stretching term $G_\omega$ (middle), and pressure source term $Q$ (bottom), while the three columns represent changing plate size: $l_p/\mathcal{L}=0.1$ (left) $0.175$ (middle) and $0.25$ (right). The lines are coded from light to dark blue depending on plate separation. }
    \label{fig:stats_ww_platesize}
\end{figure}

From the discussion above, it is clear that looking at the energy statistics is not enough to explain the origins of the fluctuation force and especially its non-monotonic character. We now turn to statistics related to the second mechanism for the generation of the force proposed in Ref.~\cite{spandan2020fluctuation}, i.e.~increased vortex stretching due to the packing of structures. In the top row of Figure \ref{fig:stats_ww_platesize}, we show the behaviour of the vorticity modulus for varying plate size and plate separation. We can observe that the flow in the slit attains the same level of vorticity as the outside flow for plate separations of around $d/\mathcal{L}=0.1$ for the small and medium plates, while for the large plate the outside vorticity level is not reached for any plate separation. To explore why this is happening, we show in the middle row of Figure \ref{fig:stats_ww_platesize} the vorticity generation through vortex stretching, defined as $G_\omega = \omega \cdot (\omega \cdot \nabla u)$. Two things can be appreciated: first, that for the small plate case and for small $d$, $G_\omega$ is higher than the average value in the slit, and it slowly drops to the baseline value as the separation increases, while for large plates at small $d$, $G_\omega$ is lower than the baseline value and slowly increases as the plate separation increases. Second, we notice that there is a small local maximum in $G_\omega$ for a selected number of cases, and that these tend to coincide with the locations where $C_F(d)$ also has a local maximum. 

This change of behaviour seems to support the idea that vorticity is the crucial driver in the generation of the force. To further examine this, we turn to the pressure source term $Q$, shown in the bottom row of Figure \ref{fig:stats_ww_platesize}. In HIT, the temporal averages of $Q$ usually have a negative bias as strain dominates vorticity \cite{hunt1988eddies,pumirpressure}. Areas which have more positive values of $Q$ indicate a greater significance of structures where vorticity dominates strain rate \cite{hunt1988eddies}. Hence, in Ref.~\cite{spandan2020fluctuation} we used this variable to analyze the effect of packing worm-shaped intense vortex structures in the slit between the plates, as in our case more positive values of $Q$ would represent areas of stronger vortex stretching, where the vortex worms were interacting in close proximity and would mutually reinforce each other and decrease the pressure. Therefore, more positive values of $Q$ can be used as a proxy for measuring the strength of the flow mechanism that further drops the average pressure between the plates and causes a non-monotonic attractive force. 

Returning to the data, we observe that the average value of $Q$ for the small plate rapidly becomes negative as plate distance increases, while it remains closer to zero for the medium plate, and can even attain positive values for the large plate. Following the reasoning in the above paragraph, this indicates the increasing significance of vortex stretching as the plates become larger. Increasing plate size means the plates pack more vortical structures between them, aligning them in preferential directions and forcing them to interact closely as hinted by the earlier visualizations. This generates more vortex stretching, further vorticity enhancement and larger pressure drops which cause a more pronounced maximum in the $C_F(d)$ curve. This enhancement in vortex stretching was also reflected on the pressure fluctuations, shown in the bottom row of Figure \ref{fig:stats_pr_platesize}. In accordance with the $Q$ data, the $p^\prime$ profiles show that the small plate affects the distribution of $p^\prime$ much less than the medium and large plates. 

\begin{figure}
    \centering
    \includegraphics[width=.50\textwidth]{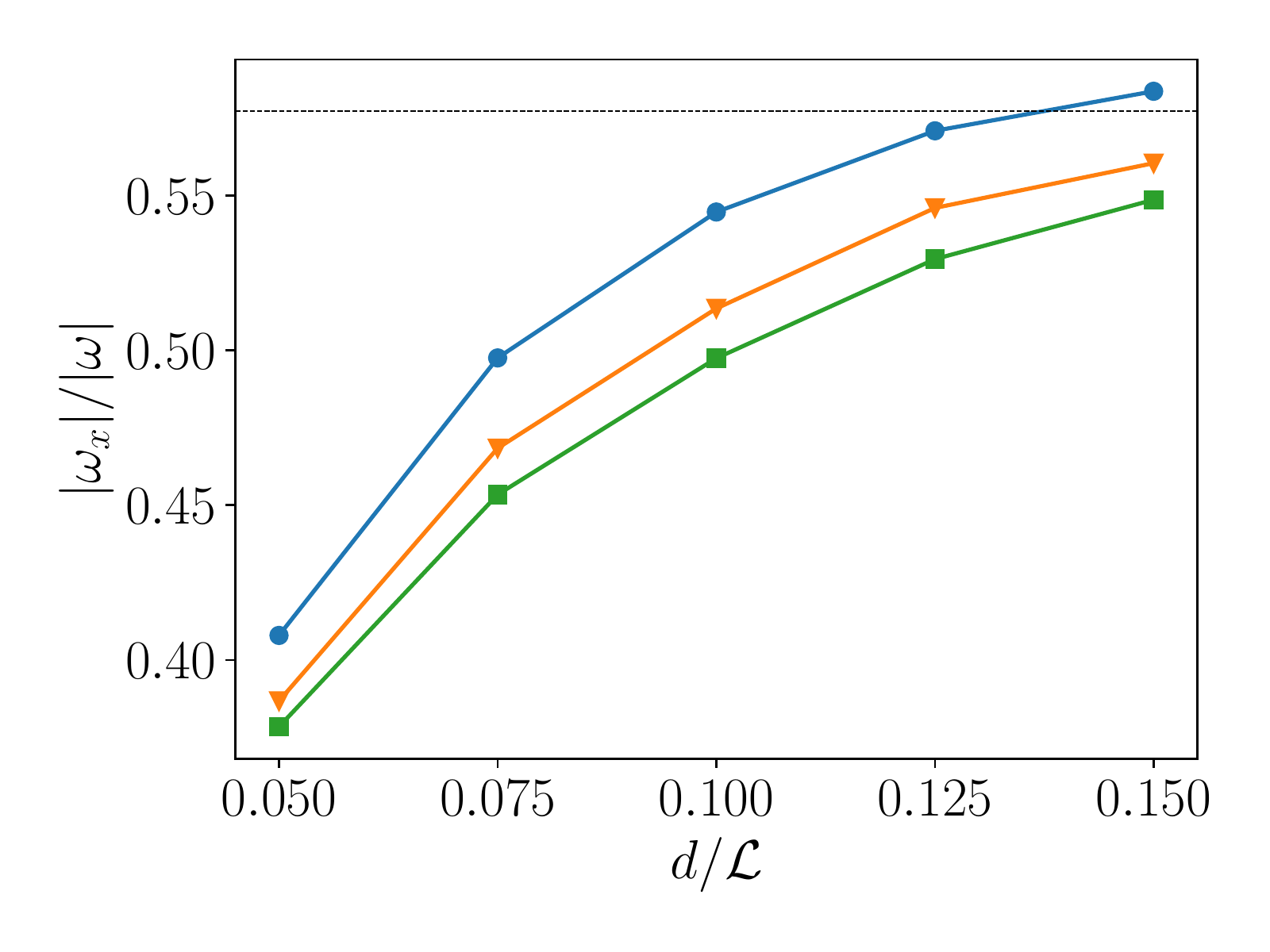}
    \caption{Vorticity anisotropy as a function of plate separation. The horizontal dashed line indicates the asymptotic value $1/\sqrt{3}$. Symbols: blue circles are $l_p/\mathcal{L}=0.1$, orange triangles are $l_p/\mathcal{L}=0.175$ and green squares are $l_p/\mathcal{L}=0.25$. }
    \label{fig:vort_aniso}
\end{figure}

For completion, we also confirm that vorticity is preferentially aligned in directions parallel to the plates by comparing the root-mean-squared value of $\omega_x$, i.e. the vorticity component normal to the plate, to the total vorticity magnitude. If vorticity is isotropic, this ratio should approach $1/\sqrt{3}$, while if it is lower, it will indicate that vorticity has some preferential alignment. Indeed, this is what the results shown in Figure \ref{fig:vort_aniso} show: vorticity is only isotropic for the largest plate separations, when the plates behave as individuals.

The flow statistics shown here lend weight to the hypothesis that the fluctuation force is mainly generated from interactions related to vorticity. However, we cannot totally rule out energy-related mechanisms, which are similar to those present in other non-equilibrium systems. 
In the next section, by progressively reducing the Reynolds number of the flow we explore what happens when vortex stretching is removed from the picture.

\subsection{The low Reynolds number limit}
\label{sec:rey}

\begin{figure}
    \centering
    \includegraphics[trim={5.5cm 0 5.5cm 0cm}, clip,width=.32\textwidth]{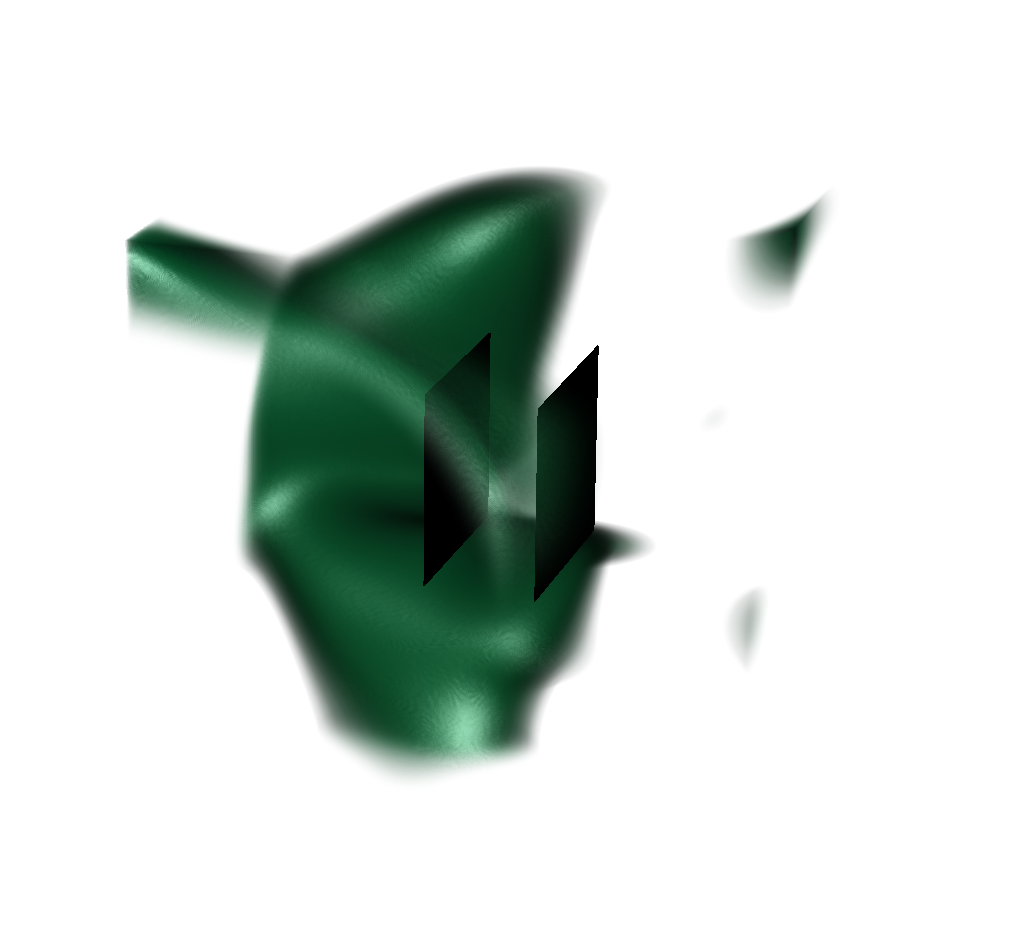}
    \includegraphics[trim={5.5cm 0 5.5cm 0cm}, clip,width=.32\textwidth]{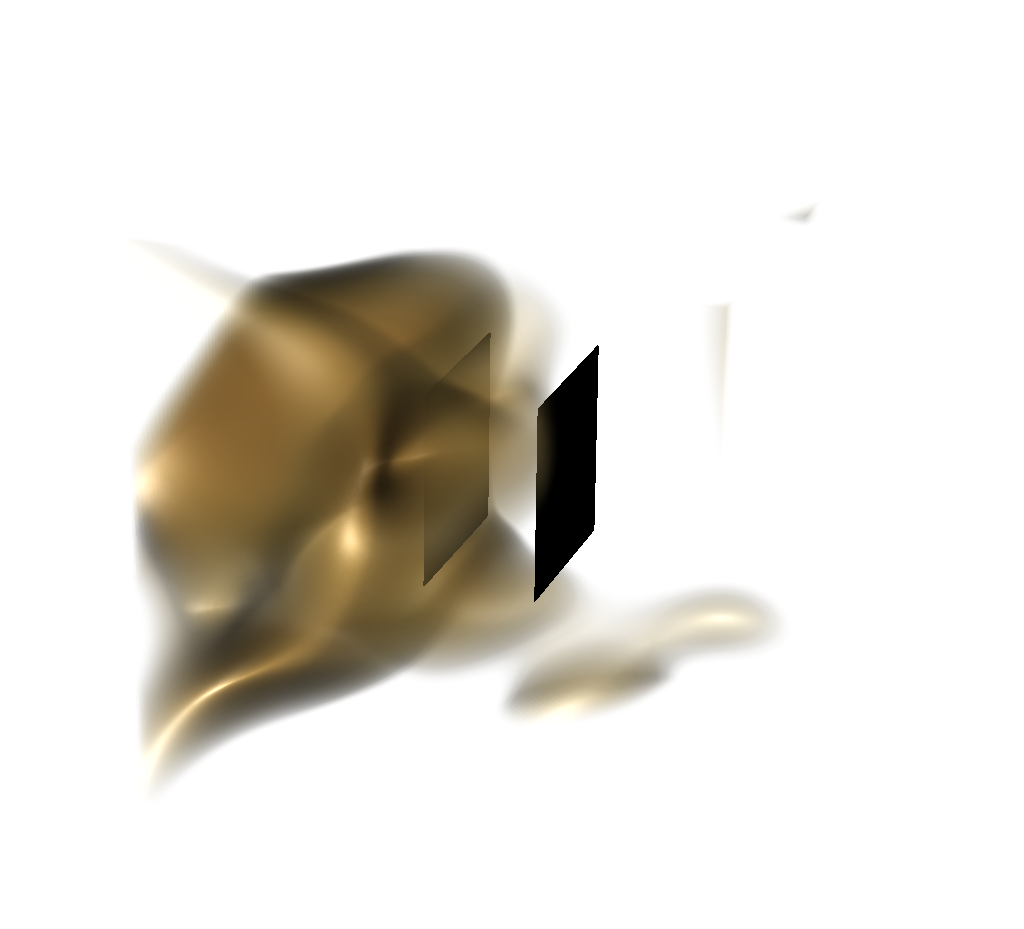}
    \caption{Volume visualization of the pressure source term $Q$ (left) and the kinetic energy $k$ (right) at $Re_\lambda=22$, $d/\mathcal{L}=0.1$ and $l_p/\mathcal{L}=0.175$. Color code as in Figure \ref{fig:vis_platesizes}. For clarity only a section of the computational domain is shown.}
    \label{fig:vis_re20}
\end{figure}

For the second set of simulations, we analyze the effect of the Reynolds number on the generation of the fluctuation force. We start by fixing the plate size to be $l_p/\mathcal{L}=0.175$, i.e.~a medium plate size, and set the plate separation at $d/\mathcal{L}=0.1$, around the force maximum. We vary $Re_\lambda$, and examine the force coefficient and other flow statistics as the flow becomes less turbulent. To better understand the changes in the flow as $Re_\lambda$ becomes small, Figure \ref{fig:vis_re20} shows an instantaneous visualization of $Q$ and $k$ for $Re_\lambda=22$. A drastically increased length-scale for the $Q$ structures can be observed, as $\eta_K$ is more than one order of magnitude larger at this $Re_\lambda$ when compared to the $Re_\lambda$ of Figure \ref{fig:vis_platesizes}. On the other hand, the energy structures remain at similarly large sizes, even if they have lost their small-scale features. 

\begin{figure}
    \centering
    \includegraphics[width=.45\textwidth]{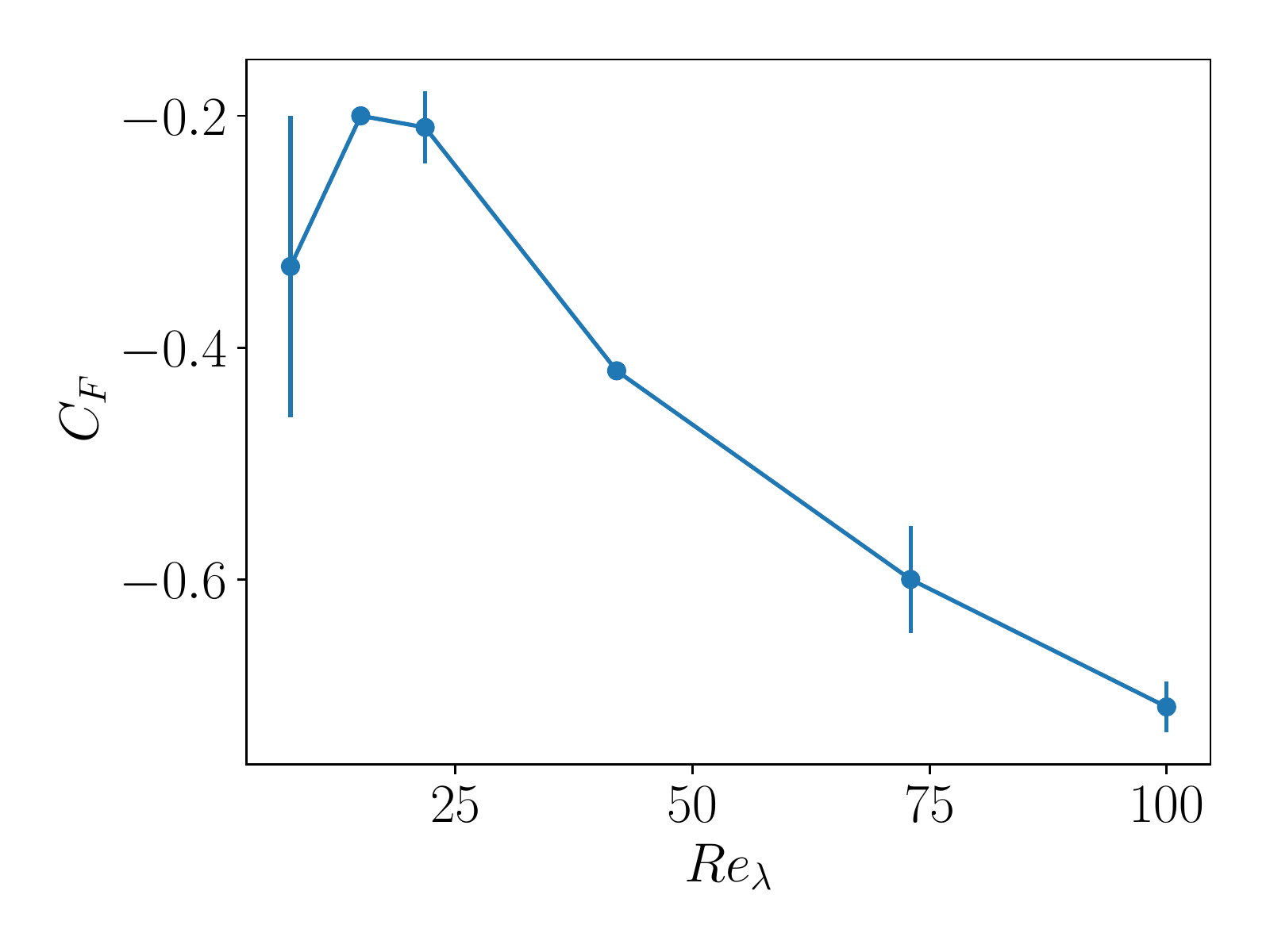}
    \includegraphics[width=.45\textwidth]{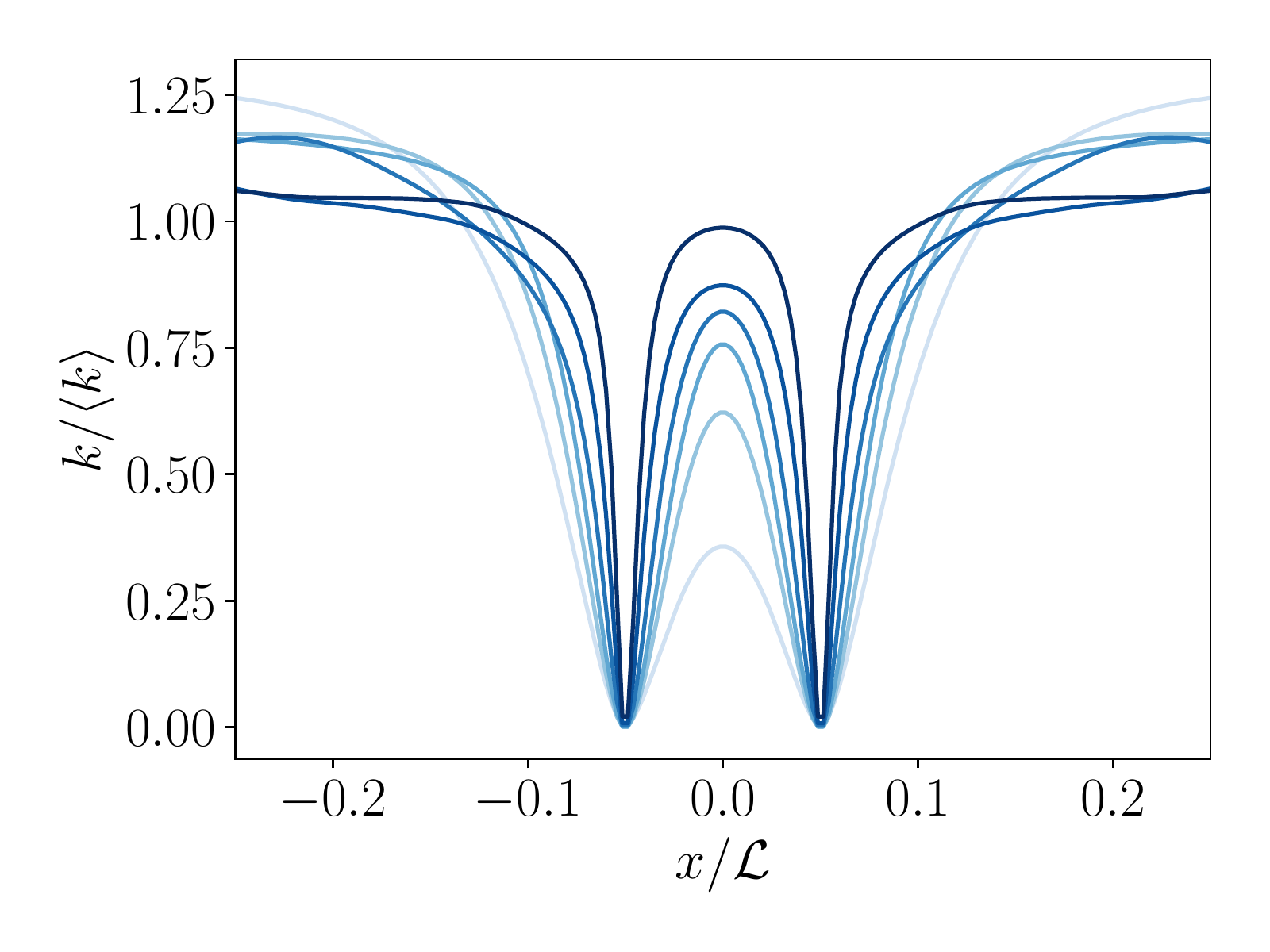}
    \includegraphics[width=.45\textwidth]{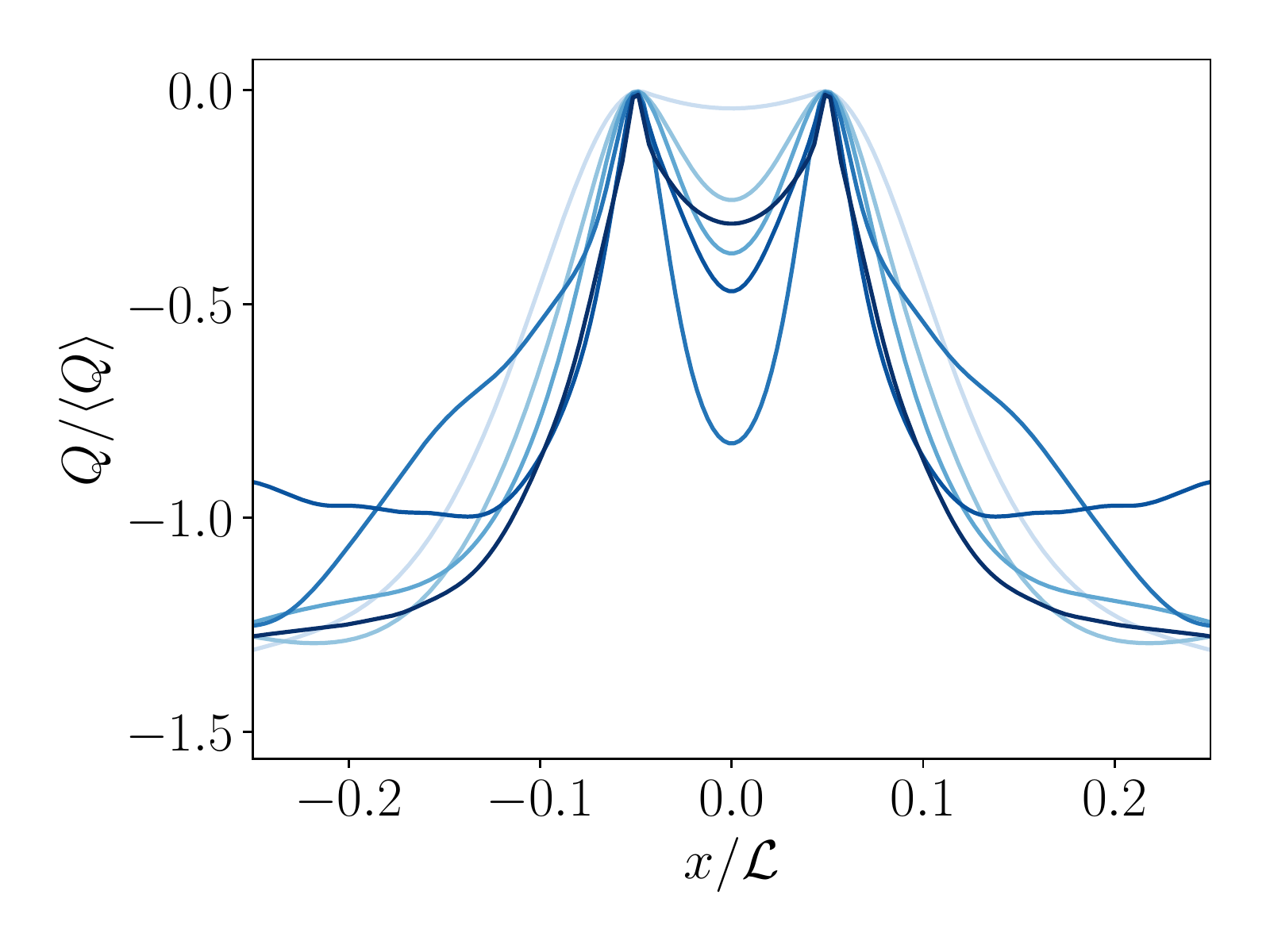}
    \includegraphics[width=.45\textwidth]{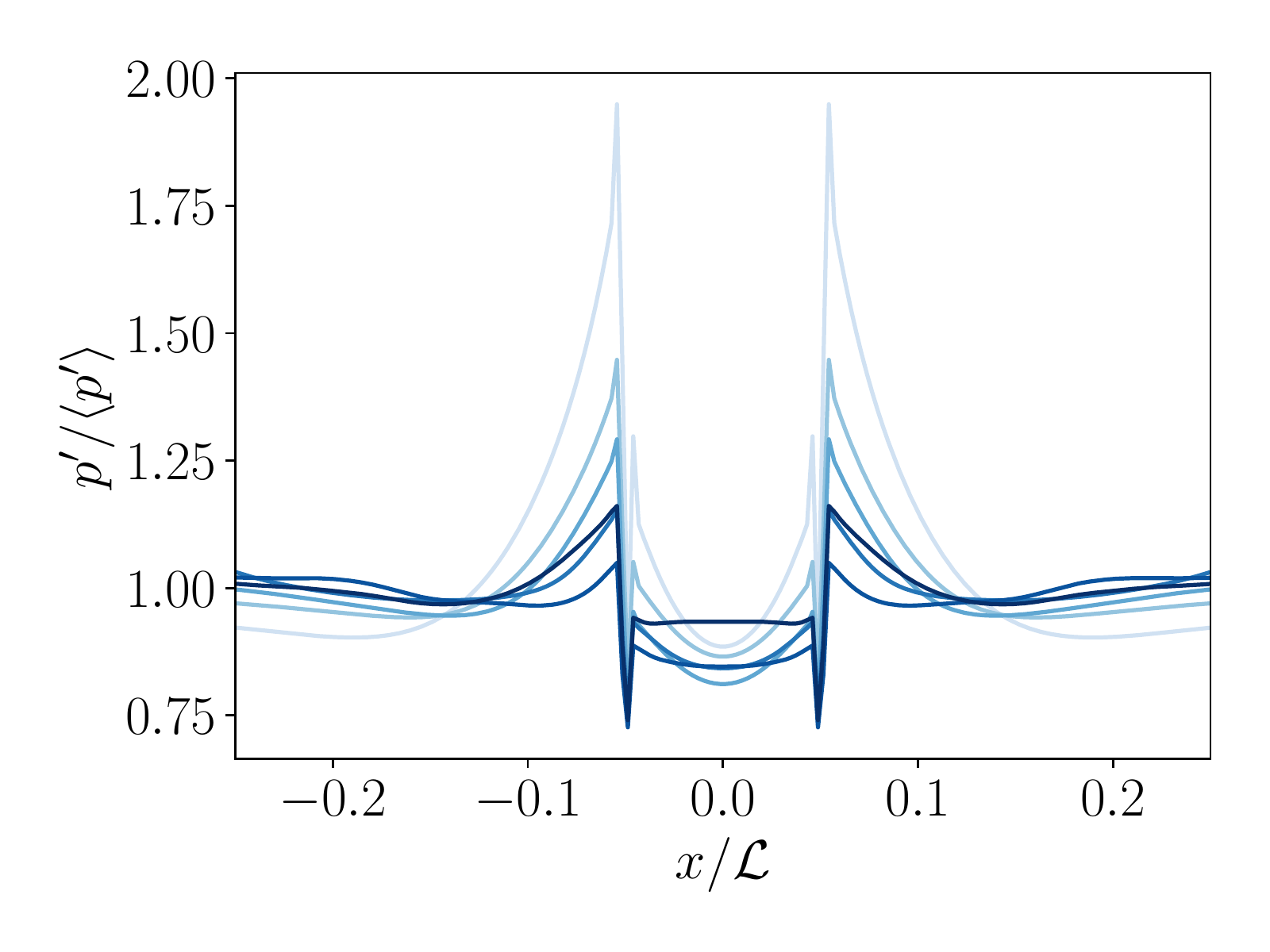}
    \caption{Top left: Non-dimensionalized force coefficient as a function of $Re_\lambda$ for $l_p/\mathcal{L}=0.175$ and  $d/\mathcal{L}=0.1$. Top right: Spatial distribution of the averaged kinetic energy for varying $Re_\lambda$ with $l_p/\mathcal{L}=0.175$ and $d/\mathcal{L}=0.1$.  The lines are coded from light to dark blue representing increasing $Re_\lambda$ as in Table \ref{tab:my_label}. Bottom left: Same as top-right for the pressure source term $Q$. Bottom right: Same as top-right for the pressure fluctuations.}
    \label{fig:cf_rel}
\end{figure}

Turning to quantitative data, in the top left panel of Figure \ref{fig:cf_rel}, we show the results for the force coefficient $C_F$ as a function of $Re_\lambda$. As we could expect from Ref.~\cite{spandan2020fluctuation}, the force becomes smaller and smaller as the Reynolds number decreases, reaching a minimum attractive force of around $C_F\approx-0.2$ at $Re_\lambda\approx 15$. For this Reynolds number, we expect no small and intense vortical structures, as hinted by Figure \ref{fig:vis_re20}. Hence, vortex stretching will have been removed as a mechanism for generating the fluctuation force. 

We also note that for the lowest Reynolds number considered ($Re_\lambda=7.6$), the force appears to increase again. However, the error bars on this are very large- even after collecting statistics for a long period. With turbulence largely absent, other physical mechanisms, or even the character of the forcing could be responsible for this increase in the attractive force. We will not further speculate on this phenomenon, as it is unclear whether it is dominated by the same mechanisms as the turbulent fluctuation force due to the lack of turbulence at this $Re_\lambda$.

To further analyze the force generation, we show in the remaining panels of Figure \ref{fig:cf_rel} the spatial behaviour of the time-averaged flow attributes analyzed in the previous section. The top right panel shows the kinetic energy $k$, which has a relatively simple behaviour which does not deviate from expectations: for the lowest $Re_\lambda=7.6$, due to the effect of viscosity, the kinetic energy is very low between the plates as the boundary layers from both plates protrude heavily into the flow. As $Re_\lambda$ increases, the plate boundary layers become thinner, the ability of the plates to exclude the energy structures becomes smaller, and the kinetic energy in the slit increases, until it almost reaches the outside level for $Re_\lambda=100$.

The behaviour of the pressure source term $Q$ and the pressure fluctuations is more complex. In the slit, $Q$ is most positive for the lowest Reynolds number considered, indicating the important role of vorticity. As $Re_\lambda$ increases, $Q$ steadily drops, meaning strain becomes predominant over vorticity, until a minimum is reached for $Re_\lambda=42$. Further increasing $Re_\lambda$ causes $Q$ to become less negative, indicating that vorticity is slowly recovering its importance in the slit. We can attribute this changing behaviour to the two different kinds of vorticity in the flow. At low Reynolds numbers, vorticity originating from the no-slip condition at the plates will be dominant, and cause more positive values of $Q$. At high Reynolds number, vorticity will come from the small vortical structures (worms) that pack the slit, and interact with each other. At medium Reynolds numbers, both effects are reduced, so $Q$ detects that vorticity is unimportant.

The two origins of vorticity mean that their consequences on the flow will be different. This is captured through the pressure fluctuations shown in the bottom right panel. For low Reynolds number, the flow regions inside and outside of the slit behave in a similar manner, with a sharp drop of fluctuations as the distance to the plate increases. However, at high Reynolds number, the pressure fluctuations do not drop sharply and can even attain a relatively flat spatial profile, similar to what was seen in Figure \ref{fig:stats_pr_platesize}. This confirms the fact that the force increases seen at low Reynolds number are caused by different mechanisms from the turbulent fluctuation force, and as such we do not explore them further. In addition, these statistics also confirm that a main driver behind the increased attractive force with increasing $Re_\lambda$ is the larger importance of the pressure drops due to vortex stretching.

\begin{figure}
    \centering
    \includegraphics[width=.45\textwidth]{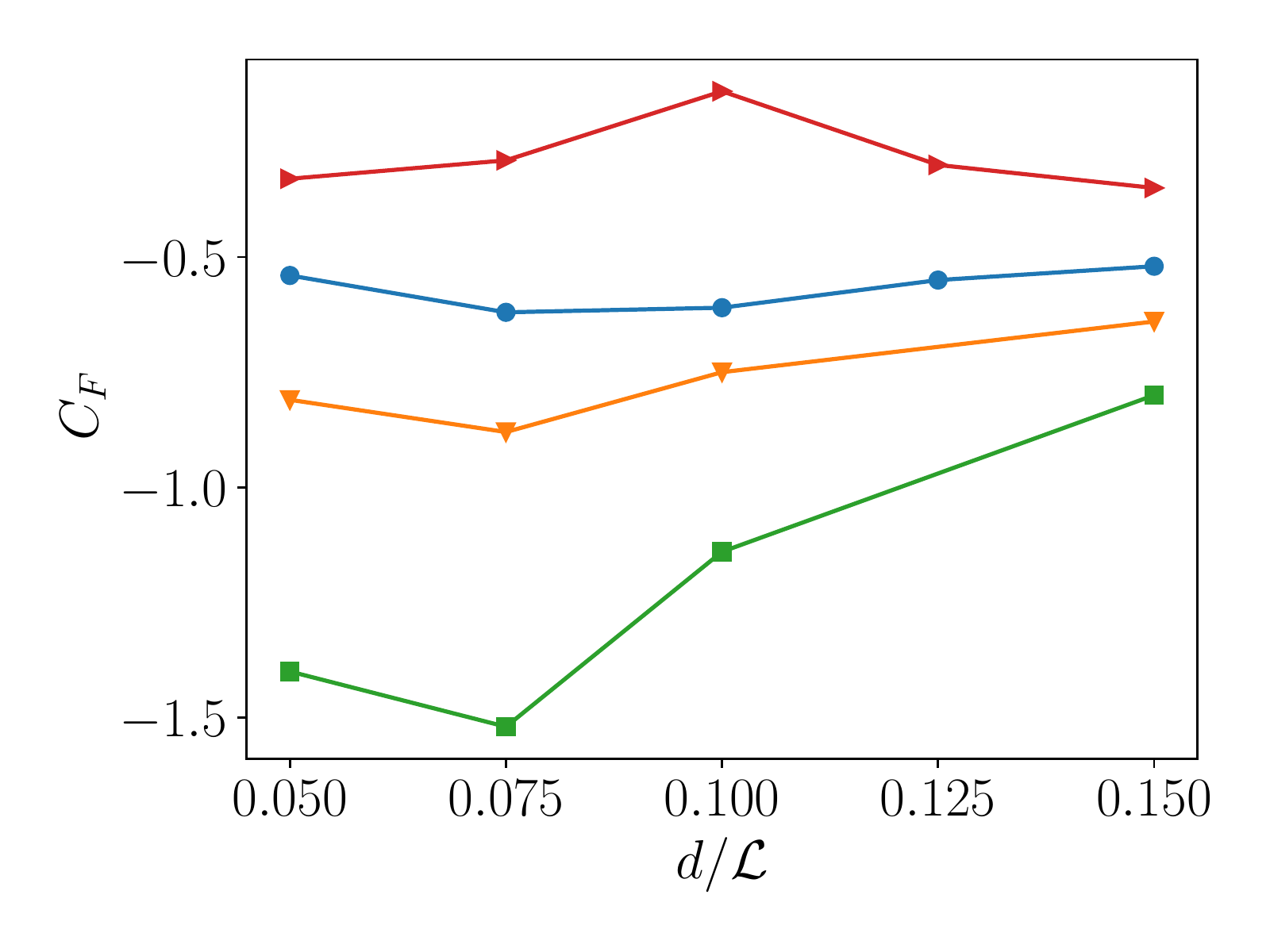}
    \includegraphics[width=.45\textwidth]{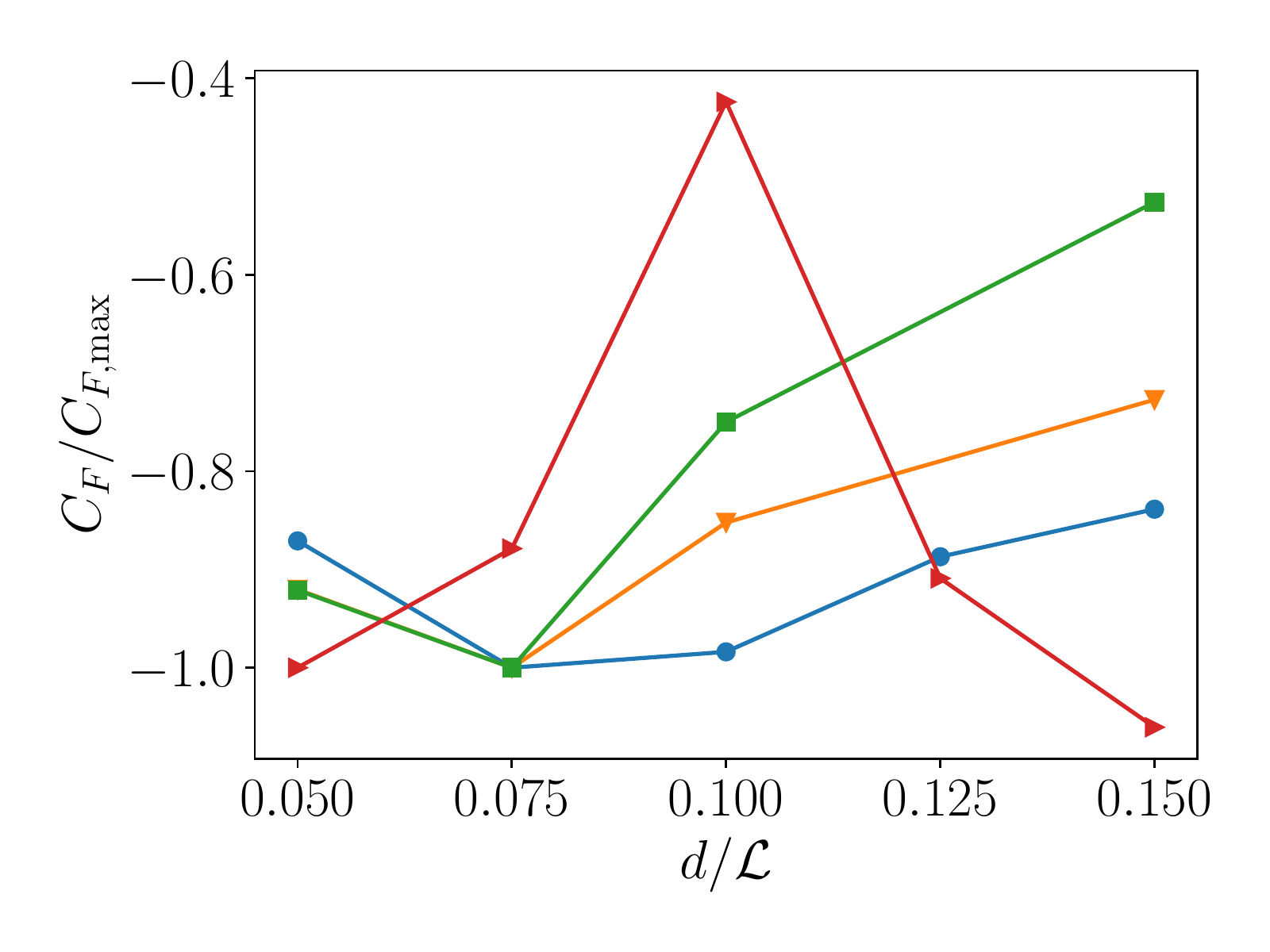}    \includegraphics[width=.32\textwidth]{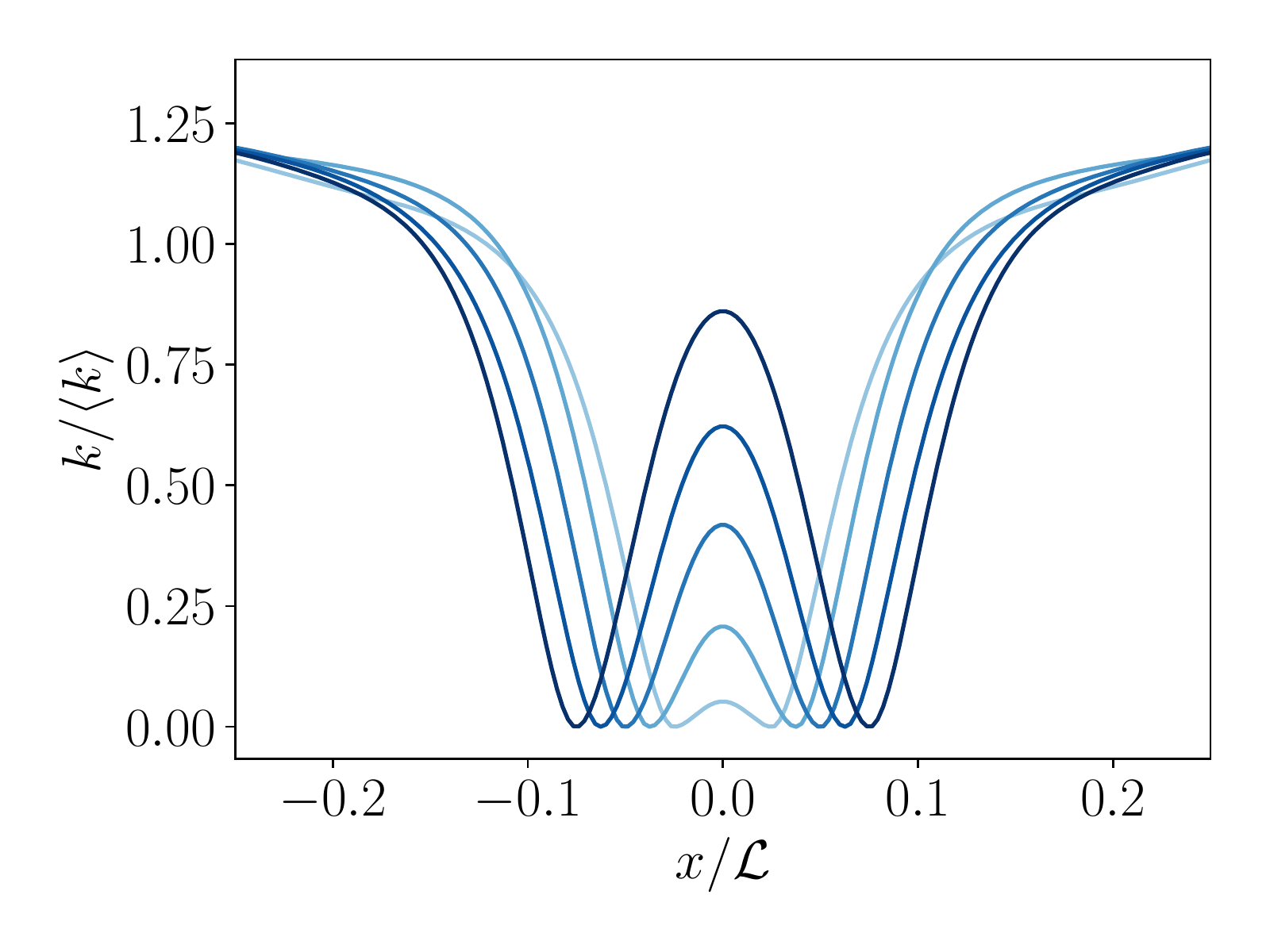}
    \includegraphics[width=.32\textwidth]{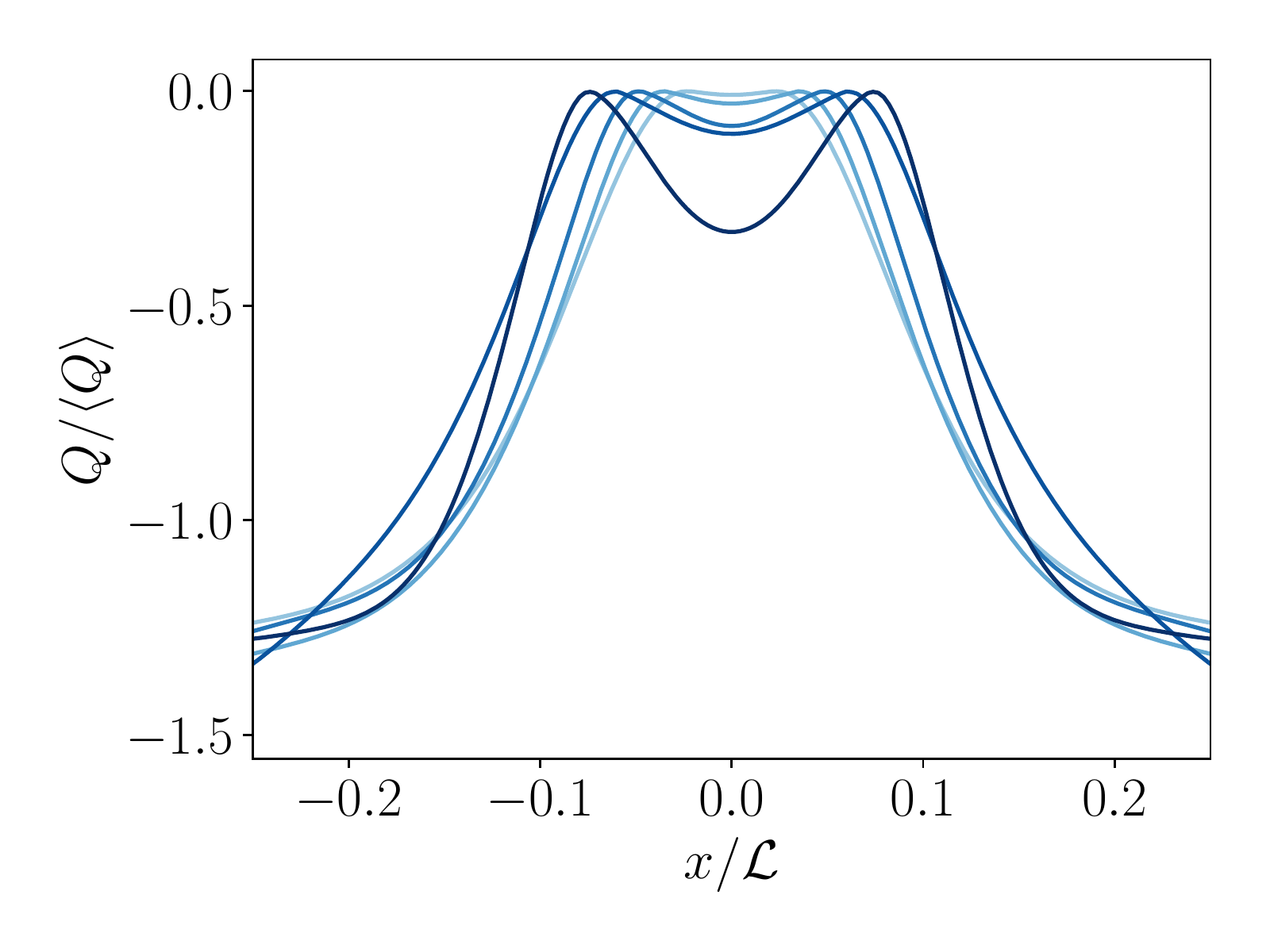}
    \includegraphics[width=.32\textwidth]{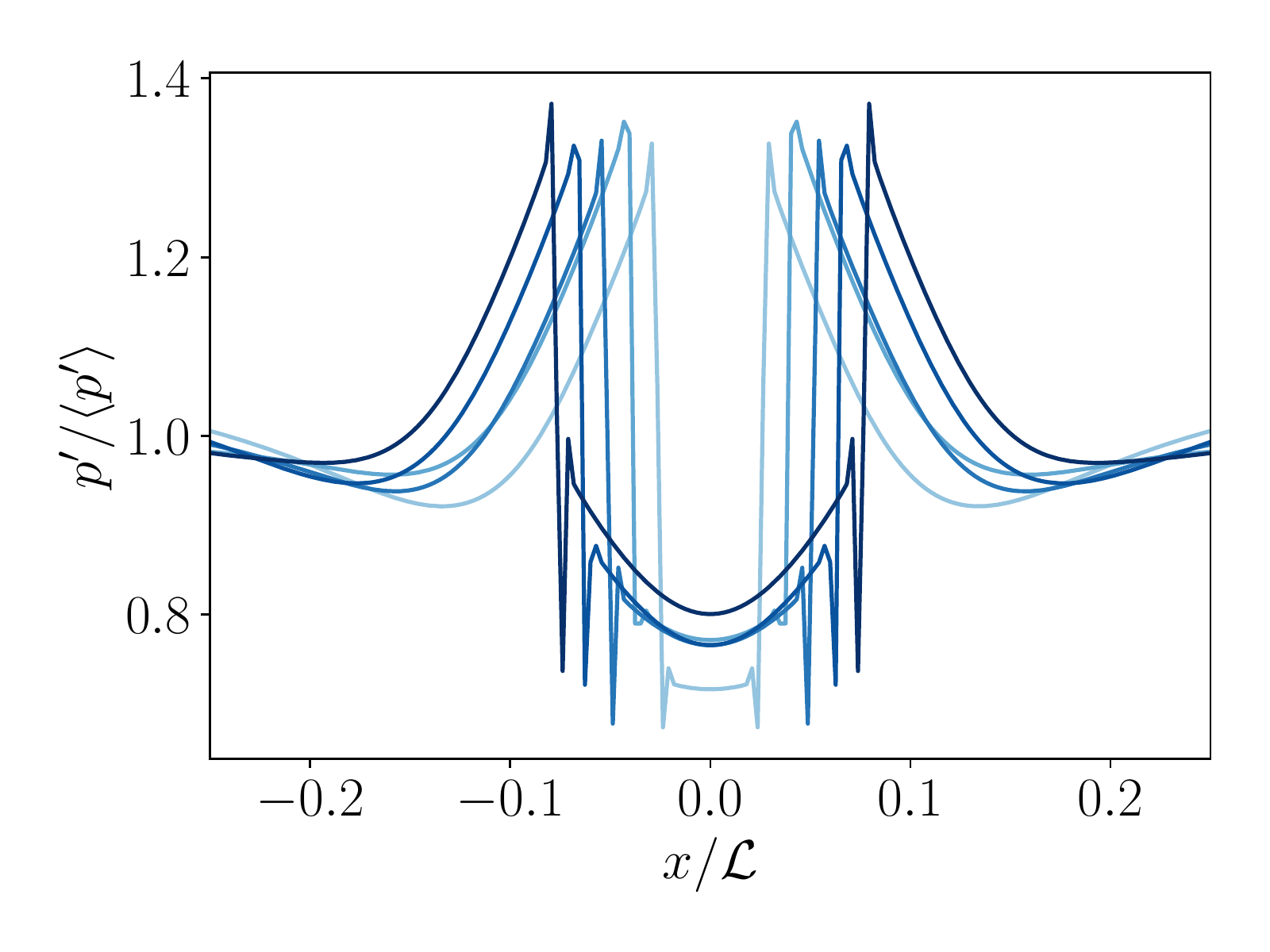}
    \caption{Top panels: Non-dimensionalized force coefficient as a function of $d/\mathcal{L}$ for $l_p/\mathcal{L}=0.25$ and $Re_\lambda=15$ (red left-facing triangle), $Re_\lambda=65$ (blue circle), $Re_\lambda=100$ (orange down-facing triangle), and $Re_\lambda=140$ (green square). Bottom panels: Spatial distribution of the averaged Kinetic energy (left), pressure source term (middle) and pressure fluctuations (right) for varying $d/\mathcal{L}$ with $l_p/\mathcal{L}=0.25$ and $Re_\lambda=15$.  The lines are coded from light to dark blue representing increasing plate separation.}
    \label{fig:cf_re15}
\end{figure}

Finally, to further prove that the non-monotonic behaviour of the force is caused by the vortex stretching, we run one additional set of simulations which set $Re_\lambda=15$, corresponding to the minimum attractive force in Figure \ref{fig:cf_rel}, fix the plate size to $l_p/\mathcal{L}=0.25$, so that data from Ref.~\cite{spandan2020fluctuation} can be used for comparison purposes, and vary $d/\mathcal{L}$. In the top-left panel of Figure \ref{fig:cf_re15} we show the behaviour of $C_F$ as a function of plate distance for the new simulation cases, as well as those cases from Ref.~\cite{spandan2020fluctuation}. We can observe that the force has lost its maximum at $d/\mathcal{L}=0.075$, and has even inverted its non-monotonic character. This is emphasized in the top-right panel, where we show the force coefficient normalized by the maximum attractive force at that $Re_\lambda$. The position of the maximum (now minimum) attractive force has also changed, and is now located at $d/\mathcal{L}=0.15$, and the attractive force drops to less than half its value at $d/\mathcal{L}=0.05$. As mentioned above, we attribute this to the loss of vortex stretching, which is not present for $Re_\lambda=15$. However, an attractive force remains due to the exclusion of energy-containing structures from the mid-gap. 

To confirm this hypothesis, we show the flow statistics in the other three panels of Figure \ref{fig:cf_rel}. We do not see any surprising behaviour for the kinetic energy: it is lower in the slit than outside, and the drop is much larger than in the cases with higher Reynolds number. As could be expected, the kinetic energy in the slit also increases with plate separation. This difference in energy inside and outside is the probable origin of the force at this $Re_\lambda$. The pressure source term behaves according to the earlier discussion: it is most positive in the slit for small separations, which is where the boundary layer dominates, and becomes more negative as the plate separation is increased. Finally, the pressure fluctuations are reduced in the slit, and they do not show the characteristics present at high $Re_\lambda$ shown in the earlier figures, which were due to vortex stretching. Instead, they show a minimum at the mid-gap. 

These results confirm that the packing of vortical structures is non-existent at this $Re_\lambda$. As this $Re_\lambda$ does not show a sharp increase in the attractive force, this confirms that the mechanism behind the increase in the attractive force at plate separations of the order $d/\mathcal{L}=0.1$ is the enhancement of vortex stretching in the slit due to the packing of intense vortical structures. The reason for the increase of the force for $d/\mathcal{L}>0.1$ is unclear, and could be either due to the complex behaviour of the energy in the slit, or due to the character of the forcing at $Re_\lambda=15$. We wish to emphasize that the forcing method was already ruled out as a possible driver of the force at $Re_\lambda=100$ in Ref.~\cite{spandan2020fluctuation}, but the details of the forcing or its anisotropic energy injection, could become more important at low $Re$. 

\section{Conclusions and Outlook}
\label{sec:conc}

We have conducted a set of additional simulations to explore the origins of the turbulent fluctuation force. Earlier work had hypothesized that there were two driving mechanisms: energy exclusion and vortex packing \cite{spandan2020fluctuation}. The force was shown to be increasing with Reynolds number in the range $Re_\lambda\in(65,140)$, and to robustly show a maximum at $d/\mathcal{L}=0.075$. However, in that study there was no quantification of the effect of plate size, and the question of what happens when the flow becomes less and less turbulent was left unanswered. 

To answer these questions, we first set $Re_\lambda=100$, and varied the plate size and plate separation. This showed that even if a fluctuation force was generated by placing the smallest plates in the flow, a minimum plate size is required for the force to show non-monotonicity. We find that the small plates show different qualitative behaviour in the statistics related to vorticity, which hints at the fact that vorticity is the crucial driver of non-monotonicity, and even force generation. We also show that the qualitative behaviour of flow statistics related to energy statistics do not show any significant dependence on plate size.

A second set of simulations probed the fluctuation force at progressively lower Reynolds numbers. By doing this, we make one of the mechanisms hypothesized to generate the turbulent fluctuation force disappear: the enhancement of vortex stretching through the packing of vortical structures. We found that if we hold the plate separation constant, as Reynolds number is lowered the force tends to decrease, until a minimum was reached at $Re_\lambda=15$. The pressure statistics confirm that vortex stretching is absent. Furthermore, by conducting simulations with varying plate separation at this $Re_\lambda$, we show that the force maximum at $d/\mathcal{L}$ disappears, and instead a force minimum appears at intermediate plate distances. This further corroborates the hypothesis of Ref.~\cite{spandan2020fluctuation} that there are two mechanisms generating the force: energy exclusion and vortex packing.

The precise origins of the force at low Reynolds numbers was left unanswered, including the question of why the force increases again at $Re_\lambda=7.6$ in Figure \ref{fig:cf_rel}(a), or why the force shows a minimum instead of a maximum at $Re_\lambda=15$ for intermediate plate distances. There could be several causes for this: for $Re_\lambda=7.6$, the result is still within error bars, or it could be due to the increasing importance of the forcing and the way it injects energy in the mid-gap. The force minimum could also be due to finite-plate size effects. We also note that other fluctuation forces have shown complicated dependencies on the energy distribution \cite{ni2015}. 

Another question left unanswered is what happens in the limit $Re_\lambda\to\infty$. The data shown in this manuscript, and Ref.~\cite{spandan2020fluctuation} which reached $Re_\lambda=140$ showed a force that continuously increased with $Re_\lambda$. While it appears reasonable to hypothesize that $C_F$ will eventually saturate, there is no clear answer to the questions of at which $Re_\lambda$ this happens, what is the asymptotic value of $C_F$, and why does this saturation take place. Further simulations, or more importantly, three-dimensional experiments, could help address this gap.

\textit{Acknowledgments:} We acknowledge the Research Computing Data Core, RCDC, at the University of Houston for providing us with computational resources and technical support. We are also indebted to Wouter Bos for remarks regarding the Extended Navier-Stokes equations which are presented in Appendix \ref{sec:app}.

\bibliography{HIT_bib}

\appendix 

\section{Active flows and the Extended Navier-Stokes equations}
\label{sec:app}

Active turbulence generated by microbial suspensions can be modelled using a Navier-Stokes equation with higher order gradient terms in the stress tensor and a scaled non-linear term \cite{wensink_meso-scale_2012,dunkel2013}. These equations are also known as the extended Navier-Stokes equations (ENSE), and have been found to reproduce some characteristics of microbial suspensions \cite{dunkel2013}. The ENSE show non-trivial energy distributions so one could expect that Casimir forces can be found in these equations. Furthermore, as these modified Navier-Stokes equations solve for the same variables, namely velocity and pressure, it appears simple to directly compare the forces in this situation to those of hydrodynamic turbulence, building a bridge between hydrodynamic forces and active matter fluctuation forces \cite{ni2015}. However, studying this proved beyond our means after some consideration. In this Appendix we explain why.

The ENS originate from  substituting the Navier-Stokes momentum equation  (\ref{eq:ns}) with an equation that originates from an extension of the Toner-Tu equations \cite{toner1995long}. The Toner-Tu were proposed as a continuum-model to understand the dynamics of large flocks of animals by generalizing existing discrete models \cite{vicsek1995}.

Following the formulation of Ref.~\cite{dunkel2013}, this equation reads: 

\begin{equation}
\displaystyle\frac{\partial\textbf{u}}{\partial t} + \lambda_0 \textbf{u}\cdot \nabla \boldsymbol{u} = -\nabla p + \lambda_1 \nabla \boldsymbol{u}^2-\beta ( \boldsymbol{u}^2-u_0^2) \boldsymbol{u} +\Gamma_0\nabla^2 \boldsymbol{u}+ \Gamma_2 (\nabla^2 )^2\boldsymbol{u}
 \label{eq:ens}
\end{equation}

\noindent where $\beta$, $\Gamma_0$, $\Gamma_2$, $\lambda_0$, $\lambda_1$ and $u_0$ are parameters whose choice can be informed physically \cite{dunkel2013}. Equation \ref{eq:ens} reduces to the Navier-Stokes equation if $\lambda_0=1$, $\beta=\lambda_1=\Gamma_2=0$ and $\Gamma_0>0$. 

The second-order term $\Gamma_0\nabla^2\textbf{u}$ can be identified with a viscous term, $\nu$ being equivalent to $\Gamma_0$. Flow forcing comes from two terms: the $\lambda_1\nabla\textbf{u}^2$, term which represents an active pressure field, and the fourth-order Swift-Hohenberg (SH) \cite{swift1977hydrodynamic} term $\Gamma_2\nabla^4\textbf{u}$,  which is designed to act as a forcing mechanism and models the bacterial forcing at intermediate and small scales. Unlike the methods usually used to force HIT, which predominantly force the large length-scales in the flow, the SH forcing steeply increases with decreasing wavenumber due to the presence of the fourth-order gradient operator. 
The two forcing mechanisms should produce a very different flow from that achieved by the random forcing used in this manuscript. However, both the active pressure field forcing, proportional to gradients of the velocity squared, and the SH term, proportional to velocity gradients, will interact with the immersed boundary method in ways we cannot adequately control. It is known that forcing methods that force a flow with a magnitude that is proportional to velocity to produce artifacts when coupled with IBM methods \cite{chouippe2015forcing}. Because we do not have experiments to compare to, the introduction of these types of forcing in our problem would make it impossible to distinguishing what is a numerical artifact and what is a physical product of these active-type forcing. 

This leaves for our consideration two more terms: the  $\beta(\textbf{u}^2-u_0^2)\textbf{u}$ term, which is meant to model the fact that if all bacteria would move in the same direction, they would achieve a collective speed $u_0$, which appears uninteresting for our purposes, and the scaled non-linear term, $\lambda_0\textbf{u}\cdot\nabla\textbf{u}$, which is meant to account for the fact that organisms such as birds or bacteria are moving through a resisting medium. The scaled non-linear term appears to be interesting enough to warrant further exploration, as in the active matter literature $\lambda_0$ is understood to be an important parameter which breaks the Galillean invariance of the equations representing the fact that bacteria need to spend energy to move. If we limit our study of the ENS to just including a scaled non-linear term in the Navier-Stokes equation, leaving the other terms as-is, the momentum equation takes the form:

\begin{equation}
     \displaystyle\frac{\partial\textbf{u}}{\partial t} + \lambda_0\textbf{u}\cdot\nabla\textbf{u} = -\rho^{-1}\nabla p +\nu\nabla^2 \textbf{u}+ \textbf{f} 
 \label{eq:ns-mod}
\end{equation}

\noindent However, no new physics is contained in this equation. Even if Eq.~\ref{eq:ns-mod} is not Galillean invariant, the Navier-Stokes equation can be recovered by re-scaling the length and time scales of the equations. Under this rescaling, the effective viscosity is reduced if $\lambda_0>1$, and it is increased if $\lambda_0<1$. Exploring the effect of $\lambda_0$ on the fluctuation force by simulating Eq.~\ref{eq:ns-mod}, where the \emph{only} modification is scaling the non-linear term, is akin to studying the same flow at a different $Re$, and this has been done in the manuscript. 

Because of the reasons above, we have chosen to limit ourselves to the study of the ``ordinary'' Navier-Stokes equations in this manuscript.

\end{document}